\documentclass[aps,prb,reprint,longbibliography]{revtex4-2}
\usepackage[T1]{fontenc}
\usepackage[english]{babel}
\usepackage{amsmath}
\usepackage{braket}
\usepackage{graphicx}
\usepackage{float}
\usepackage[colorlinks, urlcolor=blue, citecolor=blue, linkcolor=blue, pdfstartview=FitH]{hyperref}
\usepackage[all]{hypcap}
\usepackage{dcolumn}
\usepackage{bm}
\usepackage{epstopdf}
\usepackage{color}
\usepackage{xcolor}
\usepackage[normalem]{ulem}
\usepackage[toc,page,titletoc]{appendix}

\begin{document}

\def\affiIOFFE{Ioffe Institute, 194021 St.~Petersburg, Russia}

\title{Theory of optical alignment and orientation of excitons \\ in an ensemble of semiconductor nanoplatelets}

\author{O.~O.~Druzhinina}
\affiliation{\affiIOFFE}
\author{A.~V.~Rodina}
\affiliation{\affiIOFFE}

\date{23.09.2026}

\begin{abstract}
We present the theory of exciton optical alignment and optical orientation effects in a single semiconductor nanoplatelet (NPL) and in an ensemble of NPLs. The theoretical model is developed to take into account the emission from both bright and dark exciton states in the case of strong electron-hole exchange interaction and an isotropic distribution of in-plane orientations of the NPLs' edges with respect to the linear polarization axis. The conditions for the linear optical response regime on the polarized resonant excitation of the bright exciton are considered, and the steady-state and time-dependent expressions for polarized photoluminescence are presented for the low temperature case in the Faraday magnetic field at normal incidence. The symmetry of the obtained  optical response corresponds to the $C_{2}$ point group for the single NPL and the $C_{\infty}$ point group for the NPL ensemble. The bright-to-dark exciton coupling mechanism allowing one to observe the optical alignment of the dark exciton is considered. The example calculations of the effects are presented for the parameters close to those of CdSe/CdS core/shell NPLs. The effect of the fluctuations in the exciton parameters in the ensemble on the polarization dephasing in the transverse effective magnetic field is demonstrated.
\end{abstract}

\maketitle

\section{Introduction}

Semiconductor nanocrystals, since their discovery and up to our days, have been of interest in various fields of chemistry, physics, biology, and medicine, and are successfully used in various devices~\cite{Garcia2021,Efros2021}. Being synthesized from many different semiconductor materials, they can have various geometries, resulting in zero-dimensional quantum dots, one-dimensional nanorods, or two-dimensional nanoplatelets (NPLs). Semiconductor NPLs, first synthesized in 2008~\cite{Ithurria2008}, are attractive for fundamental optical studies as a model system to study physics in two dimensions, similar to quantum well heterostructures and layered materials such as graphene or transition-metal dichalcogenides. The spin and polarization properties of colloidal NPLs differ considerably from those of epitaxial quantum wells and quantum dots~\cite{Bayer2019}. 

Precise control of the NPLs' thickness manifests itself in small inhomogeneous broadening in CdSe NPLs~\cite{Ithurria2008}, as confirmed later for NPLs of different compositions~\cite{Ithurria2012,Nasilowski2016,Berends2017}. Small NPL thickness, along with the strong dielectric contrast between the NPL and the environment, enhances Coulomb interaction, which affects exciton binding energy and fine structure energy splitting, making them considerably larger than in epitaxial quantum wells~\cite{Rodina2016JETP,Shornikova2021nl}. The lowest dark exciton states in NPLs, which are optically forbidden in the dipole approximation, do contribute to the PL because of mixing with the higher-energy bright, optically active exciton states through different symmetry-lowering perturbations~\cite{Rodina2016, Shornikova2018ns, Shornikova2020nn}.  Another feature of the ensemble of NPLs, typical for nanocrystals, is the random orientation of the NPLs in solution or on the substrate, while the orientation of epitaxial quantum dots and quantum wells, as a rule, is strictly connected to the crystal lattice directions. There are always nanocrystals in the randomly oriented ensemble in which the external magnetic field additionally mixes bright and dark states, enhancing the radiative emission from the dark exciton and affecting the polarized photoluminescence (PL) from the ensemble~\cite{EfrosCh3}. 

These features make polarized PL an important tool to access the exciton fine structure and spin dynamics. Experimental techniques widely used in the spin physics of epitaxial heterostructures can be readily applied to colloidal quantum dots and NPLs. Among them are polarized PL spectroscopy methods, including optical orientation and optical alignment~\cite{ZakharchenyaBook}. 

Optical orientation consists of transferring angular momentum to the system by means of its excitation with circularly polarized light and manifests itself in the creation of unequal populations of carriers or exciton states with different projections of angular momentum. Optical alignment, in contrast, is produced by linearly polarized excitation and corresponds to the excitation of exciton states with a certain direction of oscillating dipole moment~\cite{Pikus_book1982}.
Both of these effects, as well as the conversion between linear and circular polarization, were widely presented and studied in bulk semiconductors and epitaxial heterostructures~\cite{Dzhioev1997PRB,Kusrayev2008,Koudinov2008,Shamirzaev2023}. These experimental and theoretical investigations significantly contributed to the characterization of the spin structure and spin relaxation of the excitations in semiconductors~\cite{Ekimov1971_2, Gamarts1977, Dymnikov1981,Dzhioev1973,Permogorov1983, Dzhioev1997PRB,Verbin2002}.

The theoretical description of exciton optical orientation and alignment by means of the density matrix has been developed for bulk semiconductors~\cite{Pikus_book1982, Ivchenko1977_2}, where only the radiation from the optically active (bright) exciton states has been taken into account. For low-dimensional systems realized in semiconductor heterostructures, the classical models have also mainly focused on the bright excitons formed with heavy holes in superlattices and quantum dots~\cite{Dzhioev1997PRB,Dzhioev1998,Astakhov2006,Kusrayev2008}, and more recently on indirect excitons, where the hyperfine interaction of the charge carrier and nuclear spins can be comparable to the fine structure splittings induced by the exchange interaction \cite{Smirnov2023,Smirnov2024}. 

Recently, the exciton orientation and alignment have been studied in emergent semiconducting systems: bulk perovskites~\cite{Kopteva2024,Zhiliakov2026}, colloidal perovskite nanocrystals~\cite{Nestoklon2018}, and 2D materials~\cite {Wang_2015,Yagodkin2025}.
However, the exciton fine structure of these systems is qualitatively different from that of A$_2$B$_6$ NPLs, particularly for the CdSe-based NPLs. 

Usually, large fine structure splittings and fast energy relaxation from bright to dark excitons in CdSe-based nanocrystals and NPLs prevent the experimental observation of exciton optical orientation and optical alignment. However, the so-called ``polarization memory effect''~\cite{Rodina2016JETP,EfrosCh3} or ``photoselection''~\cite{Lakowicz2006}, when the specifically oriented subensemble is excited by linearly polarized light and the same subensemble emits linearly polarized light, was observed for ensemble of colloidal nanocrystals. Recently, our group has reported the observation of exciton optical alignment and optical orientation effects in the ensemble of colloidal core-shell CdSe/CdS NPLs~\cite{Smirnova2023,Smirnova2025}. Surprisingly, not only were the optical orientation and optical alignment of bright excitons detected in the Faraday magnetic field under continuous-wave (CW) excitation~\cite{Smirnova2023}, but the optical alignment of dark excitons was also suggested and further confirmed by studying the time-resolved optical alignment dynamics under pulsed excitation~\cite{Smirnova2025}. To analyze the experiments, we developed a theoretical model in Refs.~\cite{Smirnova2023,Smirnova2025} that takes into account the lowest bright and dark exciton levels and describes the density matrix in terms of two interacting pseudospins. The experimental data analysis allowed us to estimate the energy splittings of the bright and dark exciton pairs of states in a zero magnetic field, their $g$-factors, and exciton pseudospin lifetimes within the framework of the considered model~\cite{Smirnova2023,Smirnova2025}. While the main results in Ref. ~\cite{Smirnova2023} were obtained in the approximation of isotropic pseudospin relaxation times, the time-resolved studies in Ref.~\cite{Smirnova2025} indicated their strong anisotropy.

In this paper, we extend the previous work and provide a general theoretical consideration of the polarized exciton PL in zero and external magnetic field applied in the Faraday geometry from both the bright and dark exciton states under polarized optical excitation from individual CdSe-based NPLs and from an ensemble of NPLs. We consider bright and dark excitons in terms of two interacting pseudospins and take into account the difference between longitudinal and transverse pseudospin relaxation times. We  account for the in-plane anisotropy of individual NPLs, resulting in the exchange splitting of the bright states~\cite{Goupalov1998_2} and the cubically symmetric short-range exchange interaction, resulting in the splitting of the dark exciton states~\cite{Ivchenko1995book,Semina2026arXive}. We microscopically consider the transfer of all three pseudospin components between the bright and dark exciton states and identify the conditions for the observation of the dark exciton contribution to the optical alignment on the basis of our model. Finally, we obtain the total polarization response under CW and pulsed excitation with polarized light. The model highlights the specifics of colloidal systems that include dark exciton radiative recombination, considerable bright-dark splitting that can exceed thermal energy, and exciton spin relaxation as a whole, rather than the individual electron and hole spin relaxation.

\begin{figure*}[ht]
	\centering
	\includegraphics[width=0.8\textwidth]{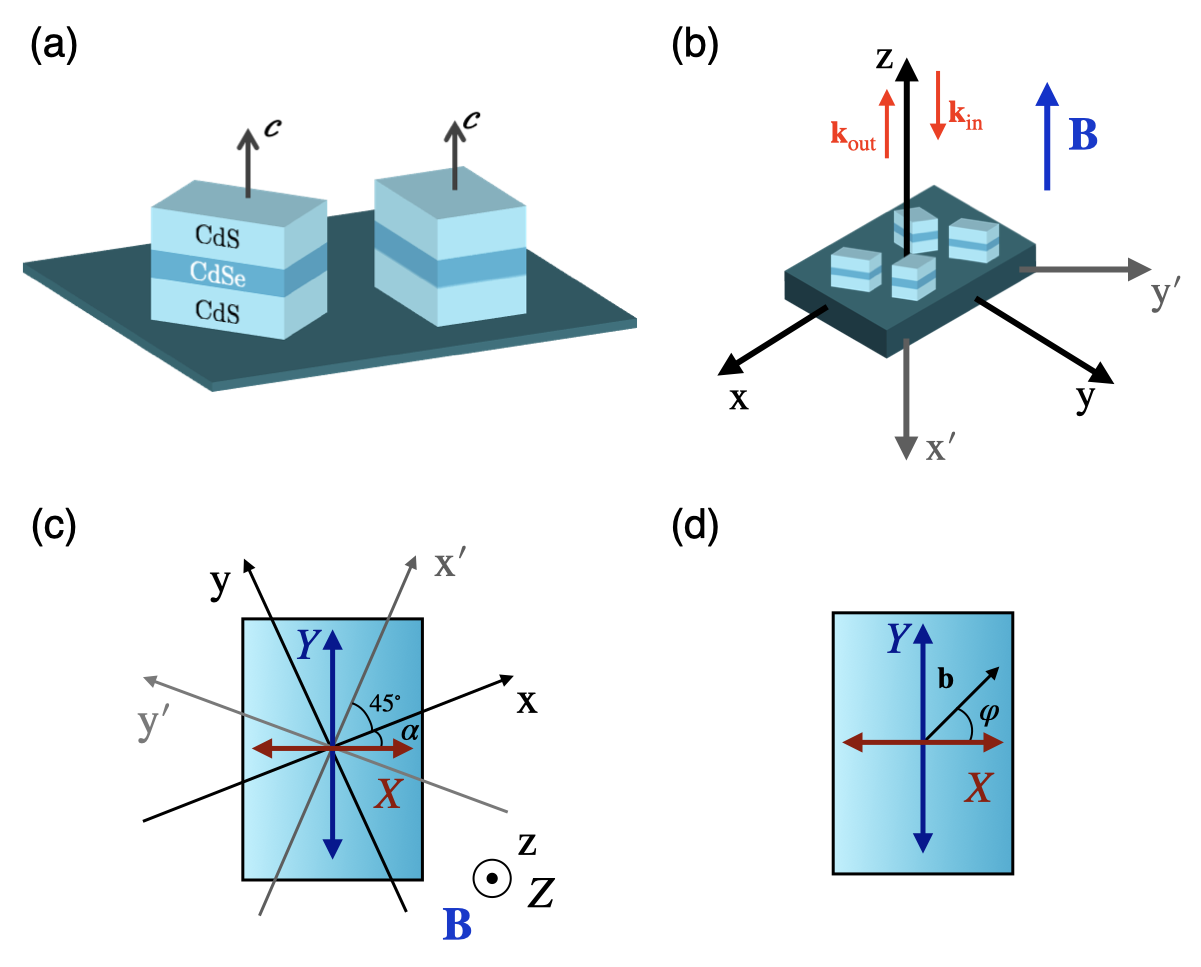}
	\caption{\label{fig:fig1} (a) Horizontally oriented NPLs  with the c-axis perpendicular to the substrate. (b) An ensemble of NPLs in the experimental geometry: $k_{\rm in}$ and $k_{\rm out}$ are the wave vectors of the incident and registered light and ${\bm B}$ is the external magnetic field directed along the $z$-axis of the laboratory frame and parallel to the c-axis. Linear polarization is registered in the axes $x,~y$ and rotated by $45^{\circ}$ around $z$-axis $x',~y'$ axes. (c) A scheme of a single NPL with its own axes  $Z \parallel z$ and  $X,~Y$ rotated by the angle $-\alpha$ with respect to $x,y$ around $z$. (d) A scheme of a single NPL with internal in-plane magnetic field $\bm b$ directed by angle  $\varphi$ with respect to the $X$-axis.}
\end{figure*}

More specifically, we assume all the NPLs lie horizontally on the substrate, with the anisotropy axis (c-axis) perpendicular to the substrate plane. Meanwhile, the in-plane orientation of  NPL edges is random (see Fig.~\ref{fig:fig1}(a)). We analyze the influence of isotropic in-plane orientation of NPLs on the substrate in the ensemble, as well as the fluctuations of the exciton parameters caused by the NPL lateral size distribution.  We also discuss the symmetry of a single NPL and the ensemble response in the Faraday magnetic field for light propagating along the anisotropy axis (Fig.~\ref{fig:fig1}(b)). 

The polarization of light propagating along the $z$ direction in the laboratory coordinate frame is characterized by three Stokes parameters. Two of them, $P_l$ and $P_{l'}$, correspond to the linear polarization degree in the laboratory $x$, $y$ axes and in the $x'$, $y'$ axes rotated by 45$^{\circ}$ degrees around the $z$-axis, respectively (see Fig.~\ref{fig:fig1}(b,c)). The third one, $P_c$, characterizes the circular polarization degree. For the light with the electric field amplitude ${\bm E}=E {\bm e}$ and the complex unit polarization vector ${\bm e}$, the three Stokes parameters are defined as:
	\begin{eqnarray} 
		P_{l} = \frac{I_x-I_y}{I_x+I_y} = |e_x|^2-|e_y|^2  \, ,  \nonumber\\
		P_{l'} = \frac{I_{x'}-I_{y'}}{I_{x'}+I_{y'}} = e_xe_y^*+e_x^*e_y \, ,   \\
		P_c = \frac{I_+-I_-}{I_++I_-} =i(e_xe_y^*-e_x^*e_y) \, . \nonumber \label{Stokes}
\end{eqnarray}
Here $I_{x(y)}$ is the intensity of the horizontally (vertically) polarized component, $I_{x'(y')}$ is the intensity of the 45$^{\circ}$ ($-45^{\circ}$) polarized component, and $I_{+(-)}$ is the intensity of $\sigma^+$ ($\sigma^-$) polarized component of light.  In the absence of both dichroism in the absorption of linearly or circularly polarized light, as well as spin-dependence in the exciton relaxation or recombination (see details in~\ref{symmetry} and \ref{dipoles}), the total emission intensity $I=I_x+I_y=I_{x'}+I_{y'}=I_++I_-=|E_x|^2+|E_y|^2=|E|^2$ is independent of the polarization of the exciting light. In this case, it is enough  to work with the Stokes vectors on the Poincar\'e sphere  comprising three  parameters in Eq.~\eqref{Stokes}  for both exciting, ${\bm P}_0 = (P^l_0,P^{l'}_0,P^c_0)$, and emitted, ${\bm P} = (P_l,P_{l'},P_c)$, light and relate them  linearly as 
\begin{eqnarray} \label{PLP0}
&&{\bm P} = {\bm P}^{\rm un} + \hat \Lambda
{\bm P}_0 \, , {\rm or}  \\ && {P}_k = P^{\rm un}_k +\sum_{k' = l,l',c} \Lambda_{kk'}P^{k'}_0\, , \quad k= l,l',c \, .  \nonumber
\end{eqnarray}
Here, the matrix $\hat \Lambda$  describes the optical response to the polarized excitation \cite{Dzhioev1997,Smirnov2023}, while the vector ${\bm P}^{\rm un}$  is independent of the polarization of the excitation, and  $|{\bm P}| \le 1$. In the absence of spin-dependent recombination, the term ${\bm P}^{\rm un}$ describes the polarization of the emitted light, which is induced by a thermodynamically equilibrium difference in the population of the exciton states. This term is often omitted, for example, in~\cite{Dzhioev1997,Smirnov2023}, when the optical orientation is considered, but it describes, for example, the magnetic field-induced circular polarization of the PL~\cite{Ivchenko2018Gross} or the linear polarization induced by the anisotropic splitting of the linearly polarized exciton states. The spin relaxation leading to the thermodynamical equilibrium between spin-split states results in an increase of ${\bm P}^{\rm un}$ and a decrease of the optical response that is proportional to the excitation polarization~\cite{ZakharchenyaBook}.

The theory we present  describes analytically ${ P}^{\rm un}_{\alpha}$ and $\Lambda_{\alpha \beta}$ for individual NPL and an ensemble of isotropically oriented NPLs, allowing us to model all experimentally observable effects $P_{\alpha}^{\beta}$ ($\alpha, \beta = c, l,l'$) for both CW and pulsed excitation. Here, as in Ref.~\cite{Smirnova2023}, $P_{\alpha}^{\beta}$ denotes the $\alpha$-polarized emission \textit{via} the $\beta$-polarized excitation $P^{\beta}_0$.  The definition of the circular polarization sign $P_0^c$ and $P_c$ is given by Eq.~\eqref{Stokes} with respect to the same direction of~$z$.

The rest of the paper is organized as follows. In Section~\ref{sec:Theory}, we introduce the theoretical model, allowing the description of the effects under study in a single NPL and in the ensemble of colloidal NPLs.  In Section~\ref{sec:Sol}, we provide the expressions for the bright and dark exciton pseudospins in a Faraday magnetic field for CW and pulsed excitation. Section~\ref{sec:Orientation} is devoted to the polarization of the exciton PL from the ensemble of isotropically oriented NPLs, including both the contributions of the bright and dark excitons. In Section~\ref{sec:Discussion}, we analyze the obtained results and model the magnetic field and time dependences of the effects under study using the exciton parameters close to the ones that were obtained in our previous works~\cite{Smirnova2023,Smirnova2025}. The role of the exciton parameters fluctuations is also discussed. In Section~\ref{sec:conclusions}, we summarize the results and provide the future prospects of the study. The additional details are provided in the following Appendices: symmetry considerations (A), dipole matrix elements (B), exciton populations (C), bright exciton pseudospin components (D), single‑NPL emission polarization (E), and dark exciton contribution (F).

\section{Theoretical model}\label{sec:Theory}

We consider zinc-blende NPLs, for example, CdSe/CdS core/shell NPLs, with the c-axis (Fig.~\ref{fig:fig1}) along the $[001]$ crystallographic direction. Their edges can be of different lengths and be directed either along $[100]$ and $[010]$ or along $[110]$ and $[1\overline{1}0]$~\cite{Yoon2021,Feng2020}.  For the sake of clarity, we consider the NPL with edges along $[100]$ and $[010]$ crystallographic directions. We choose the NPL-related coordinate frame with axes $X$ and $Y$ along the short and long NPL edges, respectively, which are rotated by the angle $-\alpha$ with respect to the laboratory axes $x$ and $y$. The $Z$-axis of the NPL frame is directed along the c-axis and coincides with the laboratory axis $z$, as shown in Fig.~\ref{fig:fig1}(b).  We consider the normal incidence of the exciting light on the sample with ${\bm k} \parallel z $ and the reverse direction of the detected light (Fig.~\ref{fig:fig1}(b)), so that the vector of the light polarization ${\bm e} = (e_x,e_y,0)$ always lays in the NPL plane. Then, the Stokes parameters of the light emitted by a single horizontally oriented NPL with $\bm k \parallel Z \parallel z$: 
\begin{eqnarray} \label{key_all}
	P_{L} = |e_X|^2-|e_Y|^2 \, , \nonumber \\
	P_{L'} = e_Xe_Y^*+e_X^*e_Y ,  \\
    P_{C} = i(e_Xe_Y^*-e_X^*e_Y) \, . \nonumber
\end{eqnarray}
Here, $e_X$ and $e_Y$ are the projections of the emitted light polarization vector ${\bm e}$ onto the $X,Y$ axes. In the laboratory frame, the registered circular polarization, $P_{c} = P_{C}$ (as well as the total intensity,  $I$), does not depend on the NPL orientation. In contrast, the linear polarization depends on the orientation of the NPL, which is rotated around the laboratory axis $z$ by an angle $-\alpha$ (Fig.~\ref{fig:fig1}(c)), and can be found as follows
\begin{gather}
P_{l}(\alpha) = P_{L} \cos (2\alpha) + P_{L'} \sin (2\alpha)\, ,  \label{Plangle}  \\
P_{l'}(\alpha) = -P_{L} \sin (2\alpha) + P_{L'} \cos (2\alpha)\, . \nonumber 
\end{gather}

In the following consideration, the external magnetic field is applied in the Faraday geometry ${\bm B} \parallel {\bm k} \parallel \bm c$.

\subsection{Bright and dark exciton contributions \\ to the PL polarization}\label{sec:Total polarization}

\begin{figure*}[ht]
	%\centering
	\includegraphics[width=0.9\linewidth]{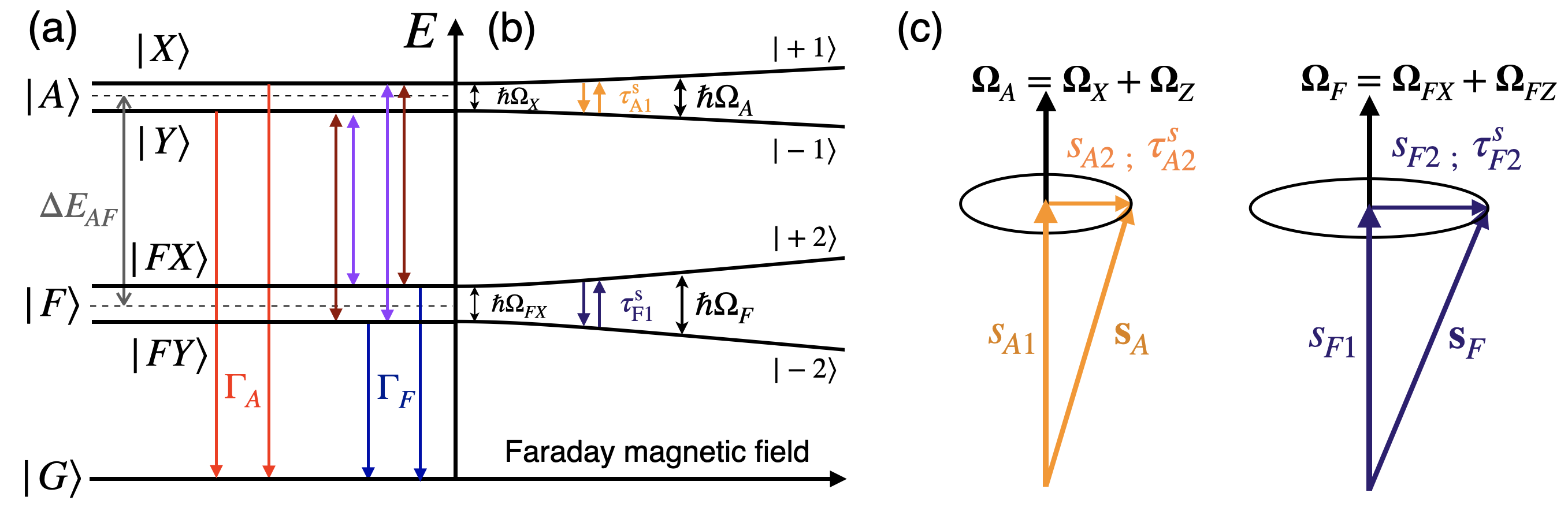}
	\caption{\label{fig:fig2} (a) A scheme of the  bright $ |A\rangle$ and dark $|F\rangle$ exciton levels split by $\Delta E_{AF}$  in zero magnetic field, with the possible  transitions between them. $\Gamma_A$ and $\Gamma_F$ denote the bright and dark exciton recombination rates.  (b) Schematic evolution of the exciton states in the Faraday magnetic field from the states $| X \rangle, |Y\rangle$ of the bright exciton (split by the energy $\hbar \Omega_X$ in zero field) and states $| FX \rangle, |FY\rangle$ of the dark exciton (split by the energy $\hbar \Omega_{FX}$) into  states $| +1 \rangle, |-1\rangle$ (split by the energy $\hbar \Omega_A$) and $|+2 \rangle, |-2\rangle$ (split by the energy $\hbar \Omega_F$), respectively, in the case of $g_A>0$ and $g_F>0$. Orange (violet) arrows show the longitudinal spin relaxation between split states of the bright (dark) exciton with characteristic times $\tau_{A1}^{\rm s}$ ($\tau_{F1}^{\rm s}$). (c) Schematic  precession of the bright and dark exciton pseudospins around effective magnetic fields characterized by the frequencies $\bm \Omega_A$ and $\bm \Omega_F$, respectively. These precessions lead to transverse spin relaxation (dephasing) with characteristic times $\tau_{A2}^{\rm s}$ and $\tau_{F2}^{\rm s}$.  }
\end{figure*}

The bright, optically active ($|A\rangle$), and dark, optically forbidden ($|F\rangle$), excitons in CdSe-based NPLs are formed from electrons with the spin projection $s_Z = \pm 1/2$ and heavy holes with the spin projection $j_Z= \pm 3/2$ on the c-axis.
In the absence of an external magnetic field and any anisotropy-related splittings, both exciton states are two-fold degenerate with respect to the total projections  $m_A=s_Z + j_Z= \pm 1$ and $m_F=s_Z + j_Z= \pm 2$ and are described by the spin wave functions $\Psi_{\pm 1}$ and $\Psi_{\pm 2}$, respectively. The isotropic electron-hole exchange interaction splits $|A\rangle$ and $|F\rangle$ exciton states by the energy $\Delta E_{AF}$ (see Fig.~\ref{fig:fig2}(a)).

The long-range exchange interaction, together with the in-plane NPL anisotropy, splits~\cite{Goupalov1998, Hu2018,Swift2024} the bright $|\pm1 \rangle$ states into linearly polarized states $|X \rangle$ and $| Y \rangle$  by the energy  $\hbar \Omega_X$ in zero magnetic field. In this case, the lower state has the oscillating dipole moment along the longest NPL edge, which is the state  $|Y\rangle$ in our model. The dark exciton $|\pm2 \rangle $ states are split by the energy $\hbar \Omega_{FX} $ due to the cubically-symmetric short-range electron-hole exchange interaction $\sim~(\hbar \Omega_{FX}/3)\sum_{\alpha=X,Y,Z} \sigma_{\alpha} J_{\alpha}^3$, where $\bm \sigma$ is a pseudovector composed of Pauli matrices and $\bm J$ is a pseudovector of matrices of the angular momentum $3/2$~\cite{Ivchenko1995book}. The resulting states~\cite{Semina2026arXive} are denoted as $| FX \rangle, | FY \rangle$ in Fig.~\ref{fig:fig2}(a). 

The exciton states are described by the wave functions $\psi_{\gamma}({\bm r}_e,{\bm r}_h) = \Psi_{\gamma}$ $\Phi_{\rm ex}({\bm r}_e,{\bm r}_h)$, with the spin exciton wave functions given by $\Psi_{\gamma}$ ($\gamma = \pm1, \pm2$) and
\begin{eqnarray} \label{psiA}
&&\Psi_{X} = \frac{\Psi_{+1}+\Psi_{-1}}{\sqrt{2}} \, , \quad \Psi_{Y} = -i \frac{\Psi_{+1}-\Psi_{-1}}{\sqrt{2}} \, ,\nonumber  \\  
&&\Psi_{FX} = \frac{\Psi_{+2}+\Psi_{-2}}{\sqrt{2}} \, , \quad \Psi_{FY} = -i \frac{\Psi_{+2}-\Psi_{-2}}{\sqrt{2}}  \, ,\\ \nonumber
\end{eqnarray}
where $\Phi_{\rm ex}({\bm r}_e,{\bm r}_h)$ is the exciton envelope wave function, assumed to be the same for all exciton states~\cite{Rodina2020}.
Note that the functions $\Psi_{FX}$ and $\Psi_{FY}$ transform as $X^2 - Y^2$ and $2 XY$, respectively.

In the basis of the $|\pm 1\rangle$, $|\pm 2\rangle$ states, the Hamiltonian including  the electron-hole exchange terms and the Zeeman field-induced term for $\bm B \parallel \bm z$  has the form:
\begin{eqnarray}\label{ham}
	&&\mathcal{H}_{AF} =\begin{pmatrix}
	\mathcal{H}_A &0\\
	0& \mathcal{H}_F\end{pmatrix} \, , \\
 &&\mathcal{H}_A = \frac{\hbar}{2} ( \Omega_X \sigma_x +\Omega_Z \sigma_z ) \, ,  \nonumber\\ 
 && \mathcal{H}_F = -\Delta E_{AF} + \frac{\hbar}{2} (\Omega_{FX} \sigma_x + \Omega_{FZ} \sigma_z  ) \, , \nonumber
 \end{eqnarray}
where $\hbar\Omega_Z = g_A \mu_B B$, $\hbar\Omega_{FZ} = g_F \mu_B B$, $g_{A}$, $g_F$ are the bright and dark exciton $g$-factors, and $\mu_B$ is the Bohr magneton.  The Hamiltonian in Eq.~\eqref{ham} does not include any perturbations that can directly mix bright and dark exciton states.  Therefore, the eigenstates of this Hamiltonian, $\Psi_A^\pm$ and $\Psi_F^\pm$ , comprise only the linear combinations of $\Psi_{\pm 1}$ and $\Psi_{\pm 2}$ functions, respectively. The energy splittings between the pairs of eigenstates are:  
\begin{eqnarray}
   && \hbar\Omega_{A} = \hbar\sqrt{\Omega_X^2+\Omega_Z^2}\ , \\
   && \hbar\Omega_{F} = \hbar\sqrt{\Omega_{FX}^2+\Omega_{FZ}^2}\ .  \nonumber
\end{eqnarray}
The evolution of the bright (dark) exciton states from $|X \rangle,|Y \rangle $ ($|FX \rangle,|FY \rangle $)  to pure $| \pm 1\rangle$ ( $| \pm 2\rangle
$) states  in an external magnetic field is shown in Fig.~\ref{fig:fig2}(b). 

The polarization of light emitted by excitons can be described using the spin density matrix formalism~\cite{Dzhioev1997PRB, Blum_book}. The strong exchange interaction leads to a large splitting between the states of the bright and dark excitons $\Delta E_{AF}\sim1\div5$~meV~\cite{Shornikova2018ns}, which is much larger than the inverse exciton lifetimes. This allows us to neglect non-diagonal block terms of the density matrix $\rho_{AF, m_A m_F}, \rho_{AF, m_F m_A} $ and to write the block-diagonal density matrix for the four exciton states~\cite{Ivchenko1995book}
\begin{gather}\label{matrix}
	\rho_{AF} = \begin{pmatrix}
		\rho_A & 0 \\
		0 & \rho_F
	\end{pmatrix}, \\
	\rho_{A} = \begin{pmatrix}
		\rho_{+1,+1} & \rho_{+1,-1} \\ 
\rho_{-1,+1} & \rho_{-1,-1} 
	\end{pmatrix}, \ \rho_{F} = \begin{pmatrix}
\rho_{+2,+2} & \rho_{+2,-2} \\
	\rho_{-2,+2} & \rho_{-2,-2} 
	\end{pmatrix},
\end{gather}
where the $2\times 2$ density matrices $\rho_{A}$ and $\rho_{F}$ characterize isolated two-level systems of the bright and dark exciton states $\{ \Psi_{+1}, \Psi_{-1}\}$ and $\{ \Psi_{+2}, \Psi_{-2}\}$, respectively. For both two-level systems, we introduce 3-dimensional bright and dark exciton pseudospins, $\bm s_A$ and $\bm s_F$, respectively, and the average pseudospins $\bm S_A = \bm s_A/N_A$ and $\bm S_F = \bm s_F/N_F$ with $S_{A}= S_F = 1/2$. The bright and dark exciton populations are given by  $N_A =  Sp [\rho_A] = \rho_{+1,+1} + \rho_{-1,-1}$ and $ N_F= Sp[\rho_F] =   \rho_{+2,+2} + \rho_{-2,-2}$, and the pseudospin components are defined as $s_{A\alpha} = Sp[\sigma_\alpha \rho_A]/2$ and $s_{F\alpha} = Sp[\sigma_\alpha \rho_F]/2$ with $\alpha=X,Y,Z$. 

In the dipole approximation, the bright exciton states $|\pm 1\rangle$ absorb and  emit circularly polarized light, while the $|X\rangle,|Y\rangle$ states interact with light that is linearly polarized along the $X,Y$ axes, respectively.  In what follows, we completely neglect any circular or linear dichroism in the absorption of light or spin-dependence in the light emission, assuming the dipole matrix elements to be equal for each pair of states (see details in ~\ref{dipoles}). Under this condition, the contribution of the bright exciton to the Stokes vector of the emission, ${\bm P}_A = (P_{LA}, P_{L'A},P_{CA}) $, can be related to its averaged pseudospin as $${\bm P}_A = 2 {\bm S}_A\ ,$$ because the pseudospin-space basis is directly related to the bright exciton dipole moment. Note that components of  ${\bm P}_A$ and ${\bm S}_A$ transform similarly under symmetry operations. 

The dark exciton states $m_F= \pm 2$  interact with light only due to the admixture of the nearest bright exciton states $m_A= \pm 1$ caused by a small perturbation. Other bright exciton states formed with the light hole are distant in energy. Thus, the interaction with them is negligible. Different mechanisms of the bright-to-dark exciton coupling resulting in dark exciton radiative recombination are considered in~\cite{Rodina2016}. Here, we consider only the small perturbation  $\hat V_e$ that provides the electron spin-flip in the exciton (see details in ~\ref{dipoles}).
Thus, the polarization of the dark exciton emission is determined by the coupling mechanism with $m_A= \pm 1$ states. This coupling allows us to relate the resulting Stokes vector of the dark exciton, ${\bm P}_F=(P_{LF},P_{L'F},P_{CF})$, with its averaged pseudospin {\it via} the  matrix $\hat G_{\beta \gamma}^{FA}$, ($\beta = X,Y,Z$, $\gamma = L, L',C$): $${\bm P}_F = 2 \hat G^{FA} {\bm S}_F\, .$$ 

We attribute the perturbation $\hat V_e$ to the presence of a real or fluctuating magnetic field perpendicular to the NPL $c$-axis $\bm b = (b\cos(\varphi), b\sin(\varphi),0)$, where $\varphi$ is the angle between $\bm b$ and the NPL $X$-axis, as shown in Fig.~\ref{fig:fig1}(d). The electron spin flip can be caused by the electron exchange interaction with dangling bond spins or other paramagnetic centers and provides the dark exciton circularly polarized emission~\cite{Rodina2016}, such that $P_{CF}=2S_{FZ}$. 
The linear polarization of the dark exciton emission depends on the angle $\varphi$.
Taking into account the presence of the field $\bm b$ acting on the electron in second-order perturbation theory, the pseudospin coupling matrix has the form
\begin{equation}\label{GFA}
    \hat{G}^{FA} (\varphi) = \begin{pmatrix}
		\cos(2\varphi)& \sin(2\varphi)& 0\\
		-\sin(2\varphi)&	\cos(2\varphi)&0\\
		0&0& 	1
	\end{pmatrix}\ .
\end{equation}

The dark exciton contribution to the PL is important at low temperatures due to its significant population. Accordingly, the total intensity from the colloidal NPLs
\begin{equation}
     I = I_A+I_F = \Gamma_A^{\rm r} N_A + \Gamma_F^{\rm r} N_F
\end{equation}
comprises the contributions from the bright and dark excitons controlled by their radiative recombination rates  $\Gamma_{A}^{\rm r}$ and $\Gamma_{F}^{\rm r}$, introduced in~\ref{dipoles}, and  populations $N_A$ and $N_F$. The dynamics of the bright and dark exciton populations obey the system of phenomenological kinetic equations, taking into account the exciton generation and recombination, as well as the energy relaxation between exciton states~\cite{Labeau2003,Shornikova2018ns} (see details in~\ref{Kinetics}).

Then, the polarization of light emitted by excitons from a single horizontally oriented NPL can be presented as: 
\begin{equation}
\bm P = {\cal A} \bm P_{A} + {\cal F} \bm P_{F} \, ,
\end{equation}
where the relative contributions from the bright, ${\cal A} = I_A/(I_A+I_F)$, and dark, ${\cal F} = I_F/(I_A+I_F)$, excitons  depend on the relation between radiative, $\Gamma_{A,F}^{\rm r}$, and nonradiative $\Gamma_{A,F}^{\rm nr}$, recombination and energy relaxation rates,  as well as on the temperature and  excitation regime  (see details in ~\ref{Kinetics}). Below we consider only the cases of resonant (or close to resonant)  CW and pulsed excitations at low temperatures $T < \Delta E_{AF}/k_B$.

\subsection{Pseudospin dynamics \\ in the Faraday magnetic field}\label{sec:Pseudospin} 

The dynamics of the density matrix $\rho_{AF}$ is described by the equation:
\begin{equation}\label{key_rho}
	\frac{\partial \rho_{AF}}{\partial t} + \frac{i}{\hbar} [ \mathcal{H}_{AF}, \rho_{AF} ] = 0 \, .
\end{equation}
The block-diagonal form of both the Hamiltonian, Eq.~\eqref{ham}, and the density matrix, Eq.~\eqref{matrix}, allows us to obtain two separate equations for the bright and dark exciton pseudospins $\bm s_{A,F}$:
\begin{gather}\label{syst_simplified}
\frac{d \bm s_A}{d t} + {\bm s_A} \times {\bm \Omega_A} = 0\, ,  \ \ 
\frac{d \bm s_F}{d t}+ \bm s_F \times \bm \Omega_F = 0\, .
\end{gather}
Each pseudospin rotates in its own effective magnetic field with a frequency $\bm \Omega_{A} = \bm \Omega_{X} +\bm \Omega_{Z}$  for the bright exciton and $\bm \Omega_{F} = \bm \Omega_{FX} +\bm \Omega_{FZ}$ for the dark exciton, which includes both the effective field directed along the $X$ pseudospin component, that causes the anisotropic splitting, and the external Faraday magnetic field. As a result, each pseudospin can be  presented as the sum of two components ${\bm s}_{A} = {\bm s}_{A1} +{\bm s}_{A2}$ and ${\bm s}_{F} = {\bm s}_{F1} +{\bm s}_{F2}$, where the longitudinal components ${\bm s}_{A1}$ and ${\bm s}_{F1}$ are directed parallel and the transverse components ${\bm s}_{A2}$ and ${\bm s}_{F2}$ -- perpendicular to the total field ${\bm \Omega}_{A}$ and ${\bm \Omega}_{F}$, respectively.

With account of the exciton generation, recombination, and energy relaxation processes, the pseudospin dynamics can be described by a system of phenomenological kinetic equations:
\begin{gather}
\frac{d\bm s_A}{dt} + {\bm s_A} \times {\bm \Omega_A} = \frac{\bm s_A^0 -\bm s_A  }{\tau_A}   - \frac{{\bm s_{A1} -\bm s_A^{\rm eq} }}{\tau^{s}_{A1}}  - \frac{{\bm s_{A2}}}{\tau^{s}_{A2}}\, , \label{sa} \\
\frac{d\bm s_F}{dt} + {\bm s_F} \times {\bm \Omega_F} = \frac{\bm s_F^0 -\bm s_F  }{\tau_F}   - \frac{{\bm s_{F1} -\bm s_F^{\rm eq} }}{\tau^{s}_{F1}}  - \frac{{\bm s_{F2}}}{\tau^{s}_{F2}}\,  . \label{sf}
\end{gather}
Here, $\tau_A$ and $\tau_F$ are the bright and dark exciton lifetimes given at low temperature by \begin{equation}\label{tau}
	\tau_A^{-1} = \Gamma_A+\gamma_0\, , \ \ \ \ \tau_F^{-1} = \Gamma_F \,  ,
\end{equation} 
where $\Gamma_{A} =\Gamma_{A}^{\rm r}+\Gamma_{A}^{\rm nr}$ and $\Gamma_{F}=\Gamma_{F}^{\rm r}+\Gamma_{F}^{\rm nr}$ are the recombination rates of the bright and dark excitons, and $\gamma_0$ is the relaxation rate from the bright state to the dark one at zero temperature (see details in~\ref{Kinetics}). 
The pseudospin generation rates $\bm s_{A,F}^0/\tau_{A,F}$ are contributed both by the laser light and by the exciton transitions between the
bright and dark states. We consider them in detail in the next subsection. 

The pseudospin relaxation processes are controlled by the bright and dark exciton longitudinal, $\tau_{A1}^s$ and  $\tau_{F1}^s$, and transverse, $\tau_{A2}^s$ and  $\tau_{F2}^s$, spin relaxation times (Fig.~\ref{fig:fig2}(b,c)). The longitudinal spin relaxation is accompanied by the energy relaxation between the spin-split energy states  of the bright and the dark exciton separately (Fig.~\ref{fig:fig2}(b)).  It tends to bring the longitudinal pseudospins ${\bm s}_{A1}$ and ${\bm s}_{F1}$ to $\bm s_{A}^{\rm eq} = N_A {\bm S}_{A}^{\rm eq} $ and $\bm s_{F}^{\rm eq}=N_F {\bm S}_{F}^{\rm eq}$, respectively, which are related to the thermodynamically equilibrium averaged pseudospins given by:
\begin{equation} \label{SAFeq}
\bm S_{A,F}^{\rm eq} = -\frac{\bm \Omega_{A,F}}{ 2\Omega_{A,F}} \tanh \frac{\hbar\Omega_{A,F}}{2k_{\rm B}T} \, . 
\end{equation}
The dephasing times  $\tau^{s}_{A2}$ and  $\tau^{s}_{F2}$ are related to the transverse pseudospin components ${\bm s}_{A2}$ and ${\bm s}_{F2}$ rotating around the effective fields ${\bm \Omega}_{A}$ and ${\bm \Omega}_{F}$ (see Fig.~\ref{fig:fig2}(c)).

\subsection{Pseudospin generation}
\label{sec:Generation}
 
In this subsection, we consider  the exciton pseudospin  generation terms in the case of CW and pulsed excitation at low temperature.

As for the bright and dark exciton populations, in the steady-state regime under resonant or quasi-resonant excitation, their general expressions shown in Eqs.~\eqref{populations_cw} are reduced to  
 \begin{equation} \label{NA0NF0}
 N_A^0=G_A \tau_A \, , \qquad  N_F^0=N_A^0 \gamma_0 \tau_F \, ,
 \end{equation}
where $G_A = \rm const$ reflects stationary bright exciton generation and is independent of the exciting light polarization as well as of the external magnetic field.  

For the case of pulsed excitation, the general expressions for $N_A(t),N_F(t)$ are also given in the~\ref{Kinetics} in Eqs.~\eqref{NAFpulsed}. At low temperature, in the case of resonant excitation  of the bright exciton, we obtain
\begin{eqnarray} \label{NAFt}
&&	N_A(t) = N_0 \exp(-t/\tau_A) \, , \\ 
&&	N_F(t)= \frac{N_0\gamma_0\tau_A\tau_F}{\tau_F -\tau_A} \left( \exp\left(-\frac{t}{\tau_F}\right) - \exp\left(-\frac{t}{\tau_A}\right)\right) \, , \nonumber
\end{eqnarray}
where $N_{A}(0)$ and $N_{F}(0)$ are the bright and dark exciton populations created by the short pulse at $t=0$, independently of the exciting light polarization and of the external magnetic field. Note that the second term in $N_F(t)$ describes the rise in the dark exciton population due to excitation transfer from the bright state.

We now consider the pseudospin generation terms ${\bm s}_{A}^0$ and ${\bm s}_{F}^0$ on the right-hand side of  the kinetic Eqs.~(\ref{sa}, \ref{sf}). We investigate the low-temperature case when the transitions from the dark exciton to the bright one are absent. Then the bright exciton pseudospin $\bm s_A^0$ is created only by the exciting light as $\bm s_A^0=N_A^0\bm S_A^0$ at CW excitation  or $\bm s_A^0=N_A(0)\bm S_A^0$ at pulsed excitation.

For the ensemble of in-plane randomly oriented NPLs, it is enough to characterize the polarization of the exciting light by two Stokes parameters, $P_{0}^l$  and $P_{0}^c$, for linearly and circularly polarized excitation, respectively. They are related to the projections $e_x^0,e_y^0$ of the light polarization vector ${\bm e}_0$ on the laboratory axes $x,y$ according to Eqs.~\eqref{Stokes}. For the NPL with $X, Y$ axes, rotated around the laboratory $z$-axis by an angle $\alpha$, the exciting light polarization generates the averaged pseudospin of the bright exciton according to
\begin{gather}
	S_{AX}^0 = \frac{\gamma_l}{2} P_0^l \cos(2\alpha)\, , \nonumber \\
	S_{AY}^0 = \frac{\gamma_l}{2} P_0^l \sin(2\alpha)\, , \label{pump1} \\
	S_{AZ}^0 =  \frac{\gamma_c}{2} P_0^c  \, . \nonumber
\end{gather}
Here, the parameters $\gamma_l \le 1$ and $\gamma_c \le 1$ are introduced to account for the possible loss of the pumped polarization during the relaxation process in the case when the excitation is quasi-resonant. Hereafter, by $P_0^l $ and $P_0^c$ we mean $\gamma_l P_0^l $ and $\gamma_c  P_0^c$, respectively. Importantly, we neglect in Eq.~\eqref{pump1} any circular or linear dichroism of the matrix elements for the absorption of light, as well as any circular or linear dichroism of the relaxation (see details in~\ref{dipoles}).

We assume that the polarized laser excitation does not create the dark exciton pseudospin $\bm s_F^0$ directly. Indeed, even taking into account for the admixture of the bright to dark exciton states, the dipole moment of the dark exciton is much smaller than that of the bright one, so it is safe to consider $G_A \gg G_F \approx 0$. However, the $\bm s_F^0$ might be created \textit{via} the relaxation of the polarized population from bright to dark exciton states. We assume that the same perturbation $\hat V_e$ associated earlier with the presence of a magnetic field $\bm b$ perpendicular to the NPL c-axis provides the electron spin-flip and is responsible not only for the radiative recombination of the dark exciton but also for the relaxation between the bright and dark excitons, accompanied by the acoustic phonons that conserve the spin projection. The ${\bm s}_F$ generation therefore takes the form: 
\begin{equation} 
    {\bm s}_F^0 =\tau_F\gamma_0 \hat G^{AF} (\varphi_0) {\bm s}_A\, ,
\end{equation}
where ${\bm s}_A $ corresponds to the steady-state solution of the Eq.~\eqref{sa} in the CW regime ${\bm s}_A^{\rm ss}$ and to ${\bm s}_A(t)$ in time after the pulsed excitation, $\hat G_{\beta \gamma}^{AF} = \hat G_{ \gamma \beta}^{FA}$ ($\beta, \gamma = X,Y,Z$, Eq.~\eqref{GFA}), however, in the case of the fluctuating field ${\bm b}$, the corresponding angle $\varphi_0$ might differ from the angle $\varphi$ (Fig.~\ref{fig:fig1}(d)) in the dark exciton generation and recombination.

\section{Exciton pseudospin in effective magnetic field}\label{sec:Sol}
\subsection{Steady state solutions at low temperature}\label{lowTcw}

We consider first the steady-state solutions of Eqs.~(\ref{sa}, \ref{sf}) under CW excitation at low temperature. In this case, the averaged pseudospins  $\bm S_{A}^{\rm ss}=\bm s_{A}^{\rm ss}/N_A^{0}$ and $\bm S_{F}^{\rm ss}=\bm s_{F}^{\rm ss}/N_F^{0}$ satisfy the same system of kinetic Eqs.~(\ref{sa}, \ref{sf}). Their longitudinal components can be written as 
\begin{eqnarray}\label{limit1}
	{\bm S}_{A1}^{\rm ss} = \frac{T_{A1}}{\tau_A} \frac{({\bm S}_A^0{\bm \Omega}_A ){\bm \Omega}_A}{{\Omega}_A^2} + \frac{T_{A1}}{\tau_{A1}^s} {\bm S}_A^{\rm eq} \, , \\
	{\bm S}_{F1}^{\rm ss} = \frac{T_{F1}}{\tau_F} \frac{({\bm S}_F^0{\bm \Omega}_F ){\bm \Omega}_F}{{\Omega}_F^2} + \frac{T_{F1}}{\tau_{F1}^s} {\bm S}_F^{\rm eq} \, 	 	
\end{eqnarray} 
with ${\bm {S}}_F^0 = \hat G^{AF} {\bm S}_A^{\rm ss} =(S_{AX}^{\rm ss} \cos\left( 2\varphi_0\right)-S_{AY}^{\rm ss} \sin\left( 2\varphi_0\right),  S_{AX}^{\rm ss} \sin\left( 2\varphi_0\right)+S_{AY}^{\rm ss} \cos\left( 2\varphi_0\right), S_{AZ}^{\rm ss})$ and the transverse components are
\begin{eqnarray}\label{limit2}
	{\bm S}_{A2}^{\rm ss} = \frac{T_{A2}}{\tau_A} \frac{T_{A2}}{1+\Omega_A^2T_{A2}^2} \left( {\bm \Omega}_A \times {\bm S}_A^0 +\frac{{\bm \Omega}_A \times [{\bm S}_A^0 \times {\bm \Omega}_A] }{T_{A2}{\Omega}_A^2} \right) \,, \nonumber \\
	\\
	{\bm S}_{F2}^{\rm ss} = \frac{T_{F2}}{\tau_F} \frac{T_{F2}}{1+\Omega_F^2T_{F2}^2} \left( {\bm \Omega}_F \times {\bm S}_F^0 +\frac{{\bm \Omega}_F \times [{\bm S}_F^0 \times {\bm \Omega}_F] }{T_{F2}{\Omega}_F^2} \right) \,, \nonumber \\
\end{eqnarray} 
Here $T_{A1,F1}$ and $T_{A2,F2}$ are the longitudinal and transverse pseudospin lifetimes, respectively:
$$\frac{1}{T_{A1} }= \frac{1}{\tau_{A}}+\frac{1}{\tau_{A1}^s},\  \frac{1}{T_{F1} }= \frac{1}{\tau_{F}}+\frac{1}{\tau_{F1}^s} \, ,$$
$$\frac{1}{T_{A2} }= \frac{1}{\tau_{A}}+\frac{1}{\tau_{A2}^s}, \ \frac{1}{T_{F2} }= \frac{1}{\tau_{F}}+\frac{1}{\tau_{F2}^s} \, .$$

One can see that in the limit $\Omega_F T_{F2} \gg 1$ and $\Omega_A T_{A2} \gg 1$, allowing many rotations of the transverse pseudospin around the effective field during its lifetime, its time-averaged component perpendicular to the effective field vanishes and the steady-state pseudospin is always directed along the effective field~\cite{Kusrayev2008, Dzhioev1997PRB}. The transverse components remain small as 
$${\bm S}_{A2}^{\rm ss} \approx  \frac{{\bm \Omega}_A \times {\bm S}_A^0}{\tau_A \Omega_A^2}\, , \quad {\bm S}_{F2}^{\rm ss} \approx  \frac{{\bm \Omega}_F \times {\bm S}_F^0}{\tau_F \Omega_F^2 } \, ,
$$
and $\tau_{A,F}\Omega_{A,F} > T_{A2,F2}\Omega_{A,F} \gg 1$.

In the opposite limit $\Omega_F T_{F2} \ll 1$ and $\Omega_A T_{A2} \ll 1$, we obtain
$${\bm S}_{A2}^{\rm ss} \approx \frac{T_{A2}}{\tau_A} \frac{{\bm \Omega}_A \times [{\bm \Omega}_A \times {\bm S}_A^0]}{ \Omega_A^2}\, , $$
$${\bm S}_{F2}^{\rm ss} \approx \frac{T_{F2}}{\tau_F}  \frac{{\bm \Omega}_F \times[{\bm \Omega}_F \times {\bm S}_F^0]}{ \Omega_F^2 } \, .
$$ One can see that the transverse components can be large in the case $T_{A2,F2} \approx \tau_{A,F}$ or vanish if $T_{A2} \approx \tau_{A2}^s \ll \tau_{A}$ and $T_{F2} \approx \tau_{F2}^s \ll \tau_{F}$, respectively.

The pseudospin components $\gamma = X,Y,Z$ can be found as 
$$
{S}_{A\gamma} = {S}_{A1\gamma} + {S}_{A2\gamma} \, , \quad {S}_{F\gamma} = {S}_{F1\gamma} + {S}_{F2\gamma} \, ,
$$
and can be written in the form 
\begin{eqnarray} 
S_{A\gamma} = \frac{T_{A1}}{\tau_{A1}^s}S_{A\gamma}^{\rm eq} + M^A_{\gamma \delta} S_{A \delta}^0 \, , \nonumber \\
S_{F\gamma} = \frac{T_{F1}}{\tau_{F1}^s}S_{F\gamma}^{\rm eq} + M^F_{\gamma \delta} S_{F \delta}^0 \, . \label{SAFM}
\end{eqnarray}
The pseudospin components $S_{A \gamma}^{\rm eq}$ and  $S_{F \gamma}^{\rm eq}$ are obtained from Eq.~\eqref{SAFeq} with  the $\gamma=X,Z$ projections of the ${\bm \Omega}_{A,F}$. The expressions for the matrix components $M^{A,F}_{\gamma\delta}$  are written in~\ref{SS_bright}.
\newline \newline

\subsection{Time-dependent solutions at low temperature}\label{lowTtime}

We look for the time-dependent solutions in the form 
\begin{eqnarray}
 {\bm s}_A(t) = {\bm S}_A(t) N_A(t)\, , \quad  {\bm s}_F(t) = {\bm S}_F(t) N_F(t)\, . 
\end{eqnarray}

For the bright exciton, its averaged pseudospin dynamics can be found as ${\bm S}_{A}(t)={\bm S}_{A1}(t)+{\bm S}_{A2}(t)$ with
\begin{widetext}
\begin{eqnarray}
	{\bm S}_{A1}(t) =  \frac{({\bm S}_A^0{\bm \Omega}_A ){\bm \Omega}_A}{{\Omega}_A^2} \exp\left(-\frac{t}{\tau_{A1}^s}\right) +  \bm S_A^{\rm eq} \left( 1 - \exp\left(-\frac{t}{\tau_{A1}^s}\right) \right)\, , \nonumber \\
{\bm S}_{A2}(t) = \left(\frac{ {\bm \Omega}_A \times {\bm S}_A^0}{{\Omega}_A} \sin ({\Omega}_A t) +\frac{{\bm \Omega}_A \times [{\bm S}_A^0 \times {\bm \Omega}_A] }{{\Omega}_A^2} \cos ({\Omega}_A t)  \right) \exp\left(-\frac{t}{\tau_{A2}^s}\right) \, . \label{SAT}
\end{eqnarray} 
The full-form expressions  for the time-dependent components proportional to the initial pumping, $S_{A\gamma}(t)  = M^A_{\gamma\delta}(t) S_{A \delta}^0$,  are written in~\ref{SS_bright}.

For the dark exciton in the case of pulsed resonant excitation of the bright exciton at low temperature in the realistic case $\tau_F \gg \tau_A$, one can be interested only in the decay of its population at a time longer than the bright exciton lifetime. In this case, $ N_F(t)$ is approximated as 
\begin{equation}
    N_F(t\gg \tau_A) \approx N_0 \gamma_0 \tau_A \exp(-t/\tau_F) \, .
\end{equation} 
The pseudospin kinetics for the dark exciton in case of $T_{F1},T_{F2} \gg T_{A1},T_{A2}$ on the timescale $t \gg \tau_A$ has the same form as for the bright one
\begin{eqnarray}
	{\bm S}_{F1}(t\gg \tau_A) =  \frac{({\bm { S}}_F^0{\bm \Omega}_F ){\bm \Omega}_F}{{\Omega}_F^2} \exp\left(-\frac{t}{\tau_{F1}^s}\right) +  \bm S_F^{\rm eq}\left( 1 - \exp\left(-\frac{t}{\tau_{F1}^s}\right) \right)\, , \nonumber \\
{\bm S}_{F2}(t\gg \tau_A) = \left(\frac{ {\bm \Omega}_F \times {\bm { S}}_F^0}{{\Omega}_F} \sin ({\Omega}_F t) +\frac{{\bm \Omega}_F \times [{\bm { S}}_F^0 \times {\bm \Omega}_F] }{{\Omega}_F^2} \cos ({\Omega}_F t)  \right) \exp\left(-\frac{t}{\tau_{F2}^s}\right) \, ,
\end{eqnarray} 
\end{widetext}
but with the same ${\bm {S}}_F^0$ as for the steady-state solutions in Sec.~\ref{lowTcw}: ${\bm {S}}_F^0 = \hat G^{AF} {\bm S}_A^{\rm ss}$.

\begin{figure*}[ht]
	%\centering
	\includegraphics[width=0.9\linewidth]{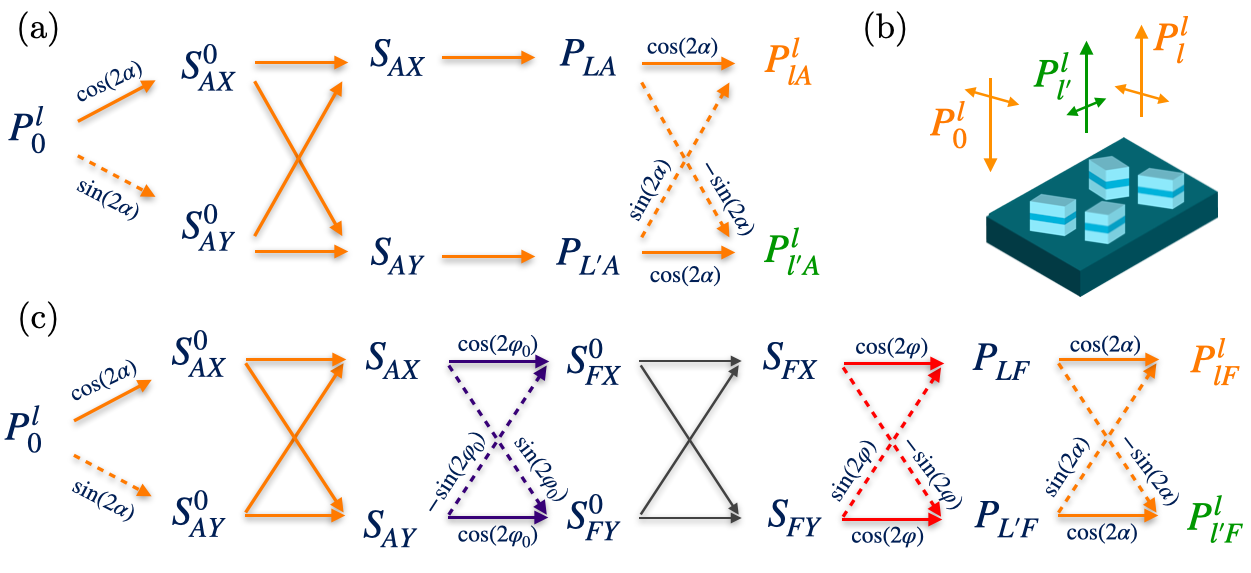}
	\caption{\label{fig:fig3}(a) A scheme of the linear polarization $P_0^l$ conversion in the bright exciton into detectable effects of optical alignment $P_l^l$ (marked in orange) and rotation of the linear polarization plane $P_{l'}^l$ (marked in green) for the NPL oriented at the angle $\alpha$ (Fig.~\ref{fig:fig1}(b)).
    (b) Geometry of the optical alignment effect registration from the isotropically oriented NPL ensemble. (c) A scheme of the transfer of the linearly polarized pseudospin components from the bright to dark exciton and its further contribution to the linearly polarized light emission. The angles $\varphi_0$ and $\varphi$ (Fig.~\ref{fig:fig1}(d)) correspond to the direction of the magnetic field $\bm b$ acting in the polarization transfer and radiation, respectively.} 
\end{figure*}

\section{PL polarization from the NPL ensemble}\label{sec:Orientation}

Recall that NPLs are isotropically oriented in the plane of the substrate (Fig.~\ref{fig:fig1}(a)). For this reason, some contributions to the polarization from single NPLs, such as the conversion of linear to circular polarization and \textit{vice versa}, or the impact connected with the thermodynamically equilibrium linear polarization, disappear upon averaging over the ensemble. Since the total intensity does not depend on the in-plane orientation of the NPL, only the averaging of the polarizations coming from each NPL over the angle $\alpha$ is needed. Below we consider the bright and dark exciton contributions to the polarization of PL from the in-plane-isotropic NPL ensemble.

\subsection{Bright exciton recombination polarization}
\label{brightLambda}

 The circular polarization of the bright exciton recombination stemming from the $S_{AZ}$ component generated by $P_0^c$ or from the  $S_{AZ}^{\rm eq}$ does not depend on the particular NPL in-plane orientation (angle $\alpha$).
 As for the linearly polarized components, the contributions coming from the bright exciton in a single NPL oriented at the angle $\alpha$ in the low-temperature regime after the initial $P_0^l$ excitation are written in~\ref{alpha}. Figure~\ref{fig:fig3}(a) shows a scheme of the linear polarization $P_0^l$ transformation in the bright exciton into $P_{lA}^l$ and $P_{l'A}^l$. Importantly, the  symmetry of the $\Lambda_{ij}^A$ matrix for the fixed angle $\alpha=0$ corresponds to the $C_2$ point group considered in~\ref{symmetry}. The averaging of the bright exciton contributions over the randomly distributed angle $\alpha$ gives the bright exciton polarizations $P_{l}^l$ and $P_{l'}^l$ (Fig.~\ref{fig:fig3}(b)) from the in-plane randomly oriented ensemble of NPLs. The components stemming from the linearly polarized excitation $P_{0}^{l'}$ can be readily obtained by replacing $\alpha$ with $\alpha + \pi/4$ without any difference after the ensemble averaging. 
The conversion of circular polarization to linear and \textit{vice versa}, as well as contributions from $S_{AX}^{\rm eq}$, vanish after $\alpha$-averaging.
The expressions for the bright exciton PL polarization induced by the pumped polarization ${\bm P}_0 = (P_0^l,0,P_0^c)$
after ensemble averaging take the form:
\begin{eqnarray}
P_{lA}^l &=& \frac{P_0^l}{2} \left[ \frac{T_{A1}}{\tau_A} \frac{\Omega_X^2}{\Omega_A^2} +  \frac{T_{A2}}{\tau_A} \left( 1+ \frac{\Omega_Z^2}{\Omega_A^2}\right) \frac{1}{(1+\Omega_A^2T_{A2}^2)} \right] , \nonumber \\  \label{PACW}
P_{l'A}^l &=& P_0^l\frac{T_{A2}}{\tau_A}\frac{\Omega_ZT_{A2}}{1+\Omega_A^2T_{A2}^2} \, , \\ \nonumber
P_{cA}^c &=& P_0^c \left[ \frac{T_{A1}}{\tau_A}  \frac{\Omega_Z^2}{\Omega_A^2} + \frac{T_{A2}}{\tau_A} \frac{\Omega_ X^2}{\Omega_A^2}\frac{1}{(1+\Omega_A^2T_{A2}^2)} \right].
\end{eqnarray} 
It is important to note that among the three Stokes parameters, only the $P_{lA}^l$ differs from $P_{lA}^l(\alpha =0)$ after averaging.

With account of the magnetic-field-induced circular polarization stemming from $S_{AZ}^{\rm eq}$, the circular polarization of the bright exciton PL  can be written as
\begin{eqnarray} \label{CWeq}
&&P_{cA} = P_{cA}^c + P_{cA}^{\rm eq} \, ,\\
&& P_{cA}^{\rm eq} =- \frac{T_{A1}}{\tau_{A1}^s} \frac{\Omega_{Z}}{\Omega_{A}} \tanh \frac{\hbar\Omega_{A}}{2k_{\rm B}T} \, .
\end{eqnarray}

By performing substitutions of the initial conditions (Eq.~\ref{pump1}) in the expressions for the pseudospin components of the bright exciton (Eq.~\ref{SAT}), collecting the corresponding components into the effects in Eqs.~\eqref{Plangle}, and averaging over the angle $\alpha$ we obtain the kinetics of the effects under study for the ensemble:
\begin{gather}
P_{lA}^l(t) = \frac{P_0^l}{2}\left[ \frac{\Omega_X^2}{\Omega_A^2} e^{-\frac{t}{\tau_{A1}^s}} +  \left( 1+ \frac{\Omega_Z^2}{\Omega_A^2}  \right) e^{-\frac{t}{\tau_{A2}^s}}\cos(\Omega_A t)\right] \, , \nonumber \\
\label{PAt} P_{l'A}^l(t) = P_0^l\frac{ \Omega_Z}{\Omega_A} \sin(\Omega_A t)e^{-\frac{t}{\tau_{A2}^s}}\, , \\
P_{cA}^c(t) =  P_0^c \left[ \frac{\Omega_Z^2}{\Omega_A^2} e^{-\frac{t}{\tau_{A1}^s}} + \frac{\Omega_X^2}{\Omega_A^2} e^{-\frac{t}{\tau_{A2}^s}}\cos(\Omega_A t)\right]\, ,  \nonumber
\end{gather}
and 
\begin{eqnarray}
&&P_{cA}(t) = P_{cA}^c(t) +P_{cA}^{\rm eq}(t) \, , \\
&&P_{cA}^{\rm eq}(t)=- \left(1-e^{-\frac{t}{\tau_{A1}^s}}\right) \frac{\Omega_{Z}}{\Omega_{A}} \tanh \frac{\hbar\Omega_{A}}{2k_{\rm B}T} \, . \nonumber
\end{eqnarray}

The results can be written in a more general form for the case of an arbitrarily initially pumped polarization ${\bm P}_0=(P_0^l,P_0^{l'},P_0^c)$ as
\begin{equation} \label{LambdaA}
{ P}_{i A} = \Lambda_{ij}^A P^0_j + P_{i A}^{\rm eq}\end{equation} 
with $i,j=l,l',c$ and $P_{iA}^{\rm eq} = \delta_{ic}P_{cA}^{\rm eq}$. Equations~(\ref{PACW},\ref{PAt}) define the components of the $\Lambda^A$ matrix as 
 $\Lambda_{ll}^A = \Lambda_{l'l'}^A = P_{lA}^l/P_0^l$, $\Lambda_{l'l}^A(B) = \Lambda_{ll'}^A(-B) = P_{l'A}^l(B)/P_0^l$, and $\Lambda_{cc}^A = P_{cA}^c/P_0^c$. Other components are vanishing upon averaging over the ensemble with an in-plane isotropic distribution of the angle $\alpha$. Importantly, the resulting symmetry of the $\Lambda_{ij}^A$ matrix for the in-plane isotropically oriented ensemble corresponds to the $C_\infty$ point group considered in~\ref{symmetry}.
	
\subsection{Dark exciton contribution to the ensemble polarization}\label{sec:Dark1}

As for the dark exciton, the crucial role is played by the perturbation activating both its radiative recombination and the transmission of polarization between the bright and dark excitons. In the presence of the magnetic field $\bm b$ in the NPL plane (Fig.~\ref{fig:fig1} (d)), these two processes depend on the direction of the magnetic field $\bm b$  and are determined by the matrices $\hat{G}^{AF}(\varphi_0)$ and $\hat{G}^{FA}(\varphi)$ (Eq.~\eqref{GFA}), respectively. A scheme reflecting the evolution of dark exciton pseudospin linearly polarized components is shown in Fig.~\ref{fig:fig3}(b).

The magnetic field ${\bm b}$ of the exchange origin fluctuates over time, so the angles $\varphi_0$ and $\varphi$ in the dark exciton excitation and recombination processes can differ. Unless there is no in-plane anisotropy of the ${\bm b}$ direction making $\varphi_0$ and $\varphi$ have preferred values (the same in all NPL), the resulting ensemble polarization should be averaged. Depending on the relation between the dark exciton lifetime and the characteristic time of the magnetic field fluctuations, the transferred polarization can be preserved or lost during the radiation. We assume that on the timescale of the dark exciton lifetime $\tau_F$, the direction of the field ${\bm b}$ is frozen. Therefore, we take $\varphi_0 = \varphi$ and perform the averaging of $\bm P_F$ over the randomly distributed angle $\varphi$. The results corresponding to the in-plane anisotropy with $\varphi_0 = \varphi = 0$ are considered in~\ref{dark}.

The circular polarization of the dark exciton recombination also does not depend on the angle $\alpha$.
As for the linearly polarized components, averaging over the $\alpha$ angle is also required. 
The conversion of circular polarization to linear and \textit{vice versa} as well as contributions from $S_{FX}^{\rm eq}$, vanish after $\alpha$-averaging.
Due to the $\varphi$-averaging,  the resulting Stokes parameters of the dark exciton polarization are directly related to those of the bright exciton polarization:
\begin{equation}
    { P}_{i F} = \Lambda_{ij}^{FA} P_{jA} + P_{i F}^{\rm eq}
\end{equation} with $i,j=l,l',c$ and $P_{iF}^{\rm eq} = \delta_{ic}P_{cF}^{\rm eq}$. 

For CW excitation, the equilibrium part of the dark exciton PL polarization in the magnetic field has the form:
\begin{equation} P_{cF}^{\rm eq} = -\frac{T_{F1}}{\tau_{F1}^s}\frac{\Omega_{FZ}}{\Omega_{F}} \tanh \frac{\hbar\Omega_{F}}{2k_{\rm B}T} \, 
\end{equation}
and the matrix $\Lambda_{ij}^{FA}$ nonzero components are:
\begin{widetext}
    \begin{eqnarray}
\Lambda_{ll}^{FA}(B) = \Lambda_{l'l'}^{FA}(B) &=& \frac{1}{2} \left[ \frac{T_{F1}}{\tau_F}   \frac{\Omega_{FX}^2}{{\Omega_{F}^2}}  + \frac{T_{F2}}{\tau_F} \frac{1}{1+\Omega_F^2 T_{F2}^2} \left( 1+   \frac{\Omega_{FZ}^2}{\Omega_F^2} \right) \right]\, , \nonumber \\  \Lambda_{ll'}^{FA}(B) = \Lambda_{l'l}^{FA}(-B) &=& -\frac{T_{F2}}{\tau_F} \frac{\Omega_{FZ}T_{F2}}{1+\Omega_F^2 T_{F2}^2}\, ,\\
    \Lambda_{cc}^{FA}(B) &=&  \frac{T_{F1}}{\tau_F} \frac{\Omega_{FZ}^2}{\Omega_{F}^2}
	+ \frac{T_{F2}}{\tau_F}\frac{\Omega_{FX}^2}{\Omega_{F}^2} \frac{1}{1+\Omega_F^2 T_{F2}^2}\, . \nonumber
\end{eqnarray}
For pulsed excitation, the equilibrium part of the dark exciton PL polarization depends on time as
\begin{equation}
P_{cF}^{\rm eq}(t) = -\left(1-e^{-\frac{t}{\tau_{F1}^s}}\right)\frac{\Omega_{FZ}}{\Omega_{F}} \tanh \frac{\hbar\Omega_{F}}{2k_{\rm B}T} \, 
\end{equation}
and $\Lambda_{ij}^{FA}$ components in the case of $T_{F1},T_{F2} \gg T_{A1},T_{A2}$ on the timescale $t \gg \tau_A$  are 
\begin{eqnarray}
    \Lambda_{ll}^{FA}(B,t \gg \tau_A) = \Lambda_{l'l'}^{FA}(B,t\gg \tau_A) &=& \frac{1}{2} \left[ \frac{\Omega_{FX}^2 }{\Omega_F^2}  e^{-t/\tau_{F1}^s} + \left(1+\frac{\Omega_{FZ}^2}{\Omega_F^2}\right) \cos(\Omega_F t)e^{-t/\tau_{F2}^s}  \right]\, , \nonumber \\  \Lambda_{ll'}^{FA}(B,t\gg \tau_A) = \Lambda_{l'l}^{FA}(-B,t\gg \tau_A) &=& -\frac{\Omega_{FZ}}{\Omega_F} \sin(\Omega_F t)e^{-t/\tau_{F2}^s}\, , \\
    \Lambda_{cc}^{FA}(B,t\gg \tau_A) &=& \frac{\Omega_{FZ}^2}{\Omega_F^2} e^{-t/\tau_{F1}^s}    + \frac{\Omega_{FX}^2}{\Omega_F^2}  \cos(\Omega_F t) e^{-t/\tau_{F2}^s}\, . \nonumber
\end{eqnarray}
\end{widetext}
However, the experiment~\cite{Smirnova2025} showed $\tau_{F2}^s\  (> T_{F2})$ shorter than $\tau_A$. Therefore, on the timescale $t \gg \tau_A$, we focus on the terms with $\tau_{F1}^s$.  
Other components are vanishing upon averaging over the ensemble with an in-plane isotropic distribution of the angles $\alpha$ and $\varphi$. 

\subsection{Final polarization}\label{sec:Final}

Finally, for the case of the low-temperature regime and resonant excitation, we can write the total polarization from the in-plane isotropically oriented NPL ensemble in the form of Eqs.~\eqref{PLP0}, which can be expressed as
\begin{eqnarray}
{P}_l^l(B)&=& \hat \Lambda_{ll}(B^2) P_0^l \nonumber  \\ &=& \hat \Lambda_{l'l'}(B^2) P_0^{l'} = {P}_{l'}^{l'}(B^2)\, ,  \\ {P}_{l'}^l(B) &=&  \hat \Lambda_{l'l}(B) P_0^l = -\Lambda_{l'l}(-B) P_0^l\nonumber \\ &=&  \hat \Lambda_{ll'}(-B) P_0^{l'} = {P}_{l}^{l'}(-B)  \, , \\ {P}_c^c(B)&=&  P_c^{\rm un}(B)+ \hat \Lambda_{cc} (B^2)  P_0^c \nonumber \\ &=& -P_c^{\rm un}(-B)+ \hat \Lambda_{cc} (B^2)  P_0^c \, . 
\end{eqnarray} 
The matrix $\Lambda_{ij} = {\cal A}\Lambda_{ij}^A+{\cal F}\Lambda_{ij}^F $ describes the polarization caused by the optical pumping of the bright exciton and possesses the symmetry considered in~\ref{symmetry} for the $C_\infty$ point group. 

In the considered case of the in-plane isotropy of the fluctuating magnetic field in the NPL ensemble, it can be written as
\begin{equation} \Lambda_{ij} = {\cal A}\Lambda_{ij}^A+{\cal F}\Lambda_{ij}^F ={\cal A}\Lambda_{ij}^A+{\cal F}\Lambda_{ik}^{FA}\Lambda_{kj}^A  \, .
\end{equation}
The vector ${\bm P}^{\rm un} = (0,0,P_c^{\rm un}(B))$ arises only in the presence of an external Faraday magnetic field, does not depend on the excitation conditions and manifests itself only in the circularly polarized PL: 
\begin{equation} \label{PcB}
P_{c}^{\rm un}(B) = {\cal A}P_{cA}^{\rm eq}(B) + {\cal F} P_{cF}^{\rm eq}(B)+ {\cal F}\Lambda_{cc}^{FA} (B) P_{cA}^{\rm eq} (B)\ 
\end{equation}
for CW excitation, and 
\begin{eqnarray} \label{PcBt}
   P_{c}^{\rm un}(B,t) & = & {\cal A}(t)P_{cA}^{\rm eq}(B,t) +  {\cal F}(t) P_{cF}^{\rm eq}(B,t) \\ 
    &+ &{\cal F} (t) \Lambda_{cc}^{FA} (B,t\gg \tau_A) P_{cA}^{\rm eq} (B)\  \nonumber
\end{eqnarray}
 for pulsed excitation.

Here, the first two terms in the time-dependent expression in Eqs.~\eqref{PcBt} are the equilibrium contributions from the bright and dark excitons that saturate over time, while the third term exponentially decays with a characteristic time $\tau_{F1}^s$.  It is induced by the transferred thermodynamically equilibrium $S_A^{\rm eq}$ component from the bright exciton to the dark one. 

\section{RESULTS AND DISCUSSION}\label{sec:Discussion}

\begin{figure}
    \centering
    \includegraphics[width=0.75\linewidth]{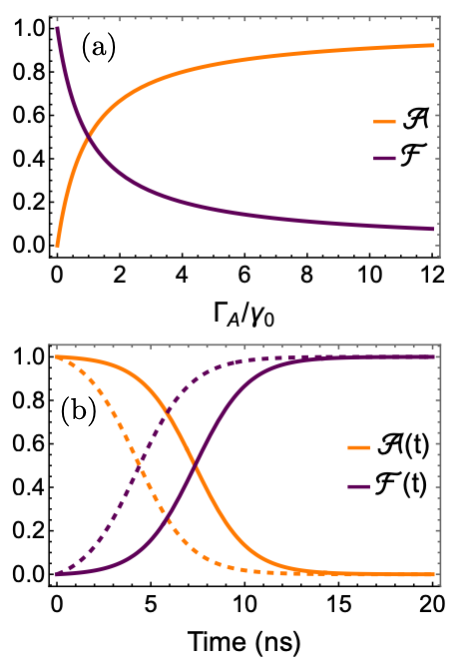}
    \caption{Bright ${\cal A}$ and dark ${\cal F}$ excitons relative intensities shown by orange  and purple lines, respectively, for the case of $\tau_A = 1.4$~ns (a) in cw regime as function of the ratio $\Gamma_A/\gamma_0$, and (b) as function of time after pulsed excitation for  $\Gamma_A/\gamma_0 = 0.3$ (dashed lines)~\cite{Shornikova2018ns} and $\Gamma_A/\gamma_0 = 2.5$ (solid lines)~\cite{Smirnova2025}.}
    \label{fig:AF}
\end{figure}

\begin{widetext}

\begin{figure*}[ht]
	\includegraphics[width=0.99\linewidth]{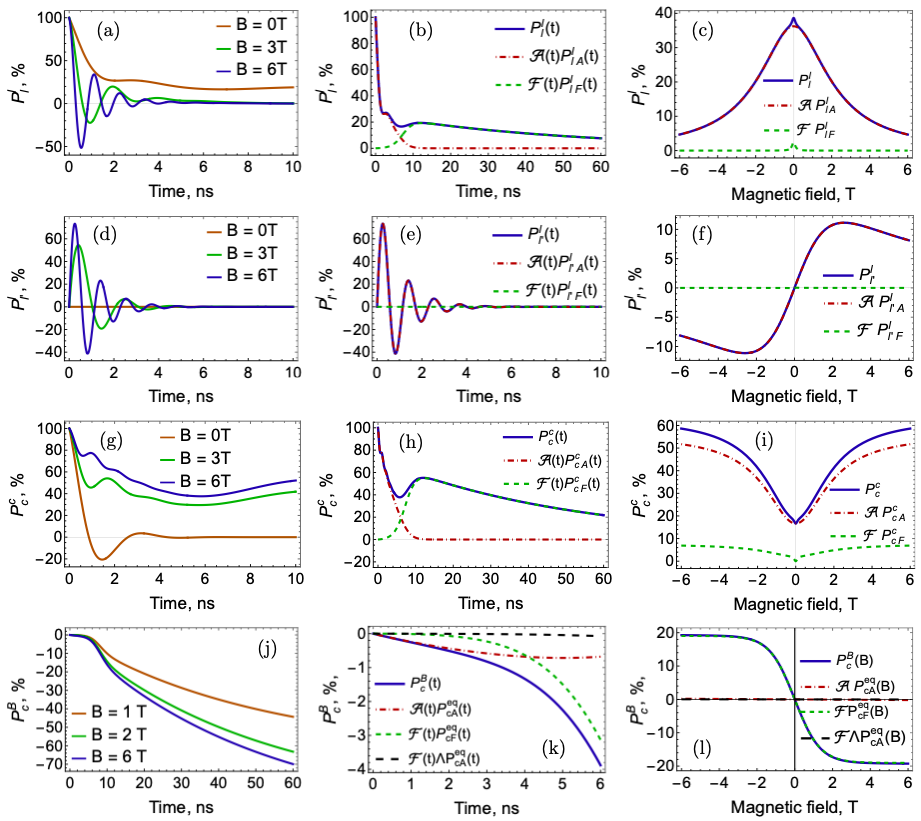}
	\caption{Time dependences in different magnetic fields of (a) the optical alignment effect $P_l^l$, (d) the linear polarization plane rotation $P_{l'}^l$, (g) the optical orientation effect $P_c^c$, (j) the magnetic-field-induced circular polarization $P_c^B$. The time-dependence decomposition into the contributions of the bright and dark excitons for (b) $P_{l}^l$ in $B = 0$~T, (e) $P_{l'}^l$ in $B = 6$~T, (h)  $P_c^c$ in $B = 6$~T, (k) $P_c^B$ in $B = 6$~T. Faraday magnetic field dependence and its analogous decomposition of (c) $P_{l}^l$, (f) $P_{l'}^l$, (i)  $P_c^c$, (l) $P_c^B$. The parameters used are listed in Table~\ref{table:1}.} 
    \label{fig:5}
\end{figure*}
    
\end{widetext}

Let us analyze the obtained expressions by using the examples of the dependences of the effects under investigation on time and on the external Faraday magnetic field in the ensemble of CdSe/CdS NPLs studied experimentally in Refs.~\cite{Smirnova2023,Smirnova2025}. The parameters used in Fig.~\ref{fig:AF} and Fig~\ref{fig:5} are taken close to those obtained in the analysis of the experimental data regarding the optical alignment kinetics after pulsed excitation~\cite{Smirnova2025} and the optical alignment and optical orientation dependences on Faraday magnetic field in the cw regime~\cite{Smirnova2023}. The time characteristics are fixed as $\tau_A = 1.4$~ns, $\tau_{A1}^s = 5$~ns ($T_{A1} = 1.1$~ns), $\tau_F = 100$~ns, and $\tau_{F1}^s = 50$~ns ($T_{F1} = 33$~ns). Other parameters are chosen as $\tau_{A2}^s = 1$~ns ($T_{A2} = 0.6$~ns), $g_A = 0.01$, $\hbar \Omega_X = 1\ \mu$eV, $g_F = 3.35$, $\hbar \Omega_{FX} = 20\ \mu$eV. As for the dark exciton pseudospin transverse relaxation time, we use $\tau_{F2}^s = 0.3$~ns ($T_{F2} = 0.3$~ns). All figures show the maximum possible polarization values for the system under consideration, \textit{i.e.}, the depolarization of excitons during excitation is neglected.
\begin{table}[ht!]
	\caption{
	Bright and dark exciton parameters used in Figs.~\ref{fig:5}~$-$ \ref{fig:6}. Ratio $\Gamma_A/\gamma_0 =2.5$.}\label{table:1}
		\begin{tabular}{|c|c|c|c|c|c|c|}		
        \hline
        $\tau_A$,&$\tau_{A1}^s$,& \textrm{$T_{A1}$},&  \textrm{$\tau_{A2}^s$},& \textrm{$T_{A2}$},& \textrm{$g_A$}& \textrm{$\hbar \Omega_X$},\\         
        $ \rm ns$ & $ \rm ns$ &$ \rm ns$ & $ \rm ns$ & $ \rm ns$ & & $ \mu \rm eV$ \\
            \hline
    1.4 & 5 &1.1&1&0.6 &0.01 & 1 \\
    \hline
    %\hline
   \textrm{$\tau_F,$}& \textrm{$\tau_{F1}^s$},& \textrm{$T_{F1},$}&\textrm{$\tau_{F2}^s,$} &\textrm{$T_{F2},$} & \textrm{$g_F$} &\textrm{$\hbar \Omega_{FX},$} \\
   $ \rm ns$ & $ \rm ns$ &$ \rm ns$ & $ \rm ns$ & $ \rm ns$ & & $ \mu \rm eV$ \\
   \hline
	100& 50& 33 & 0.3 & 0.3 & 3.35 & 20 \\
    \hline
	\end{tabular}
\end{table}
The contributions of the bright and dark excitons, due to their large difference in lifetime, manifest themselves on different timescales. Figure~\ref{fig:AF}(b) shows the time dependences of the relative intensities ${\cal A}(t),\ {\cal F}(t)$ (Eq.~\eqref{IAFt}) that reflect the corresponding timescales. For the cw regime, the bright exciton lifetime $\tau_A = 1.4$~ns and the ratio $\Gamma_A/\gamma_0 = 2.5$ give ${\cal A} = 0.7$ and ${\cal F} = 0.3$ as seen in Fig.~\ref{fig:AF}(a). In the example calculations, we neglect the nonradiative processes for the bright and dark excitons, so that $\Gamma_{F}^{\rm r}/\Gamma_{A}^{\rm r} = \Gamma_{F}/\Gamma_{A} = 0.02$ allows us to estimate the magnetic energy of the coupling field as $g_e \mu_B b = 0.2$~meV. 
We do not consider the fluctuations of the $b$ magnitudes, which would result in the redistribution of the bright and dark exciton relative  intensities ${\cal A}(t),\ {\cal F}(t)$. If other mechanisms of dark exciton recombination besides the one provided by the ${\rm \bm b}$ field are present, they should be considered nonradiative in the presented model, and they are also neglected in the example calculations.  

The optical alignment effect presented in Fig.~\ref{fig:5}(a-c) reflects that, on timescales on the order of the bright exciton lifetime, the increase in the magnetic field leads to a redistribution of exponentially decreasing and decaying oscillating contributions in the kinetics following the pulsed excitation (Fig.~\ref{fig:5}(a)). The contribution of the dark exciton becomes noticeable on a larger scale. The decomposition of the bright and dark exciton impacts for the case of $B = 0$~T (with the maximal amplitude) is depicted in Fig.~\ref{fig:5}(b) up to $60$~ns. Figure~\ref{fig:5}(c) shows the same decomposition of the optical alignment dependence on the Faraday magnetic field in the cw regime. It demonstrates the depolarization in the magnetic field, which consists of two contours, where the broad contour corresponds to the bright exciton impact, and the narrow contour corresponds to the dark exciton one. This narrow contour was also obtained by integrating the experimental data on the kinetics of the optical alignment effect over time $t \gg \tau_A$, corresponding to the dark exciton PL. 

The rotation of the linear polarization plane shown in Fig.~\ref{fig:5}(d-f) occurs in the external Faraday magnetic field. In kinetics, this effect manifests only as a decaying oscillating contribution (Fig.~\ref{fig:5}(d)). For the parameters presented above, the contribution of the dark exciton is insignificant, and the effect arises from the bright exciton, both in the kinetics shown in Fig.~\ref{fig:5}(e) for $B = 6$~T and in the magnetic field dependence shown in Fig.~\ref{fig:5}(f). 

Regarding the optical orientation effect shown in Fig.~\ref{fig:5} (g-i), in the absence of the magnetic field, only the decaying oscillating part appears in the kinetics (Fig.~\ref{fig:5}(g)). As the field increases, the exponentially decaying part becomes more prominent. At higher times, the contribution of the dark exciton appears, as shown in Fig.~\ref{fig:5}(h) for the case $B = 6$~T up to $60$~ns.
Circular polarization is restored in the applied magnetic field, as seen in Fig.~\ref{fig:5}(i). Again, the bright exciton polarization results in the broad contour. The contribution of the dark exciton is close to zero in the absence of the magnetic field and increases with the field. However, such a narrow contour was not observed in the experiment~\cite{Smirnova2023} -- there was an opposite contour that is not described within the framework of our theory. It may be a contribution of the trions and will be investigated in future studies.   

Let us turn to the $P_c^{\rm un}$, which manifests itself in a magnetic field and is presented in Fig.~\ref{fig:5}(j-l). The set of resulting curves in different magnetic fields is shown in Fig.~\ref{fig:5}(j).
Here, the main contribution comes from the dark exciton due to its substantial $g$-factor, $g_F = 3.35$, compared to the very small bright exciton $g$-factor, $g_A = 0.01$. The impact of the bright exciton is nearly negligible, as seen in Fig.~\ref{fig:5}(k). The contribution of the third term in Eq.~\eqref{PcBt} turns out to be insignificant in our case, both at short times ($\sim \tau_A$) due to ${\cal F} \ll 1$, and at long times ($\gg \tau_A$) due to the small value of $P_{cA}^{\rm eq} \ll P_{cF}^{\rm eq}$. The same is true for both the bright exciton contribution and the third term in Eq.~\eqref{PcBt} in the magnetic field dependence shown in Fig.~\ref{fig:5}(l). 

\begin{figure*}[ht]
	\centering
	\includegraphics[width=0.8\linewidth]{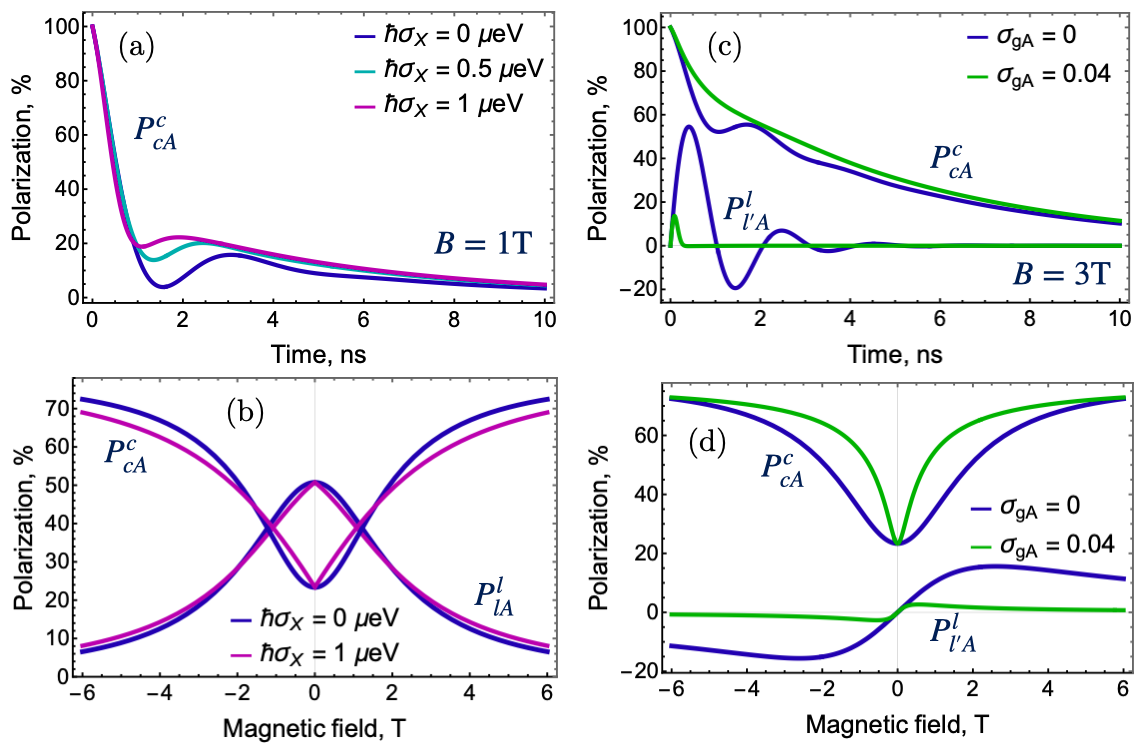}
	\caption{\label{fig:6}   Numerical calculation of the influence of the fluctuations of the anisotropic splitting $\hbar \Omega_X$ for $\hbar \Omega_{X0}=1\mu$eV on (a) kinetics of $P_c^c$ for $\hbar\sigma_X = 0.5\ \rm \mu eV$ and $\hbar\sigma_X = 1\ \rm \mu eV$ and (b) magnetic field dependences of $P_{lA}^l$ and $P_{cA}^c$ for $\hbar\sigma_X = 1\ \rm \mu eV$. The influence of the $g_A$ fluctuations for $g_{A0}=0.01$ on (c) kinetics of $P_{l'A}^l$ and $P_c^c$ for $\sigma_{gA} = 0.04$ and (d) magnetic field dependences of $P_{l'A}^l$ and $P_{cA}^c$ for $\sigma_{gA} = 0.04$. The parameters used are listed in Table~\ref{table:1}.} 
\end{figure*}

The figures are plotted for $\tau_{A2}^s = 1$~ns ($T_{A2} = T_{A1}/1.9$) and $\tau_{F2}^s = 0.3$~ns ($T_{F2} = T_{F1}/111$). However, in the experiment, oscillations were not observed for either contribution of the bright and dark excitons, which indicates fast spin dephasing in the system. This may be caused by fluctuations in the exciton parameters in the NPL ensemble, namely the anisotropic splittings and the $g$-factors. Let us take a closer look at the influence of fluctuations in the parameters of the bright exciton on its contribution to the effects under investigation. We take into account the normal distribution of the parameters for both $\hbar \Omega_X$ and $g_A$. Probability density functions of $\Omega_X$ and $\Omega_Z$ are assumed to be Gaussian:
\begin{eqnarray}
  &&  f_X (\Omega_X) = \frac{1}{\sqrt{2\pi}\sigma_{X}} e^{- \frac{(\Omega_X-\Omega_{X0})^2}{2\sigma_X^2}}\, , \\
&&    f_{Z}(\Omega_Z) = \frac{1}{\sqrt{2\pi}\sigma_{Z}}  e^{- \frac{(\Omega_Z-\Omega_{Z0})^2}{2\sigma_{Z}^2}}\, ,
\end{eqnarray}
where $\Omega_{Z0} =  g_{A0} \mu_B B/\hbar $ and $\sigma_Z = \sigma_{gA} \mu_B B/\hbar$ depend on the magnetic field. 

The suppression of oscillations for the optical alignment effect, $P_{lA}^l(t)$, in a zero magnetic field caused by fluctuations in $\hbar \Omega_{X}$ was considered in~\cite{Smirnova2025}. These fluctuations also slightly suppress oscillations in the linear polarization rotation, $P_{l'A}^l$, and in the optical orientation effect, $P_c^c$. An example is shown for $\hbar \Omega_{X0} = 1 \ \mu$eV in Fig.~\ref{fig:6}(a) for the optical orientation kinetics in $B = 1$~T for $\hbar \sigma_X = 0.5 \ \mu$eV and $\hbar \sigma_X = 1 \ \mu$eV. The influence is greatest in small magnetic fields. As the magnetic field increases, the effect becomes less pronounced. Regarding the Faraday magnetic field dependences, the contours of the effects of optical alignment, $P_{l A}^l(t)$, and optical orientation, $P_{c A}^c(t)$, become sharper in small magnetic fields, as shown in Fig.~\ref{fig:6}(b) for the case $\hbar \sigma_X = 1 \ \mu$eV comparable with $\hbar \Omega_{X0}=1 \ \mu$eV. The linear dependence of the averaged optical alignment and optical orientation on the magnetic field at small fields $\sim |\Omega_Z|$ can be obtained  analytically for the case $\hbar \Omega_{X0}=0 \ \mu$eV. The linear polarization rotation, $P_{l'A}^l(t)$, is not significantly affected. Physically, the large fluctuations of the anisotropic splitting $\hbar\sigma_X$ comparable with its mean value $\hbar \Omega_{X0}$ mean that there are NPLs with the anisotropic splitting of opposite signs in the ensemble. In our model, we have chosen the axes $X$ and $Y$ along the short and long edges of each NPL. Then, the in-plane anisotropy of the long-range electron-hole exchange interaction leads in this case to the positive anisotropic splitting $\hbar \Omega_X=E_X -E_Y>0$ in the ensemble~\cite{Hu2018,Swift2024}. However, an additional contribution to this splitting may come from the in-plane anisotropy of the cubically-symmetric short-range exchange interaction. Being of opposite sign, this contribution may partly compensate the long-range contribution or even change the sign of the resulting splitting as we  suggested in Ref.~\cite{Smirnova2025}. However, this possibility still calls for further theoretical calculations of the anisotropic splitting in semiconductor NPLs with account of both contributions. 

As for the bright exciton $g$-factor dispersion, it can be related to the fluctuations of the hole $g$-factor $g_h$. Its value is very sensitive to the mixing of the light and heavy holes, which, in turn, depends on the shape of the confining potential~\cite{Semina2021}. The values $g_A=-g_e-3g_h=0.01$ and $g_e=1.67$ correspond to $g_h=-0.56$, which is close to the experimental value for $g_h$ in thick-shell CdSe/CdS NPLs~\cite{Shornikova2018nl}. Its small variation leads to a change in the sign of $g_A$. By assuming $g_e$ fixed and only the fluctuations of $g_h$ present in the ensemble, we have $\sigma_{gA} = 3\sigma_{gh}$. Additionally, fluctuations of the effective field frequency $\bm \Omega_A$  increase with the magnetic field. Therefore, the studied effects are very sensitive to such fluctuations, especially in high magnetic fields. The strongest influence in this case is the suppression of the linear polarization plane rotation both in time and in the cw regime, which is demonstrated for $g_{A0}=\hbar \Omega_{Z0}(B)/\mu_B B=0.01$ and $B = 3$T in Fig.~\ref{fig:6}(c, d) for $\sigma_{gA} =0.04$ corresponding to $\sigma_{gh} = \sigma_{gA}/3 \approx 0.013$.  The fluctuations suppress the oscillations and affect the amplitudes of exponentially decaying components with $\tau_{A1}^s$ in the optical alignment, $P_{l A}^l$, and optical orientation, $P_{c A}^c$, effects. Their Faraday magnetic field dependences become narrower. An example for the optical orientation effect, $P_{c A}^c$, is also shown in Figs.~\ref{fig:6}(c, d).

\section{Conclusions}\label{sec:conclusions}

We present the theoretical model describing the linear optical response provided by the bright and dark exciton states in  semiconductor nanoplatelets under resonant polarized excitation of the bright exciton in the external magnetic field applied along the symmetry axis in the Faraday geometry.  The developed microscopical model is supported by symmetry considerations. We demonstrated that while the symmetry from  the individual NPL response in a magnetic field corresponds to the $C_2$  point group symmetry, the response from the in-plane isotropically oriented NPL ensemble is described by the $C_\infty$ symmetry. Importantly, in spite of the isotropic orientation of the dipoles emitting linearly polarized light, the observation of exciton optical alignment in the ensemble still allows one to detect the anisotropic splitting of excitons in individual NPLs, similar to the detection of the ``hidden'' anisotropy of the excitons localized in CdSSe solid solutions~\cite{Permogorov1983}.

The linearity of the optical response in the model presented is guaranteed by the absence of any linear or circular dichroism in the light absorption by the bright exciton, as well as by the light emission from both the bright and the dark excitons. The presence of the longitudinal spin relaxation between spin-split states in the magnetic field allows observation of the magnetic-field induced circular polarization of the bright and dark exciton emission even in the case of unpolarized excitation, while the intensity of the emitted light does not depend on the polarization of the exciting light. The linearly polarized emission under unpolarized excitation can be observed only from the individual NPL or from the in-plane oriented NPL ensemble. 

The optical alignment of the dark exciton in our model is provided by the admixture of the bright to dark exciton states by the fluctuating exchange magnetic field ``frozen'' at the lifetime of the dark exciton. Importantly, the possible in-plane anisotropy of the ``frozen'' exchange field in the NPL ensemble results in new contributions to the linearly and circularly polarized emission from the dark exciton but does not reduce the $C_\infty$ symmetry of the optical response from the ensemble.   The example calculations of all considered effects for the parameters close to those determined from the experimental data for the CdSe/CdS NPLs ensemble are presented, and the effects of the exciton parameter fluctuations in the ensemble are modelled.

\author{O.~O.~Druzhinina}
\affiliation{\affiIOFFE}
\author{A.~V.~Rodina}
\affiliation{\affiIOFFE}

\section*{Acknowledgements}
We thank M.M.~Glazov, Y.M.~Beltukov, D.S.~Smirnov, E.L.~Ivchenko, and D.R.~Yakovlev for valuable discussions. The work was supported by the Russian Science Foundation under project No. 23-12-00300-$\Pi$.

\setcounter{equation}{0}
\setcounter{figure}{0}
\setcounter{table}{0}
\renewcommand{\theequation}{A\arabic{equation}}
\renewcommand{\thefigure}{A\arabic{figure}}
\renewcommand{\thetable}{A\arabic{table}}

\renewcommand{\thesection}{}

 \section*{}

\renewcommand{\thesubsection}{Appendix A}

\subsection{Symmetry consideration.} \label{symmetry}

In general, the non-normalized polarization second-rank tensors of the exciting, $\langle E_k^0 E_n^{0*} \rangle$, and emitted, $\langle E_i E_j^*\rangle$, light are connected by the four-rank tensor as
\begin{equation} \label{EAE0}
\langle E_i E_j^* \rangle = A_{ijkn} \langle E_k^0 E_n^{0*} \rangle \, . 
\end{equation}
Here ${\bm E} = E {\bm e}$ and  ${\bm E}^0 = E_0 {\bm e_0}$ are the electric field vectors of the emitted and exciting light, and $\langle ..\rangle$ denotes the time averaging. The number of nonzero independent components of the four-rank tensor $A_{ijkn}$ depends on the symmetry of the structure. We consider the light propagating along $z$ direction, so that $i,j,k,n= x,y$, and the Faraday magnetic field ${\bm B}$ is along the $z$ direction. 

The general properties of a fourth-rank tensor describing the secondary emission or scattering of light   that do not follow from its specific point-group symmetry were considered in Ref.~\cite{Ivchenko1977} and can be applied to $A_{ijkn}$. The first restriction on the number of independent components of the tensor $A_{ijkn}$ follows directly from the definition in Eq.~\eqref{EAE0}. By considering the complex conjugate, we obtain 
$$
(*) \quad A_{jink} = A_{ijkn}^*\, , $$
so that the components $A_{xxxx}$, $A_{yyyy}$, $A_{xxyy}$ and $A_{yyxx}$ are real. The second one is  related to the time reversal operation $t \to -t$ and can be reformulated in a Faraday magnetic field as 
\begin{align}
    (**) \, A_{ijkn}(\omega,\omega_0,B) = A_{knij}(\omega_0,\omega,-B) \exp \frac{\hbar(\omega_0-\omega)}{kT}\, . \nonumber
\end{align}
Here we neglect the spatial dispersion but include the dependence on the frequencies of the emitted, $\omega$, and absorbed, $\omega_0$, light. The exponential factor reflects the relation between the intensities similar to those of the Stokes and anti-Stokes components of the scattered light \cite{long2002raman}. An accurate derivation of this property and its consequences will be considered elsewhere. 

In the following consideration, we first take into account the property (*) and consider the restrictions that follow from the point group symmetry. Then we turn to the consequence of property (**) at the end of this Section. 
We  consider the $C_{2v}$ symmetry point group for the individual NPLs (reduced to $C_2$ with ${\bm B} \parallel z$; the mirror planes are excluded)  and $C_{\infty v}$ for the ensemble (reduced to $C_\infty$ with ${\bm B} \parallel z$). 

For the $C_{2v}$ point group, if we choose the $x,y$ axes perpendicular to the $\sigma_v$ planes, all components with an odd number of $x$ and $y$ indices vanish. Other symmetry operations give $x \rightarrow - x, y \rightarrow \pm y$ and $x \rightarrow \pm x, y \rightarrow - y$. As a result, there are eight nonzero independent components:
$A_{xxxx} \ne A_{yyyy}$,  $A_{xxyy} \ne A_{yyxx}$, $A_{xyxy} = A_{yxyx}^*$, $A_{xyyx} = A_{yxxy}^*$. 
The  last four components are complex and can be rewritten in terms of the real and imaginary parts, which are related to their symmetric and antisymmetric combinations, respectively:
\begin{align}
   & A_{xyxy} = A_{\rm s} + i A_{\rm as}  \, , \, \,  
     A_{yxyx} = A_{\rm s} - i A_{\rm as}  \, , \nonumber \\ 
   & A_{xyyx} = B_{\rm s} + i B_{\rm as}  \, , \, \,  
     A_{yxxy} = B_{\rm s} - i B_{\rm as}  \, . \nonumber
\end{align}

In the ${\bm B} \parallel z$, the  $C_{2v}$ symmetry is lowered to the $C_2$  group, and the change of sign of both $x,y$ components is allowed. As a result, there are eight more complex nonzero components with an odd number of $x$ and $y$: $A_{xxxy} = A_{xxyx}^*$, $A_{xyxx} = A_{yxxx}^*$, $A_{yyyx} = A_{yyxy}^*$, $A_{yxyy} = A_{xyyy}^*$ that can be expressed in terms of symmetric and antisymmetric combinations as
 \begin{align}
&    A_{yxxx}=A_{\rm s}^{xx} + i A_{\rm as}^{xx}  \, , \, \, 
     A_{xyxx}=A_{\rm s}^{xx} - i A_{\rm as}^{xx}  \, , \nonumber \\
 &    A_{yxyy}=A_{\rm s}^{yy} + i A_{\rm as}^{yy}  \, , \, \, 
     A_{xyyy}=A_{\rm s}^{yy} - i A_{\rm as}^{yy}  \, , \\
 &     A_{xxxy}=A^{\rm s}_{xx} + i A^{\rm as}_{xx}  \, , \, \, 
     A_{xxyx}=A^{\rm s}_{xx} - i A^{\rm as}_{xx}   \, , \nonumber \\
 &    A_{yyxy}=A^{\rm s}_{yy} + i A^{\rm as}_{yy}  \, , \, \, 
        A_{yyyx}=A^{\rm s}_{yy} - i A^{\rm as}_{yy}  \, .
\end{align} The new components are the odd functions with respect to the magnetic field.

If we have full in-plane isotropy (the $C_{\infty v}$ point group in zero magnetic field and $C_{\infty}$ with ${\bm B} \parallel {\bm k} \parallel z$), the rotation around the $z$-axis by $\pi/2$, when $x \rightarrow y, y \rightarrow  -x$, gives $A_{xxxx}=A_{yyyy}$, $A_{xxyy}=A_{yyxx}$, $A_{xyyx}=A_{yxxy}$, $A_{xyxy}=A_{yxyx}$, and therefore $A_{\rm as}=B_{as}=0$ for even components. The rotational symmetry also gives the relation
$$A_{xxxx}=A_{xxyy}+A_{xyxy}+A_{xyyx}.$$
Therefore, there are only three independent components of the $A$ tensor in a zero magnetic field. In the magnetic field, for odd components, we obtain $A_s^{xx} = - A_s^{yy}$, $A_{as}^{xx} =A_{as}^{yy}$,  $A^s_{xx} = - A^s_{yy}$, and $A^{as}_{xx} =A^{as}_{yy}$  and the restriction:
$$ A_{xyyy}=A_{xxxy}+A_{xxyx}+A_{xyxx}.$$
Therefore, the tensor $A$ has three additional components.

We now want to relate the polarization vectors comprising the Stokes parameters of the exciting, ${\bm P}$,  and emitted, ${\bm P}^0$, light. First, we obtain the intensity of the emitted light from Eq.~\eqref{EAE0} as: 
\begin{eqnarray}\label{Ifull}
I =&& I^0/2  (A_{xxxx}+A_{yyyy}+A_{xxyy}+A_{yyxx})   \\
+&& I^0_l/2 (A_{xxxx}-A_{yyyy}-A_{xxyy}+A_{yyxx}) \nonumber \\
+&& I^0_{l'}/2 (A_{xxxy}+A_{xxyx}+A_{yyxy}+A_{yyyx}) \nonumber \\
+&& i I^0_c/2 (-A_{xxxy}+A_{xxyx}-A_{yyxy}+A_{yyyx}) \, , \nonumber
\end{eqnarray}
where  $I^0=|E^0|^2 =|E_x^0|^2+|E_y^0|^2$ is the intensity of the incident light. Thus, generally, the intensity $I$ depends not only on $I^0$, but also on the polarization of the exciting light via nonnormalized Stokes parameters  $I_{\beta}^0 = I^0 P_0^\beta$, $\beta=l,l',c$ of the exciting light. In this case, the intensity $I$ and nonnormalized Stokes parameters $I_l =  |E_x|^2-|E_y|^2 = P_l I$, $I_{l'} = \langle E_xE_y^* + E_x^*E_y \rangle = P_{l'} I$ and $I_c = i \langle E_xE_y^* - E_x^*E_y \rangle = P_{c} I$ of the emitted and excited light are connected by the Mueller matrix~\cite{Born1999principles}. To complete it, the expressions for $I_{l,l',c}$ can be readily obtained from Eq.~\eqref{EAE0} in addition to Eq.~\eqref{Ifull}.

The conditions to relate the normalized polarization vectors ${\bm P}$ and ${\bm P}_0$ linearly by Eq.~\eqref{PLP0} consist of the absence, in the symmetry group or in the microscopic model, of the last three terms in Eq.~\eqref{Ifull} that depend on the incident light polarization: \begin{eqnarray}
&& A_{xxxx}+A_{yyxx}=A_{yyyy}+A_{xxyy}\, , \nonumber\\
&& A^{s}_{xx}+A^{\rm s}_{yy} =0  \, , \nonumber \\
&& A^{\rm as}_{xx}+A^{\rm as}_{yy}  = 0\, . \label{cond}
\end{eqnarray}
These conditions are not automatically fulfilled in the considered symmetry groups. Indeed, the first two conditions can be violated in $C_{2v}$ point group when linear polarization dichroism in light absorption is present. The third condition can be violated even in  $C_\infty$ group because of the magnetic field induced dichroism for the absorption of circularly polarized light. We focus on the microscopic model that guarantees the fulfilment of all three conditions by neglecting any circular or linear dichroism in the absorption of light, relaxation between the exciton states before their recombination, as well as any linearly or circularly spin-dependent recombination.

Assuming the conditions in Eq.~\eqref{cond} to be fulfilled, we further obtain for the $C_2$  point symmetry group the emitted light intensity as
$$I = I^0(A_{xxxx}+A_{yyxx}) \, $$
and a linear relation between the polarization vector:
 $$P_{\alpha} = P_\alpha^{\rm un} + \Lambda_{\alpha \beta} P_0^\beta \, , \quad \alpha,\beta = l,l',c  \, , $$ 
 where $P_{\alpha} = I_{\alpha}/I$ and $P_0^{\beta} = I_{\beta}^0/I^0$.
 For $I^0=1$ and with the use of the conditions Eq.~\eqref{cond} we obtain:
\begin{eqnarray}
I_l =P_l I =&&  (A_{xxxx}-A_{yyyy}) 
+  (A_{xxxx}-A_{xxyy})P_0^l\, \nonumber \\
+&&   (A^{\rm s}_{xx} -A^{\rm s}_{yy}) P_0^{l'} 
+ (A^{\rm as}_{xx} -A^{\rm as}_{yy})P_0^c \, .  
\end{eqnarray}
\begin{eqnarray}
I_{l'}=P_{l'} I =&&  (A_{\rm s}^{xx} +A_{\rm s}^{yy})  +  (A_{\rm s}^{xx} -A_{\rm s}^{yy}) P_0^l \nonumber \\ +&& (A_{\rm s}+B_{\rm s}) P_0^{l'} +  (A_{\rm as}-B_{\rm as})  P_0^c \, .  
\end{eqnarray}
\begin{eqnarray}
I_{c}=P_c I =&&  (A_{\rm as}^{xx}+A_{\rm as}^{yy}) +  (A_{\rm as}^{xx}-A_{\rm as}^{yy}) P_0^l \nonumber \\
-&&   (A_{\rm as}+B_{\rm as}) P_0^{l'} +  (A_{\rm s}-B_{\rm s}) P_0^c\, . 
\end{eqnarray} 

Thus, in the $C_2$ group describing the individual NPL in the magnetic field along the anisotropic axis, one can observe both linearly and circularly polarized light under unpolarized excitation, even when the conditions of Eq.~\eqref{cond} are fulfilled. This polarization is related to the thermodynamically equilibrium distribution of the exciton populations. In addition, the  linear-to-circular and \textit{vice versa} polarization conversions can be expected, as well as the polarization in-plane anisotropy --  the dependence on the linear polarization of the incident light with respect to the orientation of the NPL axis.  In general, we obtained nine independent nonzero components of the matrix $\hat \Lambda$ for the $C_2$ point group symmetry.

In the $C_\infty$ point group, the linear polarizations $P_l^{\rm un} =  P_{l'}^{\rm un} = 0$ vanish, and there is only  one nonzero component $P_c^{\rm un}(B) = -P_c^{\rm un}(-B)$ describing the polarized PL in magnetic field under unpolarized excitation.  For the polarized-excitation response,  we obtain $\Lambda_{lc} =\Lambda_{cl} = \Lambda_{l'c} = \Lambda_{cl'}=0$, while $\Lambda_{ll} =\Lambda_{l'l'}$ and $\Lambda_{ll'} = -\Lambda_{l'l}$. Thus, the number of independent nonzero components of the matrix $\hat \Lambda$ is reduced to three for the $C_\infty$ point group symmetry.

We remind that we have not used the property (**) related to the time-inversion symmetry up to now. In the theory of the secondary emission, it is often possible to neglect the difference between the frequencies of the absorbed and emitted light and thus disregard the exponential factor in (**).  In such a case, which is equivalent to the regime of resonant scattering~\cite{Pikus_book1982}, it results in the following additional constraints:
$$ (i) \quad A_{as}(B) = A_{as}(-B) \, , \quad B_{as} = 0 \, ,$$
resulting into $$\Lambda_{l'c}(B)=\Lambda_{l'c}(-B) = -\Lambda_{cl'}(B)=-\Lambda_{cl'}(-B) \, ; $$
and
$$(ii) \quad A_{xx}^s=A_s^{xx} \,  , \quad A_{yy}^s=A_s^{yy}, $$ resulting into $$\Lambda_{l'l}(B)= - \Lambda_{l'l}(-B) = -\Lambda_{ll'}(B)= \Lambda_{ll'}(-B) \, ; $$
and
$$(iii) \quad A_{xx}^{as}=-A_{as}^{xx} \,  , A_{yy}^{as}=-A_{as}^{yy}, $$ resulting into $$\Lambda_{cl}(B)=  -\Lambda_{cl}(-B) = \Lambda_{lc}(B)= -\Lambda_{lc}(-B) \, , $$
so that only six of nine components of the matrix are independent in the $C_2$ point group symmetry. However, the relations (ii)-(iii) together with $A_{xxyy} = A_{yyxx}$ combined with the conditions Eq.~\eqref{cond} result into ${\bm P}^{un}=0$. We remind that under the conditions Eq.~\eqref{cond}, the nonzero components of the $P_{\alpha}^{un}$ vector appear only due to the energy relaxation of excitons to the lowest energy state during the longitudinal relaxation time. Therefore, the account of the state population reflected by the exponential factor in (**) is important for the description of ${\bm P}^{un} \ne 0$.

\setcounter{equation}{0}
\setcounter{figure}{0}
\setcounter{table}{0}

\renewcommand{\theequation}{B\arabic{equation}}
\renewcommand{\thefigure}{B\arabic{figure}}
\renewcommand{\thetable}{B\arabic{table}}
\renewcommand{\thesubsection}{Appendix B}

\subsection{Bright and dark exciton dipole matrix elements} \label{dipoles}

In the dipole approximation, the exciton interaction with light is described by the matrix elements of the perturbation operator 
$\hat V_{\rm el} = e/(m_0 \omega){\bm E} \hat {\bm p}$, where $e$ and $m_0$ are the charge and mass of the free electron, $\omega$ is the frequency of the electromagnetic field  
${\bm E} = E [ \exp(-i \omega t){\bm e} + \exp(i \omega t){\bm e^*}]$~\cite{Ivchenko2005,Anselm}. The matrix elements, both for the absorption, $<\psi_{\rm ex}|\hat V_{\rm el}|G>$,  and emission, $<G|\hat V_{\rm el}|\psi_{\rm ex}>$, of light with $\psi_{\rm ex}$ describing the exciton state and $G$ being the unexcited ground state, are proportional to the matrix elements of the momentum  operator $\hat {\bm p} = -i\hbar \nabla$. We assume that the amplitude $E_{\rm in} = E$ of the electric field inside the sample is independent on the light polarization vector ${\bm e}_0$ (for absorption) and  ${\bm e}^*$ (for emission) and that the bright exciton dipole matrix elements $|d_{+1}|=|d_{-1}|$ are~\cite{Rodina2016}
\begin{equation}
|d_{\pm 1}| \propto |\langle G | \hat p_x\pm i\hat p_y | \psi_{\pm 1}\rangle|  = \frac{|PK|}{\sqrt{2}}\,   ,  
\end{equation}
where $P$ is the matrix element taken between the valence band and conduction band Bloch functions, and $K=\int \Phi_{\rm ex}({\bm r},{\bm r}) d^3{\bm r}$, where  the exciton envelope function $\Phi_{\rm ex}({\bm r}_e,{\bm r}_h) $ is the same for all exciton states.   Note that $|d_\perp|^2=|d_{+1}|^2+ |d_{-1}|^2 =|d_{X}|^2+ |d_{Y}|^2 = |d_{X'}|^2+ |d_{Y'}|^2 $. The dipole matrix elements for the linearly polarized bright excitons also satisfy $|d_X|=|d_Y|$ and $|d_{X'}|=|d_{Y'}|$. In this approximation, the bright exciton radiative rate $\Gamma_A^{\rm r} \propto |E|^2 |d_\perp|^2$ is the same for all exciton states. Note, that it is the choice of the in-plane isotropic exciton envelope function $\Phi_{\rm ex}({\bm r}_e,{\bm r}_h)$ and the neglect of the effect of the local field anisotropy \cite{Rodina2016,Rodina2016JETP} that allow us to use the isotropic dipole matrix elements in spite of the account of the anisotropic exciton splitting in our model.

For the dark exciton states $|\pm 2>$, the dipole matrix elements vanish in first-order perturbation theory. However, the admixture of the bright exciton states by different mechanisms~\cite{Rodina2016,Rodina2018ftt,Shornikova2020nn} results in non-zero dipole matrix elements of the dark exciton and its non-zero radiative recombination rate. We consider here only one mechanism, providing the matrix element of the electron spin-flip in the exciton. For example, let us consider the magnetic field in the plane of the NPL: $\bm{b} = b_X \bm e_X + b_Y \bm e_Y$. In the basis of   $ |+1\rangle, |-1\rangle, |+2\rangle, |- 2\rangle$ exciton states, the  perturbation $V^{e} = g_e \mu_B \tilde V^{e} = g_e \mu_B( \bm \sigma \cdot \bm b)$ takes the form 
\begin{eqnarray}
\\ \tilde V_{\pm1,\pm2}^e = \begin{pmatrix}
0 & 0& b_X+ib_Y& 0\\
0&0&0 &b_X -ib_Y\\
b_X-ib_Y& 0&0& 0\\
0&b_X+ib_Y & 0& 0 \end{pmatrix}. \nonumber
\end{eqnarray} 

The direction of the field, characterized by the angle $\varphi$ between $\bm b$ and $X$ NPL axis (see Fig.~\ref{fig:fig1}(d)), determines the coupling between pseudospin components of  the bright and dark excitons. 
The magnetic field ${\bm b}$ can originate from the fluctuating exchange field created by the paramagnetic centers at the NPL surface~\cite{Rodina2015,Biadala2017,Shornikova2020nn} or core/shell interface, so that the angle $\varphi$ can fluctuate in time and be different in different individual NPLs. In the case of the fluctuation nature, the longitudinal field component $b_z$ is also present with $|b_z| = |b|/\sqrt{2}$; however, it does not mix the bright and dark exciton states.

The radiative recombination of the dark exciton is provided by the admixture of the bright and dark exciton wave functions by the perturbation $\hat V_e$. Thus, the matrix $\hat G^{FA}$ in Eq.~\eqref{GFA} describes the admixture of the bright to dark exciton pseudospin components. We assume that the amplitude of the magnetic field $|b|$ is not affected by the external magnetic field ${\bm B}$, so that $|d_{+2}|=|d_{-2}|$, $|d_{FX}|=|d_{FY}|$, $|d_{FX'}|=|d_{FY'}|$. 
The radiative rate of the dark exciton in this case is given by~\cite{Rodina2016,Rodina2018jem,Rodina2018ftt}:
\begin{equation} \label{GammaFr}
\Gamma_F^{\rm r} = \left(\frac{g_e\mu_B b}{2\Delta E_{AF}}\right)^2 \Gamma_A^{\rm r} \, \equiv \beta \, \Gamma_A^{\rm r}  \, .
\end{equation}
This result is obtained within second order perturbation theory under the condition $\beta \ll 1$.

In turn, the matrix $\hat G^{AF}$ reversed to $\hat G^{FA}$ describes the admixture of the dark exciton pseudospin to the bright exciton pseudospin components and is responsible for the energy relaxation from the bright to the dark exciton at low temperatures. If we assume that the relaxation is provided only by this admixture mechanism with the emission of the resonant acoustic phonon with the energy $E_{\rm ph} = \Delta E_{AF}$, which does not affect the exciton spin, we obtain that its rate
$\gamma_0 \propto \beta \Gamma_{\rm ph}$.
The rate of the resonant phonon emission $\Gamma_{\rm ph}$ might additionally depend on the exciton splitting  $\Delta E_{AF}$.

The microscopic conditions described above guarantee the fulfilment of the conditions Eq.~\eqref{cond} for $C_\infty$ and $C_2$ symmetry point groups with a Faraday magnetic field.

\setcounter{equation}{0}
\setcounter{figure}{0}
\setcounter{table}{0}
\renewcommand{\theequation}{C\arabic{equation}}
\renewcommand{\thefigure}{C\arabic{figure}}
\renewcommand{\thetable}{C\arabic{table}}
\renewcommand{\thesubsection}{Appendix C}

\subsection{ Bright and dark exciton populations}\label{Kinetics}

The interaction between the bright and dark excitons is realized by the mutual transitions between the states. 
With account the exciton generation and recombination processes, the bright and dark exciton populations satisfy the system of rate equations:
\begin{eqnarray}\label{Population} 
\frac{d N_{A}}{d t} &=& -\frac{N_A}{\tau_A} + \gamma_{\rm th} N_F + G_A \, ,\\
\frac{d N_{F}}{d t} &=& -\frac{N_F}{\tau_F} + (\gamma_0+\gamma_{\rm th})N_A +G_F\, ,
\end{eqnarray}
where $\tau_{A}$ and $\tau_{F}$ are the bright and dark exciton lifetimes defined as:
\begin{equation}\label{tauT}
	\tau_A^{-1} = \Gamma_A+\gamma_0+\gamma_{\rm th}\, , \ \ \ \ \tau_F^{-1} = \Gamma_F +\gamma_{\rm th}\,  .
\end{equation} 
Here $\Gamma_{A}$ and $\Gamma_{F}$ are the recombination rates of the bright and dark exciton, $\gamma_0$ is the relaxation rate from the bright state to the dark one at zero temperature, $\gamma_{\rm th}$ is the thermally activated phonon-assisted relaxation rate $\gamma_{\rm th} = \gamma_0 N_B$, where  $N_B (E)  = 1/(\exp(E/k_B T)-1)$  is the Bose–Einstein phonon occupation number at the given energy $E$ and temperature $T$, and $k_B$ is the Boltzmann constant. 
Terms $G_{A}$ and $G_{F}$ describe the exciton generation by the incident light (bright and dark exciton population created per unit time)  with $G_A + G_F = G_{0}$. 

Firstly, we are interested in stationary solutions when $d N_A/dt=d N_F/dt =0$, under constant-wave pumping with $G_A(t), G_F(t) = \rm const$. In the general case, they are given by
\begin{gather}\label{populations_cw}
    N_A^0 = \frac{\tau_A (G_A + (1-\Gamma_F \tau_F)  G_F)} {\Gamma_F \tau_F + \Gamma_A\tau_A (1-\Gamma_F \tau_F)}\, , \\
	N_F^0 = \frac{\tau_F( (1-\Gamma_A \tau_A) G_A +  G_F)}{\Gamma_F \tau_F + \Gamma_A\tau_A (1-\Gamma_F \tau_F)}\, .
\end{gather} 
In the case of resonant or quasi-resonant excitation, we can neglect the small oscillator strength of the dark exciton and assume that only $| \pm 1 \rangle$ states are excited by the laser. Therefore, $G_A = 1, G_F = 0$. In the case of non-resonant excitation, the equal population of both exciton pairs of states can be created during the energy and spin relaxation, which corresponds to the case $G_A = G_F = 1/2$.

In the case of pulsed excitation, both the duration of the pulse and the relaxation time are very short, so that it is safe to approximate $G_{A}(t)$ and $G_{F}(t)$ as $N_{A}(0) \delta(t)$ and $N_{F}(0) \delta(t)$, respectively, where $\delta(t)$ is the Dirac function and $N_{A}(0)$ and $N_{F}(0)$ are the bright and dark exciton populations created by the short pulse at $t=0$, the solutions for the bright and dark exciton populations can be written as a sum of two exponents:
\begin{eqnarray}\label{NAFpulsed}
    N_A (t) =  C_Ae^{-\Gamma_{\rm S}t}+ \left(N_A(0) - C_A \right)e^{-\Gamma_{\rm L}t}, \\
    N_F (t) = C_F  e^{-\Gamma_{\rm S}t}+\left(N_F(0) - C_F \right)e^{-\Gamma_{\rm L}t},
\end{eqnarray}
where $\Gamma_{\rm S}$ and $\Gamma_{\rm L}$ ($\Gamma_{\rm S}>\Gamma_{\rm L}$) are the rates of the short-lasting and long-lasting decays, respectively:
\begin{widetext}
\begin{equation}
\Gamma_{\rm S,L} = \frac{1}{2}\left(\Gamma_A+\Gamma_F +\gamma_0 + 2\gamma_{\rm th} \pm \sqrt{\left( \gamma_0 +\Gamma_A -\Gamma_F \right)^2 + 4\gamma_0 \gamma_{\rm th} + 4\gamma_{\rm th}^2} \right)\, .
\end{equation}
\end{widetext}
For the resonant excitation, we again assume that only the bright exciton is excited $N_A(0) = N_0, N_F(0) = 0$ and obtain
\begin{equation}
C_A= \frac{\gamma_0 +\Gamma_A - \Gamma_F + \Gamma_{\rm S}- \Gamma_{\rm L} }{2(\Gamma_{\rm S} - \Gamma_{\rm L})}, \ \ \ C_F = \frac{-\gamma_0 -\gamma_{\rm th}}{(\Gamma_{\rm S} - \Gamma_{\rm L})}\, .
\end{equation}
In the case of non-resonant excitation, after supposedly fast relaxation from the excited level higher in energy, $N_A(0) = N_F(0) = N_0/2$, we obtain the constants
\begin{equation}
C_A = \frac{\gamma_0 +\Gamma_A - \Gamma_{\rm L}}{2(\Gamma_{\rm S} - \Gamma_{\rm L})}, \ \ \ C_F = \frac{-\gamma_0 +\Gamma_F - \Gamma_{\rm L}}{2(\Gamma_{\rm S} - \Gamma_{\rm L})}\, .
\end{equation}

At the relatively large temperatures $T \sim \Delta E_{AF}/k_B$ comparable to the bright--dark exciton splitting and under the condition $\gamma_0+\gamma_{\rm th} \gg \Gamma_A, \Gamma_F$, the population of the exciton states tends to thermodynamic equilibrium. In this case
\begin{eqnarray}
  N_A^{\rm eq}(t) = N_0\frac{1}{1+\exp{(\Delta E_{AF}/k_BT)}} e^{-\Gamma_{\rm L}t}\, , \\
  N_F^{\rm eq}(t) = N_0\frac{\exp{(\Delta E_{AF}/k_BT)}}{1+\exp{(\Delta E_{AF}/k_BT)}} e^{-\Gamma_{\rm L}t}
\end{eqnarray}
at $\tau \gg \tau_A$ with
\begin{equation*}
    \Gamma_L \approx \Gamma_A\frac{1}{1+\exp{(\Delta E_{AF}/k_BT)}}+ \Gamma_F\frac{\exp{(\Delta E_{AF}/k_BT)}}{1+\exp{(\Delta E_{AF}/k_BT)}}\, . 
\end{equation*}
In the opposite case of low temperatures $T \ll \Delta E_{AF}/k_B$ so that $\gamma_{\rm th} \approx 0$ and $\gamma_0 \sim \Gamma_A$, the dynamics of $N_{A,F}(t)$ at the resonant excitation is  described by Eqs.~\eqref{NAFt} in the main text. The steady-state solutions at low temperature are described by Eq. ~\eqref{NA0NF0}  in the main text. Below, we discuss the relative intensities of the bright and dark exciton recombination at low temperatures and at the resonant CW or pulsed excitation of the bright exciton.

The intensity is related to the exciton populations as $$ I =  \Gamma_A^{\rm r} N_A + \Gamma_F^{\rm r} N_F \, , $$
where $\Gamma_{A}^{\rm r}$ and $\Gamma_{F}^{\rm r}$ are the radiative recombination rates.

At low temperature and resonant excitation, the relative contributions of the bright, ${\cal A}$, and dark, ${\cal F}$, excitons to the PL intensity and polarization are given by:
\begin{eqnarray} \label{IAF}
    {\cal F}= 1- {\cal A} = \frac{1}{1+\frac{I_A}{I_F}},  \quad
     \frac{I_A}{I_F} =\frac{\Gamma_A^{\rm  r}}{\gamma_0}\frac{ 1}{ \Gamma_F^{\rm  r}\tau_F} \approx  \frac{\Gamma_A}{\gamma_0}, 
\end{eqnarray}
where the last equality in Eq.~\eqref{IAF} is valid at low temperature in the case $\Gamma_{A,F} = \Gamma_{A,F}^{\rm  r}$. It means that the radiative recombination is substantial, even for the dark exciton ($\Gamma_F^{\rm  r} \tau_F = 1$). Therefore, 
\begin{equation}
    {\cal A} \approx \Gamma_A \tau_A ,\ {\cal F}  \approx  \gamma_0 \tau_A 
\end{equation}
and 
$$\frac{{\cal F}}{{\cal A}} \approx \frac{\gamma_0}{\Gamma_A} \propto \beta \, \frac{\Gamma_{\rm ph}}{\Gamma_A} \, .$$ The dependences of ${\cal A}$ and ${\cal F}$ on the ratio $\gamma_0/\Gamma_A$ are shown in Fig.~\ref{fig:AF}(a) for this case. 
Note that even under the condition $\beta = \Gamma_F^{\rm r}/\Gamma_A^{\rm r} = \Gamma_F/\Gamma_A \ll 1$, the situation ${\cal F}/{\cal A} > 1$ can be realized in the case $\Gamma_{\rm ph} \gg \Gamma_A$.

In the limit, when nonradiative recombination for the dark exciton dominates, $\Gamma_F^{\rm  r} \tau_F \ll 1$ and  $I_F/I_A \ll 1$, we obtain
\begin{equation}
 {\cal F} \approx  \frac{\gamma_0}{\Gamma_{A}^{\rm  r}} \Gamma_F^{\rm  r} \tau_F \propto \beta^2 \Gamma_{\rm ph}\tau_F\,  , \ \  {\cal A} \approx 1 -  \frac{\gamma_0}{\Gamma_{A}^{\rm  r}} \Gamma_F^{\rm  r} \tau_F.
\end{equation}

For pulsed excitation, in the limit $\Gamma_{A,F} = \Gamma_{A,F}^{\rm  r}$, we obtain 
\begin{eqnarray} \label{IAFt}
    {\cal F}(t)&=& 1- {\cal A}(t) = \frac{1}{1+\frac{I_A}{I_F}(t)}, \\ 
       \frac{I_A}{I_F}(t) &=& \frac{\Gamma_A^{r}}{\Gamma_F^{r}}\frac{ 1}{{\gamma_0} }\left( \frac{1}{\tau_A} - \frac{1}{\tau_F} \right) \frac{1}{e^{t\left(\frac{1}{\tau_A}-\frac{1}{\tau_F}\right)}-1}  \nonumber = \\
       &=&\frac{ \Gamma_A (\tau_F-\tau_A)}{\gamma_0 \tau_A }  \frac{1}{e^{t\left(\frac{1}{\tau_A}-\frac{1}{\tau_F}\right)}-1} \, . \nonumber
\end{eqnarray}
Note that ${\cal A} (t\gg \tau_A) = 0 , \  {\cal F} (t\gg \tau_A)= 1$, in this limit. However, the actual time from which this limit is realized might be one order of magnitude larger than $\tau_A$ and depends on the ratio $\gamma_0/\Gamma_A$ (see Fig.~\ref{fig:AF}(b)).

\setcounter{equation}{0}
\setcounter{figure}{0}
\setcounter{table}{0}
\renewcommand{\theequation}{D\arabic{equation}}
\renewcommand{\thefigure}{D\arabic{figure}}
\renewcommand{\thetable}{D\arabic{table}}
\renewcommand{\thesubsection}{Appendix D}
\subsection{Bright exciton pseudospin components in Faraday magnetic field} \label{SS_bright}

Here we present the explicit  expressions for the components of the matrix $M^{A}_{\gamma \delta}$ ($\gamma, \delta = X,Y,Z$) and of the pseudospin $S_{A\gamma}^{\rm eq}$ entering Eq.~\eqref{SAFM}. For the bright  exciton in the steady state regime $d {\bm S_A}/dt=0$ we obtain:
\begin{eqnarray}
   && M^A_{XX}= \frac{T_{A1}}{\tau_A} \frac{\Omega_X^2}{\Omega_A^2} + \frac{T_{A2}}{\tau_A} \frac{\Omega_Z^2}{\Omega_A^2 (1+T_{A2}^2\Omega_A^2)} \, , \nonumber \\  
     && M^A_{YY}= \frac{T_{A2}}{\tau_A} \frac{1}{(1+T_{A2}^2\Omega_A^2)} \, ,  \nonumber \\
    &&  M^A_{ZZ}= \frac{T_{A1}}{\tau_A} \frac{\Omega_Z^2}{\Omega_A^2} + \frac{T_{A2}}{\tau_A} \frac{\Omega_X^2}{\Omega_A^2 (1+T_{A2}^2\Omega_A^2)} \, ,  \\ 
 &&     M_{YX}^A = - M_{XY}^A = \frac{T_{A2}}{\tau_A} \frac{T_{A2} \Omega_Z }{ (1+T_{A2}^2\Omega_A^2)} \, , \nonumber \\
   &&   M_{ZY}^A = - M_{YZ}^A  = \frac{T_{A2}}{\tau_A} \frac{T_{A2} \Omega_X}{ (1+T_{A2}^2\Omega_A^2)} \, , \nonumber\\
      && M_{XZ}^A = M_{ZX}^A = \frac{T_{A1}}{\tau_A} \frac{\Omega_X \Omega_Z}{\Omega_A^2} - \frac{T_{A2}}{\tau_A} \frac{\Omega_X \Omega_Z}{\Omega_A^2 (1+T_{A2}^2\Omega_A^2)} \, . \nonumber
\end{eqnarray}
Note that the matrix $M^A_{\gamma \delta}$ ($\gamma, \delta = X,Y,Z$) has the same symmetry as $\Lambda_{\alpha \beta}$ ($\alpha, \beta = l,l',c$) considered in~\ref{symmetry} for the $C_{2v}$ symmetry point group with the magnetic field applied along the $C_2$ axis. Additionally, we obtained the relations $M^A_{YX} = - M^A_{XY}$, $ M_{ZY}^A = - M_{YZ}^A $, and $M_{XZ}^A = M_{ZX}^A $ that do not directly follow from the symmetry operations of the $C_2$ group so that the number of nonzero independent components is reduced to six. These additional symmetries reflect the time-reversal symmetry condition neglecting the difference between the exciton state energies involved in exciton  generation and pseudospin evolution in a magnetic field. However, the nonzero thermodinamically equilibrium averaged pseudospins reflecting this difference are included (see Eq.~\eqref{SAFeq}).

The components of $M^F_{\gamma \delta}$ in the steady-state regime $d {\bm S_F}/dt=0$ have the same form, with all the parameters of the bright exciton replaced by those of the dark exciton.

The time-dependent solutions for the bright exciton pseudospin components are:
\begin{eqnarray}
   && M^A_{XX} (t)=  \frac{\Omega_X^2}{\Omega_A^2} e^{-t/\tau_{A1}^s} +  \frac{\Omega_Z^2}{\Omega_A^2 } \cos(\Omega_A t )e^{-t/\tau_{A2}^s} \, ,  \\
   &&   M^A_{YY} (t)= \cos(\Omega_A t )e^{-t/\tau_{A2}^s} \, ,  \nonumber \\
    &&  M^A_{ZZ} (t)=  \frac{\Omega_Z^2}{\Omega_A^2} e^{-t/\tau_{A1}^s} +  \frac{\Omega_X^2}{\Omega_A^2} \cos(\Omega_A t )e^{-t/\tau_{A2}^s} \, , \nonumber \\ 
 &&     M_{YX}^A (t) = - M_{XY}^A (t) =  \frac{ \Omega_Z }{ \Omega_A} \sin(\Omega_A t )e^{-t/\tau_{A2}^s} \, , \nonumber \\
  &&    M_{ZY}^A(t) = - M_{YZ}^A (t)  =  \frac{\Omega_X}{ \Omega_A} \sin(\Omega_A t )e^{-t/\tau_{A2}^s}\, , \nonumber\\
      && M_{XZ}^A (t) = M_{ZX}^A (t) = \nonumber \\
      && \qquad \qquad \frac{\Omega_X \Omega_Z}{\Omega_A^2} e^{-t/\tau_{A1}^s} - \frac{\Omega_X \Omega_Z}{\Omega_A^2} \cos(\Omega_A t )e^{-t/\tau_{A2}^s}  \, .\nonumber 
\end{eqnarray}

\setcounter{equation}{0}
\setcounter{figure}{0}
\setcounter{table}{0}
\renewcommand{\theequation}{E\arabic{equation}}
\renewcommand{\thefigure}{E\arabic{figure}}
\renewcommand{\thetable}{E\arabic{table}}
\renewcommand{\thesubsection}{Appendix E}

\subsection{The bright exciton emission polarization in a single nanoplatelet.} \label{alpha}

Here we present the expression for the bright exciton polarization in the individual  NPL lying on the substrate with the short axis $X$ rotated by an angle $\alpha$ with respect to the laboratory axis $x$. The steady-state solutions are: 
\begin{gather}\label{angle}
 P_{lA}^l(\alpha)= P_0^l (M_{XX}^A \cos^2(2\alpha)+ M_{YY}^A \sin^2(2\alpha)) \, , \nonumber \\
P_{l'A}^l(\alpha)= P_0^l  \left( M_{YX}^A +\frac{1}{2}(- M_{XX}^A + M_{YY}^A) \sin(4\alpha)\right)\, , \nonumber \\
P_{cA}^l(\alpha)= P_0^l (M_{XZ}^A \cos(2\alpha)+ M_{ZY}^A \sin(2\alpha)) \, , \\
P_{lA}^c(\alpha)= P_0^c (M_{XZ}^A \cos(2\alpha)- M_{ZY}^A \sin(2\alpha)) \, , \nonumber\\
P_{l'A}^c(\alpha)= P_0^c (-M_{ZY}^A \cos(2\alpha)- M_{XZ}^A \sin(2\alpha)) \,  \nonumber
\end{gather} 
for the excitation-dependent polarization. Time-dependent solutions can be presented in the same form with $M_{\gamma \delta}^A \rightarrow M_{\gamma \delta}^A (t)$.

The results for the excitation-independent polarization terms are:
\begin{gather}
P_{lA}^{\rm eq}(\alpha) = -\frac{ T_{A1} }{\tau_{A1}^s} \frac{\Omega_X}{\Omega_A} \tanh \left( \frac{\hbar \Omega_A}{2 k_B T}\right) \cos(2\alpha) \, ,  \\
P_{lA}^{\rm eq}(t,\alpha) = -(1 - e^{-\frac{t}{\tau_{A1}^s}}) \frac{\Omega_X}{\Omega_A} \tanh \left( \frac{\hbar \Omega_A}{2 k_B T}\right) \cos(2\alpha) \, ,  \nonumber \\
P_{l'A}^{\rm eq}(\alpha) = \frac{ T_{A1} }{\tau_{A1}^s} \frac{\Omega_X}{\Omega_A} \tanh \left( \frac{\hbar \Omega_A}{2 k_B T}\right) \sin(2\alpha) \, ,  \\
P_{l'A}^{\rm eq}(t,\alpha) =  (1 - e^{-\frac{t}{\tau_{A1}^s}}) \frac{\Omega_X}{\Omega_A} \tanh \left( \frac{\hbar \Omega_A}{2 k_B T}\right) \sin(2\alpha) \, . \nonumber
\end{gather}

The polarizations proportional to the $P_0^{l'}$ excitation can be obtained by replacing $\alpha$ by $\alpha+ \pi/4$ in all the effects for the single NPL.

Integration of the above expressions over the arbitrary angle $0<\alpha<2\pi$ results in bright exciton contributions to the ensemble polarization presented in Sec.~\ref{brightLambda} and defines the components of the matrix $\Lambda_{\gamma \delta}^A$ ($\gamma,\delta = l,l',c$). The property $\Lambda_{ll}^A=\Lambda_{l'l'}^A$ corresponding to the $C_\infty$ symmetry can be easily proved by replacing $\alpha$ with $\alpha+ \pi/4$ in all the effects for the single NPL before the angular averaging.

\setcounter{equation}{0}
\setcounter{figure}{0}
\setcounter{table}{0}
\renewcommand{\theequation}{F\arabic{equation}}
\renewcommand{\thefigure}{F\arabic{figure}}
\renewcommand{\thetable}{F\arabic{table}}
\renewcommand{\thesubsection}{Appendix F}

\subsection{Dark exciton contribution to the PL polarization} \label{dark}

Here, we present the expressions for the effects related to the dark exciton contribution.
As it is shown in Fig.~\ref{fig:fig3}(b) in the main text, the transfer of the excited by the linearly polarized light bright exciton pseudospin components $S_{AX}$ and $S_{AY}$ to the dark exciton components $S_{FX}$ and $S_{FY}$ allows for the appearance of the dark exciton contribution to the optical alignment and linear polarization rotation effects. The condition for its possible manifestation is that $\varphi = \varphi_0$, which means that the direction of the effective magnetic field mediating the state mixing is preserved on the timescale of the dark exciton lifetime, so that we have nearly the same magnetic field direction acting in the transfer and radiation processes. One can see from Fig.~\ref{fig:fig3}(b), that, upon  averaging over the angle $\varphi$ arbitrarily oriented in the ensemble of NPLs, only the channels proportional to the $\cos^2(2\varphi)$ and $\sin^2(2\varphi)$ contribute to the effects, while the contributions proportional to  $\cos(2\varphi)\sin(2\varphi)$ average to zero. The respective results are presented in the main text.

\begin{figure*}[ht!]
	\includegraphics[width=0.8\linewidth]{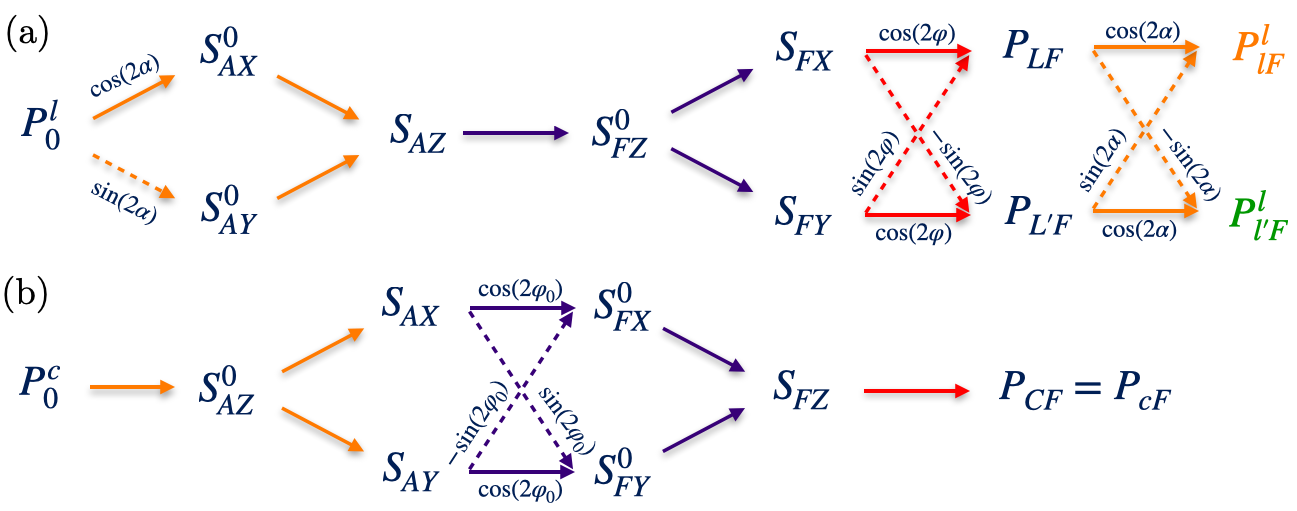}
	\caption{\label{fig:darklin} The contribution of the dark exciton to the (a) linear polarization, $P_{lF}^l,\ P_{l'F}^l$, and to the (b) optical orientation effect, $P_{cF}^c$, caused by the two consecutive linear-to-circular and circular-to-linear conversions of the bright and dark exciton pseudospin components in (a) and \textit{vice versa} in (b). All these contributions vanish after averaging over $\varphi$ in the case of an isotropically oriented exchange field $\bm b$ in the NPL.}
\end{figure*}

In this section, we discuss additional contributions that are nonzero if the angle $\varphi$ is not arbitrary but fixed because of some symmetry-breaking perturbations, and  we present the expressions for a certain $\varphi_0 = \varphi$ after averaging over $\alpha$. Thus,  the channels shown in Fig.~\ref{fig:fig3}(b) and proportional to the $\cos(2\varphi)\sin(2\varphi)$ also contribute to the $P_{l'F}^l$ effect.  

 For the linear polarization coming after the excitation with linearly polarized light, there are also contributions that are related to the transfer of the circular bright exciton pseudospin component $S_{AZ} \rightarrow S_{FZ}^0$ and linear-to-circular conversion in the bright exciton followed by reversal circular-to-linear conversion in the dark exciton. This results in contributions to both $P_{lF}^l$ and $P_{l'F}^l$ for the given $\varphi$ that we characterize as a double conversion, as schematically shown in Fig.~\ref{fig:darklin}(a).  These contributions are proportional to $\cos(2\varphi)$ and $\sin(2\varphi)$ and vanish after $\varphi$ averaging.
Thus, after averaging the expressions over the ensemble while allowing only for arbitrarily angle $\alpha$, we obtain: 
\begin{widetext}
\begin{eqnarray}
P_{lF}^l =&&\frac{P_0^l}{2}[( M_{XX}^A M_{XX}^F + M_{YY}^AM_{YY}^F)  \cos^2(2\varphi) + (M_{XX}^A M_{YY}^F + M_{YY}^AM_{XX}^F) \sin^2(2\varphi) + 2M_{XY}^AM_{YX}^F + \nonumber \\
+  &&(M_{ZX}^AM_{XZ}^F + M_{ZY}^AM_{YZ}^F)\cos(2\varphi) + (M_{ZX}^A M_{YZ}^F - M_{ZY}^AM_{XZ}^F) \sin(2\varphi)  ] \, ,\\
P_{l'F}^l && = \frac{P_0^l}{2} [(M_{XX}^A+M_{YY}^A) M_{YX}^F + M_{YX}^A(M_{XX}^F+M_{YY}^F)+ (-M_{XX}^A+M_{YY}^A)(M_{XX}^F-M_{YY}^F)\sin(2\varphi)\cos(2\varphi) + \nonumber \\
&& -  (M_{ZX}^A M_{XZ}^F + M_{ZY}^A M_{YZ}^F) \sin(2\varphi) + (M_{ZX}^A M_{YZ}^F - M_{ZY}^A M_{XZ}^F) \cos(2\varphi)]\, . \nonumber
\end{eqnarray}
The analogously obtained expressions for the dark exciton contribution to the linear polarization on the timescale $t \gg \tau_A$ differ from those in the CW regime by changing $M_i^F \rightarrow M_i^F(t)$ and have the following form:
\begin{eqnarray}
P_{lF}^l(t) &&=\frac{P_0^l}{2}[( M_{XX}^A M_{XX}^F(t) + M_{YY}^AM_{YY}^F(t))  \cos^2(2\varphi) + (M_{XX}^A M_{YY}^F(t) + M_{YY}^AM_{XX}^F(t)) \sin^2(2\varphi) +  \\
&& +  2M_{XY}^AM_{YX}^F(t) +  (M_{ZX}^AM_{XZ}^F(t) + M_{ZY}^AM_{YZ}^F(t))\cos(2\varphi) + (M_{ZX}^A M_{YZ}^F(t) - M_{ZY}^AM_{XZ}^F(t)) \sin(2\varphi)  ] \, , \nonumber \\
P_{l'F}^l(t) && = \frac{P_0^l}{2} [(M_{XX}^A+M_{YY}^A) M_{YX}^F(t) + M_{YX}^A(M_{XX}^F(t)+M_{YY}^F(t))+ \nonumber \\
&& + (-M_{XX}^A+M_{YY}^A)(M_{XX}^F(t)-M_{YY}^F(t))\sin(2\varphi)\cos(2\varphi) + \nonumber \\
&& -  (M_{ZX}^A M_{XZ}^F(t) + M_{ZY}^A M_{YZ}^F(t)) \sin(2\varphi) + (M_{ZX}^A M_{YZ}^F(t) - M_{ZY}^A M_{XZ}^F(t)) \cos(2\varphi)] \, .\nonumber 
\end{eqnarray}
Double conversion contributions to the optical orientation effect involve circular-to-linear conversion in the bright exciton followed by reversal linear-to-circular conversion in the dark exciton, as shown in Fig.~\ref{fig:darklin}(b). The averaging over $\varphi$ also leads to their vanishing, but with the certain $\varphi$, we get:
\begin{eqnarray}
    P_c^c = P_0^c [ M_{ZZ}^A M_{ZZ}^F + (M_{XZ}^AM_{ZX}^F + M_{YZ}M_{ZY}^F) \cos(2\varphi) +(- M_{YZ}^AM_{ZX}^F + M_{XZ}^A M_{ZY}^F)\sin(2\varphi) ]
\end{eqnarray}
for the CW regime and
\begin{eqnarray}
    P_c^c(t) = P_0^c [ M_{ZZ}^A M_{ZZ}^F(t) + (M_{XZ}^AM_{ZX}^F(t) + M_{YZ}M_{ZY}^F(t)) \cos(2\varphi) +(- M_{YZ}^AM_{ZX}^F(t) + M_{XZ}^A M_{ZY}^F(t))\sin(2\varphi) ]
\end{eqnarray}
for the kinetics at $t\gg \tau_A$.

\end{widetext}


\begin{thebibliography}{62}%
\makeatletter
\providecommand \@ifxundefined [1]{%
 \@ifx{#1\undefined}
}%
\providecommand \@ifnum [1]{%
 \ifnum #1\expandafter \@firstoftwo
 \else \expandafter \@secondoftwo
 \fi
}%
\providecommand \@ifx [1]{%
 \ifx #1\expandafter \@firstoftwo
 \else \expandafter \@secondoftwo
 \fi
}%
\providecommand \natexlab [1]{#1}%
\providecommand \enquote  [1]{``#1''}%
\providecommand \bibnamefont  [1]{#1}%
\providecommand \bibfnamefont [1]{#1}%
\providecommand \citenamefont [1]{#1}%
\providecommand \href@noop [0]{\@secondoftwo}%
\providecommand \href [0]{\begingroup \@sanitize@url \@href}%
\providecommand \@href[1]{\@@startlink{#1}\@@href}%
\providecommand \@@href[1]{\endgroup#1\@@endlink}%
\providecommand \@sanitize@url [0]{\catcode `\\12\catcode `\$12\catcode
  `\&12\catcode `\#12\catcode `\^12\catcode `\_12\catcode `\%12\relax}%
\providecommand \@@startlink[1]{}%
\providecommand \@@endlink[0]{}%
\providecommand \url  [0]{\begingroup\@sanitize@url \@url }%
\providecommand \@url [1]{\endgroup\@href {#1}{\urlprefix }}%
\providecommand \urlprefix  [0]{URL }%
\providecommand \Eprint [0]{\href }%
\providecommand \doibase [0]{https://doi.org/}%
\providecommand \selectlanguage [0]{\@gobble}%
\providecommand \bibinfo  [0]{\@secondoftwo}%
\providecommand \bibfield  [0]{\@secondoftwo}%
\providecommand \translation [1]{[#1]}%
\providecommand \BibitemOpen [0]{}%
\providecommand \bibitemStop [0]{}%
\providecommand \bibitemNoStop [0]{.\EOS\space}%
\providecommand \EOS [0]{\spacefactor3000\relax}%
\providecommand \BibitemShut  [1]{\csname bibitem#1\endcsname}%
\let\auto@bib@innerbib\@empty
%</preamble>
\bibitem [{\citenamefont {García~de Arquer}\ \emph {et~al.}(2021)\citenamefont
  {García~de Arquer}, \citenamefont {Talapin}, \citenamefont {Klimov},
  \citenamefont {Arakawa}, \citenamefont {Bayer},\ and\ \citenamefont
  {Sargent}}]{Garcia2021}%
  \BibitemOpen
  \bibfield  {author} {\bibinfo {author} {\bibfnamefont {F.~P.}\ \bibnamefont
  {García~de Arquer}}, \bibinfo {author} {\bibfnamefont {D.~V.}\ \bibnamefont
  {Talapin}}, \bibinfo {author} {\bibfnamefont {V.~I.}\ \bibnamefont {Klimov}},
  \bibinfo {author} {\bibfnamefont {Y.}~\bibnamefont {Arakawa}}, \bibinfo
  {author} {\bibfnamefont {M.}~\bibnamefont {Bayer}},\ and\ \bibinfo {author}
  {\bibfnamefont {E.~H.}\ \bibnamefont {Sargent}},\ }\bibfield  {title}
  {\bibinfo {title} {Semiconductor quantum dots: Technological progress and
  future challenges},\ }\href {https://doi.org/10.1126/science.aaz8541}
  {\bibfield  {journal} {\bibinfo  {journal} {Science}\ }\textbf {\bibinfo
  {volume} {373}},\ \bibinfo {pages} {eaaz8541} (\bibinfo {year}
  {2021})}\BibitemShut {NoStop}%
\bibitem [{\citenamefont {Efros}\ and\ \citenamefont {Brus}(2021)}]{Efros2021}%
  \BibitemOpen
  \bibfield  {author} {\bibinfo {author} {\bibfnamefont {Al.~L.}\ \bibnamefont
  {Efros}}\ and\ \bibinfo {author} {\bibfnamefont {L.~E.}\ \bibnamefont
  {Brus}},\ }\bibfield  {title} {\bibinfo {title} {Nanocrystal quantum dots:
  From discovery to modern development},\ }\href
  {https://doi.org/10.1021/acsnano.1c01399} {\bibfield  {journal} {\bibinfo
  {journal} {ACS Nano}\ }\textbf {\bibinfo {volume} {15}},\ \bibinfo {pages}
  {6192} (\bibinfo {year} {2021})}\BibitemShut {NoStop}%
\bibitem [{\citenamefont {Ithurria}\ and\ \citenamefont
  {Dubertret}(2008)}]{Ithurria2008}%
  \BibitemOpen
  \bibfield  {author} {\bibinfo {author} {\bibfnamefont {S.}~\bibnamefont
  {Ithurria}}\ and\ \bibinfo {author} {\bibfnamefont {B.}~\bibnamefont
  {Dubertret}},\ }\bibfield  {title} {\bibinfo {title} {Quasi 2d colloidal cdse
  platelets with thicknesses controlled at the atomic level},\ }\href
  {https://doi.org/10.1021/ja807724e} {\bibfield  {journal} {\bibinfo
  {journal} {J. Am. Chem. Soc.}\ }\textbf {\bibinfo {volume} {130}},\ \bibinfo
  {pages} {16504} (\bibinfo {year} {2008})}\BibitemShut {NoStop}%
\bibitem [{\citenamefont {Bayer}(2019)}]{Bayer2019}%
  \BibitemOpen
  \bibfield  {author} {\bibinfo {author} {\bibfnamefont {M.}~\bibnamefont
  {Bayer}},\ }\bibfield  {title} {\bibinfo {title} {Bridging two worlds:
  Colloidal versus epitaxial quantum dots},\ }\href
  {https://doi.org/https://doi.org/10.1002/andp.201900039} {\bibfield
  {journal} {\bibinfo  {journal} {Ann. Phys.}\ }\textbf {\bibinfo {volume}
  {531}},\ \bibinfo {pages} {1900039} (\bibinfo {year} {2019})}\BibitemShut
  {NoStop}%
\bibitem [{\citenamefont {Ithurria}\ and\ \citenamefont
  {Talapin}(2012)}]{Ithurria2012}%
  \BibitemOpen
  \bibfield  {author} {\bibinfo {author} {\bibfnamefont {S.}~\bibnamefont
  {Ithurria}}\ and\ \bibinfo {author} {\bibfnamefont {D.~V.}\ \bibnamefont
  {Talapin}},\ }\bibfield  {title} {\bibinfo {title} {Colloidal atomic layer
  deposition (c-{ALD}) using self-limiting reactions at nanocrystal surface
  coupled to phase transfer between polar and nonpolar media},\ }\href
  {https://doi.org/10.1021/ja308088d} {\bibfield  {journal} {\bibinfo
  {journal} {J. Am. Chem. Soc.}\ }\textbf {\bibinfo {volume} {134}},\ \bibinfo
  {pages} {18585} (\bibinfo {year} {2012})}\BibitemShut {NoStop}%
\bibitem [{\citenamefont {Nasilowski}\ \emph {et~al.}(2016)\citenamefont
  {Nasilowski}, \citenamefont {Mahler}, \citenamefont {Lhuillier},
  \citenamefont {Ithurria},\ and\ \citenamefont {Dubertret}}]{Nasilowski2016}%
  \BibitemOpen
  \bibfield  {author} {\bibinfo {author} {\bibfnamefont {M.}~\bibnamefont
  {Nasilowski}}, \bibinfo {author} {\bibfnamefont {B.}~\bibnamefont {Mahler}},
  \bibinfo {author} {\bibfnamefont {E.}~\bibnamefont {Lhuillier}}, \bibinfo
  {author} {\bibfnamefont {S.}~\bibnamefont {Ithurria}},\ and\ \bibinfo
  {author} {\bibfnamefont {B.}~\bibnamefont {Dubertret}},\ }\bibfield  {title}
  {\bibinfo {title} {Two-dimensional colloidal nanocrystals},\ }\href
  {https://doi.org/10.1021/acs.chemrev.6b00164} {\bibfield  {journal} {\bibinfo
   {journal} {Chem. Rev.}\ }\textbf {\bibinfo {volume} {116}},\ \bibinfo
  {pages} {10934} (\bibinfo {year} {2016})}\BibitemShut {NoStop}%
\bibitem [{\citenamefont {Berends}\ and\ \citenamefont
  {de~Mello~Donega}(2017)}]{Berends2017}%
  \BibitemOpen
  \bibfield  {author} {\bibinfo {author} {\bibfnamefont {A.~C.}\ \bibnamefont
  {Berends}}\ and\ \bibinfo {author} {\bibfnamefont {C.}~\bibnamefont
  {de~Mello~Donega}},\ }\bibfield  {title} {\bibinfo {title} {Ultrathin one-
  and two-dimensional colloidal semiconductor nanocrystals: Pushing quantum
  confinement to the limit},\ }\href
  {https://doi.org/10.1021/acs.jpclett.7b01640} {\bibfield  {journal} {\bibinfo
   {journal} {J. Phys. Chem. Lett.}\ }\textbf {\bibinfo {volume} {8}},\
  \bibinfo {pages} {4077} (\bibinfo {year} {2017})}\BibitemShut {NoStop}%
\bibitem [{\citenamefont {Rodina}\ and\ \citenamefont
  {Efros}(2016{\natexlab{a}})}]{Rodina2016JETP}%
  \BibitemOpen
  \bibfield  {author} {\bibinfo {author} {\bibfnamefont {A.~V.}\ \bibnamefont
  {Rodina}}\ and\ \bibinfo {author} {\bibfnamefont {Al.~L.}\ \bibnamefont
  {Efros}},\ }\bibfield  {title} {\bibinfo {title} {Effect of dielectric
  confinement on optical properties of colloidal nanostructures},\ }\href
  {https://doi.org/10.7868/S0044451016030159} {\bibfield  {journal} {\bibinfo
  {journal} {JETP}\ }\textbf {\bibinfo {volume} {149}},\ \bibinfo {pages} {555}
  (\bibinfo {year} {2016}{\natexlab{a}})}\BibitemShut {NoStop}%
\bibitem [{\citenamefont {Shornikova}\ \emph {et~al.}(2021)\citenamefont
  {Shornikova}, \citenamefont {Yakovlev}, \citenamefont {Gippius},
  \citenamefont {Qiang}, \citenamefont {Dubertret}, \citenamefont {Khan},
  \citenamefont {Di~Giacomo}, \citenamefont {Moreels},\ and\ \citenamefont
  {Bayer}}]{Shornikova2021nl}%
  \BibitemOpen
  \bibfield  {author} {\bibinfo {author} {\bibfnamefont {E.~V.}\ \bibnamefont
  {Shornikova}}, \bibinfo {author} {\bibfnamefont {D.~R.}\ \bibnamefont
  {Yakovlev}}, \bibinfo {author} {\bibfnamefont {N.~A.}\ \bibnamefont
  {Gippius}}, \bibinfo {author} {\bibfnamefont {G.}~\bibnamefont {Qiang}},
  \bibinfo {author} {\bibfnamefont {B.}~\bibnamefont {Dubertret}}, \bibinfo
  {author} {\bibfnamefont {A.~H.}\ \bibnamefont {Khan}}, \bibinfo {author}
  {\bibfnamefont {A.}~\bibnamefont {Di~Giacomo}}, \bibinfo {author}
  {\bibfnamefont {I.}~\bibnamefont {Moreels}},\ and\ \bibinfo {author}
  {\bibfnamefont {M.}~\bibnamefont {Bayer}},\ }\bibfield  {title} {\bibinfo
  {title} {Exciton binding energy in {CdSe} nanoplatelets measured by one- and
  two-photon absorption},\ }\href
  {https://doi.org/10.1021/acs.nanolett.1c04159} {\bibfield  {journal}
  {\bibinfo  {journal} {Nano Letters}\ }\textbf {\bibinfo {volume} {21}},\
  \bibinfo {pages} {10525} (\bibinfo {year} {2021})}\BibitemShut {NoStop}%
\bibitem [{\citenamefont {Rodina}\ and\ \citenamefont
  {Efros}(2016{\natexlab{b}})}]{Rodina2016}%
  \BibitemOpen
  \bibfield  {author} {\bibinfo {author} {\bibfnamefont {A.~V.}\ \bibnamefont
  {Rodina}}\ and\ \bibinfo {author} {\bibfnamefont {Al.~L.}\ \bibnamefont
  {Efros}},\ }\bibfield  {title} {\bibinfo {title} {Radiative recombination
  from dark excitons in nanocrystals: Activation mechanisms and polarization
  properties},\ }\href {https://doi.org/10.1103/PhysRevB.93.155427} {\bibfield
  {journal} {\bibinfo  {journal} {Phys. Rev. B}\ }\textbf {\bibinfo {volume}
  {93}},\ \bibinfo {pages} {155427} (\bibinfo {year}
  {2016}{\natexlab{b}})}\BibitemShut {NoStop}%
\bibitem [{\citenamefont {Shornikova}\ \emph
  {et~al.}(2018{\natexlab{a}})\citenamefont {Shornikova}, \citenamefont
  {Biadala}, \citenamefont {Yakovlev}, \citenamefont {Sapega}, \citenamefont
  {Kusrayev}, \citenamefont {Mitioglu}, \citenamefont {Ballottin},
  \citenamefont {Christianen}, \citenamefont {Belykh}, \citenamefont {Kochiev},
  \citenamefont {Sibeldin}, \citenamefont {Golovatenko}, \citenamefont
  {Rodina}, \citenamefont {Gippius}, \citenamefont {Kuntzmann}, \citenamefont
  {Jiang}, \citenamefont {Nasilowski}, \citenamefont {Dubertret},\ and\
  \citenamefont {Bayer}}]{Shornikova2018ns}%
  \BibitemOpen
  \bibfield  {author} {\bibinfo {author} {\bibfnamefont {E.~V.}\ \bibnamefont
  {Shornikova}}, \bibinfo {author} {\bibfnamefont {L.}~\bibnamefont {Biadala}},
  \bibinfo {author} {\bibfnamefont {D.~R.}\ \bibnamefont {Yakovlev}}, \bibinfo
  {author} {\bibfnamefont {V.~F.}\ \bibnamefont {Sapega}}, \bibinfo {author}
  {\bibfnamefont {Y.~G.}\ \bibnamefont {Kusrayev}}, \bibinfo {author}
  {\bibfnamefont {A.~A.}\ \bibnamefont {Mitioglu}}, \bibinfo {author}
  {\bibfnamefont {M.~V.}\ \bibnamefont {Ballottin}}, \bibinfo {author}
  {\bibfnamefont {P.~C.~M.}\ \bibnamefont {Christianen}}, \bibinfo {author}
  {\bibfnamefont {V.~V.}\ \bibnamefont {Belykh}}, \bibinfo {author}
  {\bibfnamefont {M.~V.}\ \bibnamefont {Kochiev}}, \bibinfo {author}
  {\bibfnamefont {N.~N.}\ \bibnamefont {Sibeldin}}, \bibinfo {author}
  {\bibfnamefont {A.~A.}\ \bibnamefont {Golovatenko}}, \bibinfo {author}
  {\bibfnamefont {A.~V.}\ \bibnamefont {Rodina}}, \bibinfo {author}
  {\bibfnamefont {N.~A.}\ \bibnamefont {Gippius}}, \bibinfo {author}
  {\bibfnamefont {A.}~\bibnamefont {Kuntzmann}}, \bibinfo {author}
  {\bibfnamefont {Y.}~\bibnamefont {Jiang}}, \bibinfo {author} {\bibfnamefont
  {M.}~\bibnamefont {Nasilowski}}, \bibinfo {author} {\bibfnamefont
  {B.}~\bibnamefont {Dubertret}},\ and\ \bibinfo {author} {\bibfnamefont
  {M.}~\bibnamefont {Bayer}},\ }\bibfield  {title} {\bibinfo {title}
  {{{Addressing}} the exciton fine structure in colloidal nanocrystals: The
  case of {{CdSe}} nanoplatelets},\ }\href {https://doi.org/10.1039/C7NR07206F}
  {\bibfield  {journal} {\bibinfo  {journal} {Nanoscale}\ }\textbf {\bibinfo
  {volume} {10}},\ \bibinfo {pages} {646} (\bibinfo {year}
  {2018}{\natexlab{a}})}\BibitemShut {NoStop}%
\bibitem [{\citenamefont {Shornikova}\ \emph {et~al.}(2020)\citenamefont
  {Shornikova}, \citenamefont {Golovatenko}, \citenamefont {Yakovlev},
  \citenamefont {Rodina}, \citenamefont {Biadala}, \citenamefont {Qiang},
  \citenamefont {Kuntzmann}, \citenamefont {Nasilowski}, \citenamefont
  {Dubertret}, \citenamefont {Polovitsyn}, \citenamefont {Moreels},\ and\
  \citenamefont {Bayer}}]{Shornikova2020nn}%
  \BibitemOpen
  \bibfield  {author} {\bibinfo {author} {\bibfnamefont {E.~V.}\ \bibnamefont
  {Shornikova}}, \bibinfo {author} {\bibfnamefont {A.~A.}\ \bibnamefont
  {Golovatenko}}, \bibinfo {author} {\bibfnamefont {D.~R.}\ \bibnamefont
  {Yakovlev}}, \bibinfo {author} {\bibfnamefont {A.~V.}\ \bibnamefont
  {Rodina}}, \bibinfo {author} {\bibfnamefont {L.}~\bibnamefont {Biadala}},
  \bibinfo {author} {\bibfnamefont {G.}~\bibnamefont {Qiang}}, \bibinfo
  {author} {\bibfnamefont {A.}~\bibnamefont {Kuntzmann}}, \bibinfo {author}
  {\bibfnamefont {M.}~\bibnamefont {Nasilowski}}, \bibinfo {author}
  {\bibfnamefont {B.}~\bibnamefont {Dubertret}}, \bibinfo {author}
  {\bibfnamefont {A.}~\bibnamefont {Polovitsyn}}, \bibinfo {author}
  {\bibfnamefont {I.}~\bibnamefont {Moreels}},\ and\ \bibinfo {author}
  {\bibfnamefont {M.}~\bibnamefont {Bayer}},\ }\bibfield  {title} {\bibinfo
  {title} {Surface spin magnetism controls the polarized exciton emission from
  {CdSe} nanoplatelets},\ }\href {https://doi.org/10.1038/s41565-019-0631-7}
  {\bibfield  {journal} {\bibinfo  {journal} {Nat. Nanotechnol.}\ }\textbf
  {\bibinfo {volume} {15}},\ \bibinfo {pages} {277} (\bibinfo {year}
  {2020})}\BibitemShut {NoStop}%
\bibitem [{\citenamefont {Efros}(2003)}]{EfrosCh3}%
  \BibitemOpen
  \bibfield  {author} {\bibinfo {author} {\bibfnamefont {Al.~L.}\ \bibnamefont
  {Efros}},\ }\bibfield  {title} {\bibinfo {title} {{Fine Structure and
  Polarization Properties of Band-Edge Excitons in Semiconductor
  Nanocrystals}},\ }in\ \href@noop {} {\emph {\bibinfo {booktitle}
  {Semiconductor and Metal Nanocrystals: Synthesis and Electronic and Optical
  Properties}}},\ \bibinfo {editor} {edited by\ \bibinfo {editor} {\bibnamefont
  {{V. I. Klimov and M. Dekker}}}}\ (\bibinfo  {publisher} {New York},\
  \bibinfo {year} {2003})\ Chap.~\bibinfo {chapter} {3}\BibitemShut {NoStop}%
\bibitem [{\citenamefont {Meier}\ and\ \citenamefont
  {Zakharchenya}(1984)}]{ZakharchenyaBook}%
  \BibitemOpen
  \bibinfo {editor} {\bibfnamefont {F.}~\bibnamefont {Meier}}\ and\ \bibinfo
  {editor} {\bibfnamefont {B.~P.}\ \bibnamefont {Zakharchenya}},\ eds.,\
  \href@noop {} {\emph {\bibinfo {title} {Optical Orientation}}}\ (\bibinfo
  {publisher} {North-Holland},\ \bibinfo {address} {Amsterdam},\ \bibinfo
  {year} {1984})\BibitemShut {NoStop}%
\bibitem [{\citenamefont {Pikus}\ and\ \citenamefont
  {Ivchenko}(1982)}]{Pikus_book1982}%
  \BibitemOpen
  \bibfield  {author} {\bibinfo {author} {\bibfnamefont {G.~E.}\ \bibnamefont
  {Pikus}}\ and\ \bibinfo {author} {\bibfnamefont {E.~L.}\ \bibnamefont
  {Ivchenko}},\ }\bibfield  {title} {\bibinfo {title} {Optical orientation and
  polarized luminescence of excitons in semiconductors},\ }in\ \href@noop {}
  {\emph {\bibinfo {booktitle} {Excitons}}},\ \bibinfo {editor} {edited by\
  \bibinfo {editor} {\bibfnamefont {E.~I.}\ \bibnamefont {Rashba}}\ and\
  \bibinfo {editor} {\bibfnamefont {M.~D.}\ \bibnamefont {Sturge}}}\ (\bibinfo
  {publisher} {North-Holland},\ \bibinfo {address} {Amsterdam},\ \bibinfo
  {year} {1982})\ Chap.~\bibinfo {chapter} {6}\BibitemShut {NoStop}%
\bibitem [{\citenamefont {Dzhioev}\ \emph
  {et~al.}(1997{\natexlab{a}})\citenamefont {Dzhioev}, \citenamefont {Gibbs},
  \citenamefont {Ivchenko}, \citenamefont {Khitrova}, \citenamefont {Korenev},
  \citenamefont {Tkachuk},\ and\ \citenamefont
  {Zakharchenya}}]{Dzhioev1997PRB}%
  \BibitemOpen
  \bibfield  {author} {\bibinfo {author} {\bibfnamefont {R.~I.}\ \bibnamefont
  {Dzhioev}}, \bibinfo {author} {\bibfnamefont {H.~M.}\ \bibnamefont {Gibbs}},
  \bibinfo {author} {\bibfnamefont {E.~L.}\ \bibnamefont {Ivchenko}}, \bibinfo
  {author} {\bibfnamefont {G.}~\bibnamefont {Khitrova}}, \bibinfo {author}
  {\bibfnamefont {V.~L.}\ \bibnamefont {Korenev}}, \bibinfo {author}
  {\bibfnamefont {M.~N.}\ \bibnamefont {Tkachuk}},\ and\ \bibinfo {author}
  {\bibfnamefont {B.~P.}\ \bibnamefont {Zakharchenya}},\ }\bibfield  {title}
  {\bibinfo {title} {Determination of interface preference by observation of
  linear-to-circular polarization conversion under optical orientation of
  excitons in \mbox{type-II} {GaAs/AlAs} superlattices},\ }\href
  {https://doi.org/10.1103/PhysRevB.56.13405} {\bibfield  {journal} {\bibinfo
  {journal} {Phys. Rev. B}\ }\textbf {\bibinfo {volume} {56}},\ \bibinfo
  {pages} {13405} (\bibinfo {year} {1997}{\natexlab{a}})}\BibitemShut {NoStop}%
\bibitem [{\citenamefont {Kusrayev}(2008)}]{Kusrayev2008}%
  \BibitemOpen
  \bibfield  {author} {\bibinfo {author} {\bibfnamefont {Y.~G.}\ \bibnamefont
  {Kusrayev}},\ }\bibfield  {title} {\bibinfo {title} {{Optical orientation of
  excitons and carriers in quantum dots}},\ }\href
  {https://doi.org/10.1088/0268-1242/23/11/114013} {\bibfield  {journal}
  {\bibinfo  {journal} {Semicond. Sci. Technol.}\ }\textbf {\bibinfo {volume}
  {23}},\ \bibinfo {pages} {114013} (\bibinfo {year} {2008})}\BibitemShut
  {NoStop}%
\bibitem [{\citenamefont {Koudinov}\ \emph {et~al.}(2008)\citenamefont
  {Koudinov}, \citenamefont {Namozov}, \citenamefont {Kusrayev}, \citenamefont
  {Lee}, \citenamefont {Dobrowolska},\ and\ \citenamefont
  {Furdyna}}]{Koudinov2008}%
  \BibitemOpen
  \bibfield  {author} {\bibinfo {author} {\bibfnamefont {A.~V.}\ \bibnamefont
  {Koudinov}}, \bibinfo {author} {\bibfnamefont {B.~R.}\ \bibnamefont
  {Namozov}}, \bibinfo {author} {\bibfnamefont {Y.~G.}\ \bibnamefont
  {Kusrayev}}, \bibinfo {author} {\bibfnamefont {S.}~\bibnamefont {Lee}},
  \bibinfo {author} {\bibfnamefont {M.}~\bibnamefont {Dobrowolska}},\ and\
  \bibinfo {author} {\bibfnamefont {J.~K.}\ \bibnamefont {Furdyna}},\
  }\bibfield  {title} {\bibinfo {title} {Two-step versus one-step model of the
  interpolarization conversion and statistics of {CdSe/ZnSe} quantum dot
  elongations},\ }\href {https://doi.org/10.1103/PhysRevB.78.045309} {\bibfield
   {journal} {\bibinfo  {journal} {Phys. Rev. B}\ }\textbf {\bibinfo {volume}
  {78}},\ \bibinfo {pages} {045309} (\bibinfo {year} {2008})}\BibitemShut
  {NoStop}%
\bibitem [{\citenamefont {Shamirzaev}\ \emph {et~al.}(2023)\citenamefont
  {Shamirzaev}, \citenamefont {Shumilin}, \citenamefont {Smirnov},
  \citenamefont {Kudlacik}, \citenamefont {Nekrasov}, \citenamefont {Kusrayev},
  \citenamefont {Yakovlev},\ and\ \citenamefont {Bayer}}]{Shamirzaev2023}%
  \BibitemOpen
  \bibfield  {author} {\bibinfo {author} {\bibfnamefont {T.~S.}\ \bibnamefont
  {Shamirzaev}}, \bibinfo {author} {\bibfnamefont {A.~V.}\ \bibnamefont
  {Shumilin}}, \bibinfo {author} {\bibfnamefont {D.~S.}\ \bibnamefont
  {Smirnov}}, \bibinfo {author} {\bibfnamefont {D.}~\bibnamefont {Kudlacik}},
  \bibinfo {author} {\bibfnamefont {S.~V.}\ \bibnamefont {Nekrasov}}, \bibinfo
  {author} {\bibfnamefont {Y.~G.}\ \bibnamefont {Kusrayev}}, \bibinfo {author}
  {\bibfnamefont {D.~R.}\ \bibnamefont {Yakovlev}},\ and\ \bibinfo {author}
  {\bibfnamefont {M.}~\bibnamefont {Bayer}},\ }\bibfield  {title} {\bibinfo
  {title} {Optical orientation of excitons in a longitudinal magnetic field in
  indirect-band-gap {(In,Al)As/AlAs} quantum dots with \mbox{type-I} band
  alignment},\ }\href {https://doi.org/10.3390/nano13040729} {\bibfield
  {journal} {\bibinfo  {journal} {Nanomaterials}\ }\textbf {\bibinfo {volume}
  {13}},\ \bibinfo {pages} {729} (\bibinfo {year} {2023})}\BibitemShut
  {NoStop}%
\bibitem [{\citenamefont {Ekimov}\ and\ \citenamefont
  {Safarov}(1971)}]{Ekimov1971_2}%
  \BibitemOpen
  \bibfield  {author} {\bibinfo {author} {\bibfnamefont {A.~I.}\ \bibnamefont
  {Ekimov}}\ and\ \bibinfo {author} {\bibfnamefont {V.~I.}\ \bibnamefont
  {Safarov}},\ }\bibfield  {title} {\bibinfo {title} {Influence of spin
  relaxation of "hot" electrons in the effectiveness of optical orientation in
  semiconductors},\ }\href@noop {} {\bibfield  {journal} {\bibinfo  {journal}
  {JETP Lett.}\ }\textbf {\bibinfo {volume} {13}},\ \bibinfo {pages} {495}
  (\bibinfo {year} {1971})}\BibitemShut {NoStop}%
\bibitem [{\citenamefont {Gamarts}\ \emph {et~al.}(1977)\citenamefont
  {Gamarts}, \citenamefont {Ivchenko}, \citenamefont {Karaman}, \citenamefont
  {Mushinskii}, \citenamefont {Pikus}, \citenamefont {Razbirin},\ and\
  \citenamefont {Starukhin}}]{Gamarts1977}%
  \BibitemOpen
  \bibfield  {author} {\bibinfo {author} {\bibfnamefont {E.}~\bibnamefont
  {Gamarts}}, \bibinfo {author} {\bibfnamefont {E.}~\bibnamefont {Ivchenko}},
  \bibinfo {author} {\bibfnamefont {M.}~\bibnamefont {Karaman}}, \bibinfo
  {author} {\bibfnamefont {V.}~\bibnamefont {Mushinskii}}, \bibinfo {author}
  {\bibfnamefont {G.}~\bibnamefont {Pikus}}, \bibinfo {author} {\bibfnamefont
  {B.}~\bibnamefont {Razbirin}},\ and\ \bibinfo {author} {\bibfnamefont
  {A.}~\bibnamefont {Starukhin}},\ }\bibfield  {title} {\bibinfo {title}
  {Optical orientation and alignment of free excitons in {GaSe} during
  resonance excitation. {Experiment}},\ }\href@noop {} {\bibfield  {journal}
  {\bibinfo  {journal} {JETP}\ }\textbf {\bibinfo {volume} {46}},\ \bibinfo
  {pages} {590} (\bibinfo {year} {1977})}\BibitemShut {NoStop}%
\bibitem [{\citenamefont {Dymnikov}\ \emph {et~al.}(1981)\citenamefont
  {Dymnikov}, \citenamefont {Mirlin}, \citenamefont {Nikitin}, \citenamefont
  {Perel'}, \citenamefont {Reshina},\ and\ \citenamefont
  {Sapega}}]{Dymnikov1981}%
  \BibitemOpen
  \bibfield  {author} {\bibinfo {author} {\bibfnamefont {V.~D.}\ \bibnamefont
  {Dymnikov}}, \bibinfo {author} {\bibfnamefont {D.~N.}\ \bibnamefont
  {Mirlin}}, \bibinfo {author} {\bibfnamefont {L.~P.}\ \bibnamefont {Nikitin}},
  \bibinfo {author} {\bibfnamefont {V.~I.}\ \bibnamefont {Perel'}}, \bibinfo
  {author} {\bibfnamefont {I.~I.}\ \bibnamefont {Reshina}},\ and\ \bibinfo
  {author} {\bibfnamefont {V.~F.}\ \bibnamefont {Sapega}},\ }\bibfield  {title}
  {\bibinfo {title} {Depolarization of hot photoluminescence of gallium
  arsenide crystals in a magnetic field. {Determination} of the energy
  relaxation times of hot electrons},\ }\href@noop {} {\bibfield  {journal}
  {\bibinfo  {journal} {JETP}\ }\textbf {\bibinfo {volume} {53}},\ \bibinfo
  {pages} {912} (\bibinfo {year} {1981})}\BibitemShut {NoStop}%
\bibitem [{\citenamefont {Dzhioev}\ \emph {et~al.}(1973)\citenamefont
  {Dzhioev}, \citenamefont {Zakharchenya},\ and\ \citenamefont
  {Fleisher}}]{Dzhioev1973}%
  \BibitemOpen
  \bibfield  {author} {\bibinfo {author} {\bibfnamefont {R.~I.}\ \bibnamefont
  {Dzhioev}}, \bibinfo {author} {\bibfnamefont {B.~P.}\ \bibnamefont
  {Zakharchenya}},\ and\ \bibinfo {author} {\bibfnamefont {V.~G.}\ \bibnamefont
  {Fleisher}},\ }\bibfield  {title} {\bibinfo {title} {Investigation of
  semiconductor paramagnetism using luminescence polarization in a weak
  magnetic field},\ }\href@noop {} {\bibfield  {journal} {\bibinfo  {journal}
  {JETP Lett.}\ }\textbf {\bibinfo {volume} {17}},\ \bibinfo {pages} {174}
  (\bibinfo {year} {1973})}\BibitemShut {NoStop}%
\bibitem [{\citenamefont {Permogorov}\ \emph {et~al.}(1983)\citenamefont
  {Permogorov}, \citenamefont {Reznitsky}, \citenamefont {Verbin},\ and\
  \citenamefont {Lysenko}}]{Permogorov1983}%
  \BibitemOpen
  \bibfield  {author} {\bibinfo {author} {\bibfnamefont {S.}~\bibnamefont
  {Permogorov}}, \bibinfo {author} {\bibfnamefont {A.}~\bibnamefont
  {Reznitsky}}, \bibinfo {author} {\bibfnamefont {S.}~\bibnamefont {Verbin}},\
  and\ \bibinfo {author} {\bibfnamefont {V.}~\bibnamefont {Lysenko}},\
  }\bibfield  {title} {\bibinfo {title} {Exciton mobility edge in
  {CdS}$_{1-x}${Se}$_x$ solid solutions},\ }\href
  {https://doi.org/10.1016/0038-1098(83)90084-4} {\bibfield  {journal}
  {\bibinfo  {journal} {Solid State Communications}\ }\textbf {\bibinfo
  {volume} {47}},\ \bibinfo {pages} {5} (\bibinfo {year} {1983})}\BibitemShut
  {NoStop}%
\bibitem [{\citenamefont {Verbin}\ \emph {et~al.}(2003)\citenamefont {Verbin},
  \citenamefont {Efimov}, \citenamefont {Petrov}, \citenamefont {Ignatiev},
  \citenamefont {Dolgikh}, \citenamefont {Eliseev}, \citenamefont {Gerlovin},
  \citenamefont {Ovsyankin},\ and\ \citenamefont {Masumoto}}]{Verbin2002}%
  \BibitemOpen
  \bibfield  {author} {\bibinfo {author} {\bibfnamefont {S.~Y.}\ \bibnamefont
  {Verbin}}, \bibinfo {author} {\bibfnamefont {Y.~P.}\ \bibnamefont {Efimov}},
  \bibinfo {author} {\bibfnamefont {V.~M.}\ \bibnamefont {Petrov}}, \bibinfo
  {author} {\bibfnamefont {I.~V.}\ \bibnamefont {Ignatiev}}, \bibinfo {author}
  {\bibfnamefont {Y.~K.}\ \bibnamefont {Dolgikh}}, \bibinfo {author}
  {\bibfnamefont {S.~A.}\ \bibnamefont {Eliseev}}, \bibinfo {author}
  {\bibfnamefont {I.~Y.}\ \bibnamefont {Gerlovin}}, \bibinfo {author}
  {\bibfnamefont {V.~V.}\ \bibnamefont {Ovsyankin}},\ and\ \bibinfo {author}
  {\bibfnamefont {Y.}~\bibnamefont {Masumoto}},\ }\bibfield  {title} {\bibinfo
  {title} {{Spin memory in the n-doped GaAs/AlGaAs quantum wells}},\ }in\ \href
  {https://doi.org/10.1117/12.514466} {\emph {\bibinfo {booktitle} {10th
  International Symposium on Nanostructures: Physics and Technology}}},\ Vol.\
  \bibinfo {volume} {5023},\ \bibinfo {editor} {edited by\ \bibinfo {editor}
  {\bibfnamefont {Z.~I.}\ \bibnamefont {Alferov}}\ and\ \bibinfo {editor}
  {\bibfnamefont {L.}~\bibnamefont {Esaki}}},\ \bibinfo {organization}
  {International Society for Optics and Photonics}\ (\bibinfo  {publisher}
  {SPIE},\ \bibinfo {year} {2003})\ pp.\ \bibinfo {pages} {432 --
  435}\BibitemShut {NoStop}%
\bibitem [{\citenamefont {Ivchenko}\ \emph {et~al.}(1977)\citenamefont
  {Ivchenko}, \citenamefont {Pikus}, \citenamefont {Razbirin},\ and\
  \citenamefont {Starukhin}}]{Ivchenko1977_2}%
  \BibitemOpen
  \bibfield  {author} {\bibinfo {author} {\bibfnamefont {E.~L.}\ \bibnamefont
  {Ivchenko}}, \bibinfo {author} {\bibfnamefont {G.~E.}\ \bibnamefont {Pikus}},
  \bibinfo {author} {\bibfnamefont {B.~S.}\ \bibnamefont {Razbirin}},\ and\
  \bibinfo {author} {\bibfnamefont {A.~I.}\ \bibnamefont {Starukhin}},\
  }\bibfield  {title} {\bibinfo {title} {Optical orientation and alignment of
  free excitations in {GaSe} under resonant excitation. {Theory}},\ }\href@noop
  {} {\bibfield  {journal} {\bibinfo  {journal} {JETP}\ }\textbf {\bibinfo
  {volume} {45}},\ \bibinfo {pages} {1172} (\bibinfo {year}
  {1977})}\BibitemShut {NoStop}%
\bibitem [{\citenamefont {Dzhioev}\ \emph {et~al.}(1998)\citenamefont
  {Dzhioev}, \citenamefont {Zakharchenya}, \citenamefont {Ivchenko},
  \citenamefont {Korenev}, \citenamefont {Kusraev}, \citenamefont {Ledentsov},
  \citenamefont {Ustinov}, \citenamefont {Zhukov},\ and\ \citenamefont
  {Tsatsul'nikov}}]{Dzhioev1998}%
  \BibitemOpen
  \bibfield  {author} {\bibinfo {author} {\bibfnamefont {R.~I.}\ \bibnamefont
  {Dzhioev}}, \bibinfo {author} {\bibfnamefont {B.~P.}\ \bibnamefont
  {Zakharchenya}}, \bibinfo {author} {\bibfnamefont {E.~L.}\ \bibnamefont
  {Ivchenko}}, \bibinfo {author} {\bibfnamefont {V.~L.}\ \bibnamefont
  {Korenev}}, \bibinfo {author} {\bibfnamefont {Y.~G.}\ \bibnamefont
  {Kusraev}}, \bibinfo {author} {\bibfnamefont {N.~N.}\ \bibnamefont
  {Ledentsov}}, \bibinfo {author} {\bibfnamefont {V.~M.}\ \bibnamefont
  {Ustinov}}, \bibinfo {author} {\bibfnamefont {A.~E.}\ \bibnamefont
  {Zhukov}},\ and\ \bibinfo {author} {\bibfnamefont {A.~F.}\ \bibnamefont
  {Tsatsul'nikov}},\ }\bibfield  {title} {\bibinfo {title} {Optical orientation
  and alignment of excitons in quantum dots},\ }\href
  {https://doi.org/10.1134/1.1130397} {\bibfield  {journal} {\bibinfo
  {journal} {Phys. Solid State}\ }\textbf {\bibinfo {volume} {40}},\ \bibinfo
  {pages} {790} (\bibinfo {year} {1998})}\BibitemShut {NoStop}%
\bibitem [{\citenamefont {Astakhov}\ \emph {et~al.}(2006)\citenamefont
  {Astakhov}, \citenamefont {Kiessling}, \citenamefont {Platonov},
  \citenamefont {Slobodskyy}, \citenamefont {Mahapatra}, \citenamefont {Ossau},
  \citenamefont {Schmidt}, \citenamefont {Brunner},\ and\ \citenamefont
  {Molenkamp}}]{Astakhov2006}%
  \BibitemOpen
  \bibfield  {author} {\bibinfo {author} {\bibfnamefont {G.~V.}\ \bibnamefont
  {Astakhov}}, \bibinfo {author} {\bibfnamefont {T.}~\bibnamefont {Kiessling}},
  \bibinfo {author} {\bibfnamefont {A.~V.}\ \bibnamefont {Platonov}}, \bibinfo
  {author} {\bibfnamefont {T.}~\bibnamefont {Slobodskyy}}, \bibinfo {author}
  {\bibfnamefont {S.}~\bibnamefont {Mahapatra}}, \bibinfo {author}
  {\bibfnamefont {W.}~\bibnamefont {Ossau}}, \bibinfo {author} {\bibfnamefont
  {G.}~\bibnamefont {Schmidt}}, \bibinfo {author} {\bibfnamefont
  {K.}~\bibnamefont {Brunner}},\ and\ \bibinfo {author} {\bibfnamefont {L.~W.}\
  \bibnamefont {Molenkamp}},\ }\bibfield  {title} {\bibinfo {title}
  {Circular-to-linear and linear-to-circular conversion of optical polarization
  by semiconductor quantum dots},\ }\href
  {https://doi.org/10.1103/PhysRevLett.96.027402} {\bibfield  {journal}
  {\bibinfo  {journal} {Phys. Rev. Lett.}\ }\textbf {\bibinfo {volume} {96}},\
  \bibinfo {pages} {027402} (\bibinfo {year} {2006})}\BibitemShut {NoStop}%
\bibitem [{\citenamefont {Smirnov}\ and\ \citenamefont
  {Ivchenko}(2023)}]{Smirnov2023}%
  \BibitemOpen
  \bibfield  {author} {\bibinfo {author} {\bibfnamefont {D.~S.}\ \bibnamefont
  {Smirnov}}\ and\ \bibinfo {author} {\bibfnamefont {E.~L.}\ \bibnamefont
  {Ivchenko}},\ }\bibfield  {title} {\bibinfo {title} {Theory of polarized
  photoluminescence of indirect band gap excitons in \mbox{type-I} quantum
  dots},\ }\href {https://doi.org/10.1103/PhysRevB.108.195432} {\bibfield
  {journal} {\bibinfo  {journal} {Phys. Rev. B}\ }\textbf {\bibinfo {volume}
  {108}},\ \bibinfo {pages} {195432} (\bibinfo {year} {2023})}\BibitemShut
  {NoStop}%
\bibitem [{\citenamefont {Smirnov}\ and\ \citenamefont
  {Ivchenko}(2024)}]{Smirnov2024}%
  \BibitemOpen
  \bibfield  {author} {\bibinfo {author} {\bibfnamefont {D.~S.}\ \bibnamefont
  {Smirnov}}\ and\ \bibinfo {author} {\bibfnamefont {E.~L.}\ \bibnamefont
  {Ivchenko}},\ }\bibfield  {title} {\bibinfo {title} {Interplay between
  hyperfine and anisotropic exchange interactions in exciton luminescence of
  quantum dots},\ }\href {https://doi.org/10.61011/OS.2024.08.59033.6714-24}
  {\bibfield  {journal} {\bibinfo  {journal} {Opt. Spectrosc.}\ }\textbf
  {\bibinfo {volume} {132}},\ \bibinfo {pages} {864} (\bibinfo {year}
  {2024})}\BibitemShut {NoStop}%
\bibitem [{\citenamefont {Kopteva}\ \emph {et~al.}(2024)\citenamefont
  {Kopteva}, \citenamefont {Yakovlev}, \citenamefont {Yalcin}, \citenamefont
  {Akimov}, \citenamefont {Nestoklon}, \citenamefont {Glazov}, \citenamefont
  {Kotur}, \citenamefont {Kudlacik}, \citenamefont {Zhukov}, \citenamefont
  {Kirstein}, \citenamefont {Hordiichuk}, \citenamefont {Dirin}, \citenamefont
  {Kovalenko},\ and\ \citenamefont {Bayer}}]{Kopteva2024}%
  \BibitemOpen
  \bibfield  {author} {\bibinfo {author} {\bibfnamefont {N.~E.}\ \bibnamefont
  {Kopteva}}, \bibinfo {author} {\bibfnamefont {D.~R.}\ \bibnamefont
  {Yakovlev}}, \bibinfo {author} {\bibfnamefont {E.}~\bibnamefont {Yalcin}},
  \bibinfo {author} {\bibfnamefont {I.~A.}\ \bibnamefont {Akimov}}, \bibinfo
  {author} {\bibfnamefont {M.~O.}\ \bibnamefont {Nestoklon}}, \bibinfo {author}
  {\bibfnamefont {M.~M.}\ \bibnamefont {Glazov}}, \bibinfo {author}
  {\bibfnamefont {M.}~\bibnamefont {Kotur}}, \bibinfo {author} {\bibfnamefont
  {D.}~\bibnamefont {Kudlacik}}, \bibinfo {author} {\bibfnamefont {E.~A.}\
  \bibnamefont {Zhukov}}, \bibinfo {author} {\bibfnamefont {E.}~\bibnamefont
  {Kirstein}}, \bibinfo {author} {\bibfnamefont {O.}~\bibnamefont
  {Hordiichuk}}, \bibinfo {author} {\bibfnamefont {D.~N.}\ \bibnamefont
  {Dirin}}, \bibinfo {author} {\bibfnamefont {M.~V.}\ \bibnamefont
  {Kovalenko}},\ and\ \bibinfo {author} {\bibfnamefont {M.}~\bibnamefont
  {Bayer}},\ }\bibfield  {title} {\bibinfo {title} {Highly-polarized emission
  provided by giant optical orientation of exciton spins in lead halide
  perovskite crystals},\ }\href
  {https://doi.org/https://doi.org/10.1002/advs.202403691} {\bibfield
  {journal} {\bibinfo  {journal} {Adv. Sci.}\ }\textbf {\bibinfo {volume}
  {11}},\ \bibinfo {pages} {2403691} (\bibinfo {year} {2024})}\BibitemShut
  {NoStop}%
\bibitem [{\citenamefont {Zhiliakov}\ \emph {et~al.}(2026)\citenamefont
  {Zhiliakov}, \citenamefont {Kopteva}, \citenamefont {Yugova}, \citenamefont
  {Yakovlev}, \citenamefont {Akimov},\ and\ \citenamefont
  {Bayer}}]{Zhiliakov2026}%
  \BibitemOpen
  \bibfield  {author} {\bibinfo {author} {\bibfnamefont {V.~L.}\ \bibnamefont
  {Zhiliakov}}, \bibinfo {author} {\bibfnamefont {N.~E.}\ \bibnamefont
  {Kopteva}}, \bibinfo {author} {\bibfnamefont {I.~A.}\ \bibnamefont {Yugova}},
  \bibinfo {author} {\bibfnamefont {D.~R.}\ \bibnamefont {Yakovlev}}, \bibinfo
  {author} {\bibfnamefont {I.~A.}\ \bibnamefont {Akimov}},\ and\ \bibinfo
  {author} {\bibfnamefont {M.}~\bibnamefont {Bayer}},\ }\bibfield  {title}
  {\bibinfo {title} {Exciton spin structure in lead halide perovskite
  semiconductors explored via spin dynamics in a magnetic field},\ }\href
  {https://doi.org/10.1103/mwk6-grms} {\bibfield  {journal} {\bibinfo
  {journal} {Phys. Rev. B}\ }\textbf {\bibinfo {volume} {113}},\ \bibinfo
  {pages} {155206} (\bibinfo {year} {2026})}\BibitemShut {NoStop}%
\bibitem [{\citenamefont {Nestoklon}\ \emph {et~al.}(2018)\citenamefont
  {Nestoklon}, \citenamefont {Goupalov}, \citenamefont {Dzhioev}, \citenamefont
  {Ken}, \citenamefont {Korenev}, \citenamefont {Kusrayev}, \citenamefont
  {Sapega}, \citenamefont {de~Weerd}, \citenamefont {Gomez}, \citenamefont
  {Gregorkiewicz}, \citenamefont {Lin}, \citenamefont {Suenaga}, \citenamefont
  {Fujiwara}, \citenamefont {Matyushkin},\ and\ \citenamefont
  {Yassievich}}]{Nestoklon2018}%
  \BibitemOpen
  \bibfield  {author} {\bibinfo {author} {\bibfnamefont {M.~O.}\ \bibnamefont
  {Nestoklon}}, \bibinfo {author} {\bibfnamefont {S.~V.}\ \bibnamefont
  {Goupalov}}, \bibinfo {author} {\bibfnamefont {R.~I.}\ \bibnamefont
  {Dzhioev}}, \bibinfo {author} {\bibfnamefont {O.~S.}\ \bibnamefont {Ken}},
  \bibinfo {author} {\bibfnamefont {V.~L.}\ \bibnamefont {Korenev}}, \bibinfo
  {author} {\bibfnamefont {Y.~G.}\ \bibnamefont {Kusrayev}}, \bibinfo {author}
  {\bibfnamefont {V.~F.}\ \bibnamefont {Sapega}}, \bibinfo {author}
  {\bibfnamefont {C.}~\bibnamefont {de~Weerd}}, \bibinfo {author}
  {\bibfnamefont {L.}~\bibnamefont {Gomez}}, \bibinfo {author} {\bibfnamefont
  {T.}~\bibnamefont {Gregorkiewicz}}, \bibinfo {author} {\bibfnamefont
  {J.}~\bibnamefont {Lin}}, \bibinfo {author} {\bibfnamefont {K.}~\bibnamefont
  {Suenaga}}, \bibinfo {author} {\bibfnamefont {Y.}~\bibnamefont {Fujiwara}},
  \bibinfo {author} {\bibfnamefont {L.~B.}\ \bibnamefont {Matyushkin}},\ and\
  \bibinfo {author} {\bibfnamefont {I.~N.}\ \bibnamefont {Yassievich}},\
  }\bibfield  {title} {\bibinfo {title} {Optical orientation and alignment of
  excitons in ensembles of inorganic perovskite nanocrystals},\ }\href
  {https://doi.org/10.1103/PhysRevB.97.235304} {\bibfield  {journal} {\bibinfo
  {journal} {Phys. Rev. B}\ }\textbf {\bibinfo {volume} {97}},\ \bibinfo
  {pages} {235304} (\bibinfo {year} {2018})}\BibitemShut {NoStop}%
\bibitem [{\citenamefont {Wang}\ \emph {et~al.}(2015)\citenamefont {Wang},
  \citenamefont {Bouet}, \citenamefont {Glazov}, \citenamefont {Amand},
  \citenamefont {Ivchenko}, \citenamefont {Palleau}, \citenamefont {Marie},\
  and\ \citenamefont {Urbaszek}}]{Wang_2015}%
  \BibitemOpen
  \bibfield  {author} {\bibinfo {author} {\bibfnamefont {G.}~\bibnamefont
  {Wang}}, \bibinfo {author} {\bibfnamefont {L.}~\bibnamefont {Bouet}},
  \bibinfo {author} {\bibfnamefont {M.~M.}\ \bibnamefont {Glazov}}, \bibinfo
  {author} {\bibfnamefont {T.}~\bibnamefont {Amand}}, \bibinfo {author}
  {\bibfnamefont {E.~L.}\ \bibnamefont {Ivchenko}}, \bibinfo {author}
  {\bibfnamefont {E.}~\bibnamefont {Palleau}}, \bibinfo {author} {\bibfnamefont
  {X.}~\bibnamefont {Marie}},\ and\ \bibinfo {author} {\bibfnamefont
  {B.}~\bibnamefont {Urbaszek}},\ }\bibfield  {title} {\bibinfo {title}
  {Magneto-optics in transition metal diselenide monolayers},\ }\href
  {https://doi.org/10.1088/2053-1583/2/3/034002} {\bibfield  {journal}
  {\bibinfo  {journal} {2D Materials}\ }\textbf {\bibinfo {volume} {2}},\
  \bibinfo {pages} {034002} (\bibinfo {year} {2015})}\BibitemShut {NoStop}%
\bibitem [{\citenamefont {Yagodkin}\ \emph {et~al.}(2025)\citenamefont
  {Yagodkin}, \citenamefont {Burfeindt}, \citenamefont {Iakovlev},
  \citenamefont {Kumar}, \citenamefont {Dewambrechies}, \citenamefont {Yücel},
  \citenamefont {Höfer}, \citenamefont {Gahl}, \citenamefont {Glazov},\ and\
  \citenamefont {Bolotin}}]{Yagodkin2025}%
  \BibitemOpen
  \bibfield  {author} {\bibinfo {author} {\bibfnamefont {D.}~\bibnamefont
  {Yagodkin}}, \bibinfo {author} {\bibfnamefont {K.}~\bibnamefont {Burfeindt}},
  \bibinfo {author} {\bibfnamefont {Z.~A.}\ \bibnamefont {Iakovlev}}, \bibinfo
  {author} {\bibfnamefont {A.~M.}\ \bibnamefont {Kumar}}, \bibinfo {author}
  {\bibfnamefont {A.}~\bibnamefont {Dewambrechies}}, \bibinfo {author}
  {\bibfnamefont {O.}~\bibnamefont {Yücel}}, \bibinfo {author} {\bibfnamefont
  {B.}~\bibnamefont {Höfer}}, \bibinfo {author} {\bibfnamefont
  {C.}~\bibnamefont {Gahl}}, \bibinfo {author} {\bibfnamefont {M.~M.}\
  \bibnamefont {Glazov}},\ and\ \bibinfo {author} {\bibfnamefont {K.~I.}\
  \bibnamefont {Bolotin}},\ }\bibfield  {title} {\bibinfo {title} {Fermi
  polarons under strain-induced pseudomagnetic fields},\ }\href
  {https://doi.org/10.1038/s41467-025-66192-y} {\bibfield  {journal} {\bibinfo
  {journal} {Nat. Commun.}\ }\textbf {\bibinfo {volume} {16}},\ \bibinfo
  {pages} {10232} (\bibinfo {year} {2025})}\BibitemShut {NoStop}%
\bibitem [{\citenamefont {Lakowicz}(2006)}]{Lakowicz2006}%
  \BibitemOpen
  \bibfield  {author} {\bibinfo {author} {\bibfnamefont {J.~R.}\ \bibnamefont
  {Lakowicz}},\ }\href@noop {} {\emph {\bibinfo {title} {Principles of
  Fluorescence Spectroscopy}}},\ \bibinfo {edition} {3rd}\ ed.\ (\bibinfo
  {publisher} {Springer},\ \bibinfo {address} {New York, NY},\ \bibinfo {year}
  {2006})\ pp.\ \bibinfo {pages} {XXVI, 954}\BibitemShut {NoStop}%
\bibitem [{\citenamefont {Smirnova}\ \emph {et~al.}(2023)\citenamefont
  {Smirnova}, \citenamefont {Kalitukha}, \citenamefont {Rodina}, \citenamefont
  {Dimitriev}, \citenamefont {Sapega}, \citenamefont {Ken}, \citenamefont
  {Korenev}, \citenamefont {Kozyrev}, \citenamefont {Nekrasov}, \citenamefont
  {Kusrayev}, \citenamefont {Yakovlev}, \citenamefont {Dubertret},\ and\
  \citenamefont {Bayer}}]{Smirnova2023}%
  \BibitemOpen
  \bibfield  {author} {\bibinfo {author} {\bibfnamefont {O.~O.}\ \bibnamefont
  {Smirnova}}, \bibinfo {author} {\bibfnamefont {I.~V.}\ \bibnamefont
  {Kalitukha}}, \bibinfo {author} {\bibfnamefont {A.~V.}\ \bibnamefont
  {Rodina}}, \bibinfo {author} {\bibfnamefont {G.~S.}\ \bibnamefont
  {Dimitriev}}, \bibinfo {author} {\bibfnamefont {V.~F.}\ \bibnamefont
  {Sapega}}, \bibinfo {author} {\bibfnamefont {O.~S.}\ \bibnamefont {Ken}},
  \bibinfo {author} {\bibfnamefont {V.~L.}\ \bibnamefont {Korenev}}, \bibinfo
  {author} {\bibfnamefont {N.~V.}\ \bibnamefont {Kozyrev}}, \bibinfo {author}
  {\bibfnamefont {S.~V.}\ \bibnamefont {Nekrasov}}, \bibinfo {author}
  {\bibfnamefont {Y.~G.}\ \bibnamefont {Kusrayev}}, \bibinfo {author}
  {\bibfnamefont {D.~R.}\ \bibnamefont {Yakovlev}}, \bibinfo {author}
  {\bibfnamefont {B.}~\bibnamefont {Dubertret}},\ and\ \bibinfo {author}
  {\bibfnamefont {M.}~\bibnamefont {Bayer}},\ }\bibfield  {title} {\bibinfo
  {title} {Optical alignment and optical orientation of excitons in {CdSe/CdS}
  colloidal nanoplatelets},\ }\href {https://doi.org/10.3390/nano13172402}
  {\bibfield  {journal} {\bibinfo  {journal} {Nanomaterials}\ }\textbf
  {\bibinfo {volume} {13}},\ \bibinfo {pages} {2402} (\bibinfo {year}
  {2023})}\BibitemShut {NoStop}%
\bibitem [{\citenamefont {Smirnova}\ \emph {et~al.}(2025)\citenamefont
  {Smirnova}, \citenamefont {Kozyrev}, \citenamefont {Nekrasov}, \citenamefont
  {Kalitukha}, \citenamefont {Dubertret},\ and\ \citenamefont
  {Rodina}}]{Smirnova2025}%
  \BibitemOpen
  \bibfield  {author} {\bibinfo {author} {\bibfnamefont {O.~O.}\ \bibnamefont
  {Smirnova}}, \bibinfo {author} {\bibfnamefont {N.~V.}\ \bibnamefont
  {Kozyrev}}, \bibinfo {author} {\bibfnamefont {S.~V.}\ \bibnamefont
  {Nekrasov}}, \bibinfo {author} {\bibfnamefont {I.~V.}\ \bibnamefont
  {Kalitukha}}, \bibinfo {author} {\bibfnamefont {B.}~\bibnamefont
  {Dubertret}},\ and\ \bibinfo {author} {\bibfnamefont {A.~V.}\ \bibnamefont
  {Rodina}},\ }\bibfield  {title} {\bibinfo {title} {Time-resolved exciton
  photoluminescence and its linear polarization of an ensemble of {CdSe/CdS}
  colloidal nanoplatelets},\ }\href
  {https://doi.org/https://doi.org/10.1016/j.optmat.2024.116469} {\bibfield
  {journal} {\bibinfo  {journal} {Opt. Mater.}\ }\textbf {\bibinfo {volume}
  {159}},\ \bibinfo {pages} {116469} (\bibinfo {year} {2025})}\BibitemShut
  {NoStop}%
\bibitem [{\citenamefont {Gupalov}\ \emph {et~al.}(1998)\citenamefont
  {Gupalov}, \citenamefont {Ivchenko},\ and\ \citenamefont
  {Kavokin}}]{Goupalov1998_2}%
  \BibitemOpen
  \bibfield  {author} {\bibinfo {author} {\bibfnamefont {S.~V.}\ \bibnamefont
  {Gupalov}}, \bibinfo {author} {\bibfnamefont {E.~L.}\ \bibnamefont
  {Ivchenko}},\ and\ \bibinfo {author} {\bibfnamefont {A.~V.}\ \bibnamefont
  {Kavokin}},\ }\bibfield  {title} {\bibinfo {title} {Fine structure of
  localized exciton levels in quantum wells},\ }\href
  {https://doi.org/10.1134/1.558441} {\bibfield  {journal} {\bibinfo  {journal}
  {J. Exp. Theor. Phys.}\ }\textbf {\bibinfo {volume} {86}},\ \bibinfo {pages}
  {388} (\bibinfo {year} {1998})}\BibitemShut {NoStop}%
\bibitem [{\citenamefont {Ivchenko}\ and\ \citenamefont
  {Pikus}(1995)}]{Ivchenko1995book}%
  \BibitemOpen
  \bibfield  {author} {\bibinfo {author} {\bibfnamefont {E.~I.}\ \bibnamefont
  {Ivchenko}}\ and\ \bibinfo {author} {\bibfnamefont {G.~E.}\ \bibnamefont
  {Pikus}},\ }\href@noop {} {\emph {\bibinfo {title} {{Superlattices and other
  heterostucture}}}}\ (\bibinfo  {publisher} {Springer},\ \bibinfo {address}
  {Berlin},\ \bibinfo {year} {1995})\BibitemShut {NoStop}%
\bibitem [{\citenamefont {Semina}\ \emph {et~al.}(2026)\citenamefont {Semina},
  \citenamefont {Druzhinina}, \citenamefont {Golovatentko},\ and\ \citenamefont
  {Rodina}}]{Semina2026arXive}%
  \BibitemOpen
  \bibfield  {author} {\bibinfo {author} {\bibfnamefont {M.~A.}\ \bibnamefont
  {Semina}}, \bibinfo {author} {\bibfnamefont {O.~O.}\ \bibnamefont
  {Druzhinina}}, \bibinfo {author} {\bibfnamefont {A.~A.}\ \bibnamefont
  {Golovatentko}},\ and\ \bibinfo {author} {\bibfnamefont {A.~V.}\ \bibnamefont
  {Rodina}},\ }\href@noop {} {\bibinfo {title} {Exciton fine structure in
  nanocrystals: effect of cuboidal and spheroidal shapes}} (\bibinfo {year}
  {2026}),\ \Eprint {https://arxiv.org/abs/2608.24663} {arXiv:2608.24663}
  \BibitemShut {NoStop}%
\bibitem [{\citenamefont {Dzhioev}\ \emph
  {et~al.}(1997{\natexlab{b}})\citenamefont {Dzhioev}, \citenamefont
  {Zakharchenya}, \citenamefont {Ivchenko}, \citenamefont {Korenev},
  \citenamefont {Kusraev}, \citenamefont {Ledentsov}, \citenamefont {Ustinov},
  \citenamefont {Zhukov},\ and\ \citenamefont {Tsatsul'nikov}}]{Dzhioev1997}%
  \BibitemOpen
  \bibfield  {author} {\bibinfo {author} {\bibfnamefont {R.~I.}\ \bibnamefont
  {Dzhioev}}, \bibinfo {author} {\bibfnamefont {B.~P.}\ \bibnamefont
  {Zakharchenya}}, \bibinfo {author} {\bibfnamefont {E.~L.}\ \bibnamefont
  {Ivchenko}}, \bibinfo {author} {\bibfnamefont {V.~L.}\ \bibnamefont
  {Korenev}}, \bibinfo {author} {\bibfnamefont {Y.~G.}\ \bibnamefont
  {Kusraev}}, \bibinfo {author} {\bibfnamefont {N.~N.}\ \bibnamefont
  {Ledentsov}}, \bibinfo {author} {\bibfnamefont {V.~M.}\ \bibnamefont
  {Ustinov}}, \bibinfo {author} {\bibfnamefont {A.~E.}\ \bibnamefont
  {Zhukov}},\ and\ \bibinfo {author} {\bibfnamefont {A.~F.}\ \bibnamefont
  {Tsatsul'nikov}},\ }\bibfield  {title} {\bibinfo {title} {Fine structure of
  excitonic levels in quantum dots},\ }\href {https://doi.org/10.1134/1.567429}
  {\bibfield  {journal} {\bibinfo  {journal} {JETP Lett.}\ }\textbf {\bibinfo
  {volume} {65}},\ \bibinfo {pages} {804} (\bibinfo {year}
  {1997}{\natexlab{b}})}\BibitemShut {NoStop}%
\bibitem [{\citenamefont {Ivchenko}(2018)}]{Ivchenko2018Gross}%
  \BibitemOpen
  \bibfield  {author} {\bibinfo {author} {\bibfnamefont {E.~L.}\ \bibnamefont
  {Ivchenko}},\ }\bibfield  {title} {\bibinfo {title} {Magnetic circular
  polarization of exciton photoluminescence},\ }\href
  {https://doi.org/10.1134/S1063783418080127} {\bibfield  {journal} {\bibinfo
  {journal} {Phys. Solid State}\ }\textbf {\bibinfo {volume} {60}},\ \bibinfo
  {pages} {1514} (\bibinfo {year} {2018})}\BibitemShut {NoStop}%
\bibitem [{\citenamefont {Yoon}\ \emph {et~al.}(2021)\citenamefont {Yoon},
  \citenamefont {Lee}, \citenamefont {Yeo}, \citenamefont {Ryou}, \citenamefont
  {Lee}, \citenamefont {Kim},\ and\ \citenamefont {Lee}}]{Yoon2021}%
  \BibitemOpen
  \bibfield  {author} {\bibinfo {author} {\bibfnamefont {D.-E.}\ \bibnamefont
  {Yoon}}, \bibinfo {author} {\bibfnamefont {J.}~\bibnamefont {Lee}}, \bibinfo
  {author} {\bibfnamefont {H.}~\bibnamefont {Yeo}}, \bibinfo {author}
  {\bibfnamefont {J.}~\bibnamefont {Ryou}}, \bibinfo {author} {\bibfnamefont
  {Y.~K.}\ \bibnamefont {Lee}}, \bibinfo {author} {\bibfnamefont {Y.-H.}\
  \bibnamefont {Kim}},\ and\ \bibinfo {author} {\bibfnamefont {D.~C.}\
  \bibnamefont {Lee}},\ }\bibfield  {title} {\bibinfo {title} {Atomistics of
  asymmetric lateral growth of colloidal zincblende {CdSe} nanoplatelets},\
  }\href {https://doi.org/10.1021/acs.chemmater.1c00563} {\bibfield  {journal}
  {\bibinfo  {journal} {Chem. Mater.}\ }\textbf {\bibinfo {volume} {33}},\
  \bibinfo {pages} {4813} (\bibinfo {year} {2021})}\BibitemShut {NoStop}%
\bibitem [{\citenamefont {Feng}\ \emph {et~al.}(2020)\citenamefont {Feng},
  \citenamefont {Yakovlev}, \citenamefont {Dubertret},\ and\ \citenamefont
  {Bayer}}]{Feng2020}%
  \BibitemOpen
  \bibfield  {author} {\bibinfo {author} {\bibfnamefont {D.}~\bibnamefont
  {Feng}}, \bibinfo {author} {\bibfnamefont {D.~R.}\ \bibnamefont {Yakovlev}},
  \bibinfo {author} {\bibfnamefont {B.}~\bibnamefont {Dubertret}},\ and\
  \bibinfo {author} {\bibfnamefont {M.}~\bibnamefont {Bayer}},\ }\bibfield
  {title} {\bibinfo {title} {Charge separation dynamics in {CdSe/CdS}
  core/shell nanoplatelets addressed by coherent electron spin precession},\
  }\href {https://doi.org/10.1021/acsnano.0c02402} {\bibfield  {journal}
  {\bibinfo  {journal} {ACS Nano}\ }\textbf {\bibinfo {volume} {14}},\ \bibinfo
  {pages} {7237} (\bibinfo {year} {2020})}\BibitemShut {NoStop}%
\bibitem [{\citenamefont {Goupalov}\ and\ \citenamefont
  {Ivchenko}(1998)}]{Goupalov1998}%
  \BibitemOpen
  \bibfield  {author} {\bibinfo {author} {\bibfnamefont {S.~V.}\ \bibnamefont
  {Goupalov}}\ and\ \bibinfo {author} {\bibfnamefont {E.~L.}\ \bibnamefont
  {Ivchenko}},\ }\bibfield  {title} {\bibinfo {title} {{Electron--hole
  long-range exchange interaction in semiconductor quantum dots}},\ }\href
  {https://doi.org/10.1016/S0022-0248(98)80083-3} {\bibfield  {journal}
  {\bibinfo  {journal} {J. Cryst. Growth}\ }\textbf {\bibinfo {volume}
  {184-185}},\ \bibinfo {pages} {393} (\bibinfo {year} {1998})}\BibitemShut
  {NoStop}%
\bibitem [{\citenamefont {Hu}\ \emph {et~al.}(2018)\citenamefont {Hu},
  \citenamefont {Singh}, \citenamefont {Goupalov}, \citenamefont
  {Hollingsworth},\ and\ \citenamefont {Htoon}}]{Hu2018}%
  \BibitemOpen
  \bibfield  {author} {\bibinfo {author} {\bibfnamefont {Z.}~\bibnamefont
  {Hu}}, \bibinfo {author} {\bibfnamefont {A.}~\bibnamefont {Singh}}, \bibinfo
  {author} {\bibfnamefont {S.~V.}\ \bibnamefont {Goupalov}}, \bibinfo {author}
  {\bibfnamefont {J.~A.}\ \bibnamefont {Hollingsworth}},\ and\ \bibinfo
  {author} {\bibfnamefont {H.}~\bibnamefont {Htoon}},\ }\bibfield  {title}
  {\bibinfo {title} {Influence of morphology on the blinking mechanisms and the
  excitonic fine structure of single colloidal nanoplatelets},\ }\href
  {https://doi.org/10.1039/c8nr06234j} {\bibfield  {journal} {\bibinfo
  {journal} {Nanoscale}\ }\textbf {\bibinfo {volume} {10}},\ \bibinfo {pages}
  {22861} (\bibinfo {year} {2018})}\BibitemShut {NoStop}%
\bibitem [{\citenamefont {Swift}\ \emph {et~al.}(2024)\citenamefont {Swift},
  \citenamefont {Efros},\ and\ \citenamefont {Erwin}}]{Swift2024}%
  \BibitemOpen
  \bibfield  {author} {\bibinfo {author} {\bibfnamefont {M.~W.}\ \bibnamefont
  {Swift}}, \bibinfo {author} {\bibfnamefont {Al.~L.}\ \bibnamefont {Efros}},\
  and\ \bibinfo {author} {\bibfnamefont {S.~C.}\ \bibnamefont {Erwin}},\
  }\bibfield  {title} {\bibinfo {title} {Controlling light emission from
  semiconductor nanoplatelets using surface chemistry},\ }\href
  {https://doi.org/10.1038/s41467-024-51842-4} {\bibfield  {journal} {\bibinfo
  {journal} {Nat. Commun.}\ }\textbf {\bibinfo {volume} {15}},\ \bibinfo
  {pages} {7737} (\bibinfo {year} {2024})}\BibitemShut {NoStop}%
\bibitem [{\citenamefont {Rodina}\ and\ \citenamefont
  {Ivchenko}(2020)}]{Rodina2020}%
  \BibitemOpen
  \bibfield  {author} {\bibinfo {author} {\bibfnamefont {A.~V.}\ \bibnamefont
  {Rodina}}\ and\ \bibinfo {author} {\bibfnamefont {E.~L.}\ \bibnamefont
  {Ivchenko}},\ }\bibfield  {title} {\bibinfo {title} {Theory of single and
  double electron spin-flip raman scattering in semiconductor nanoplatelets},\
  }\href {https://doi.org/10.1103/PhysRevB.102.235432} {\bibfield  {journal}
  {\bibinfo  {journal} {Phys. Rev. B}\ }\textbf {\bibinfo {volume} {102}},\
  \bibinfo {pages} {235432} (\bibinfo {year} {2020})}\BibitemShut {NoStop}%
\bibitem [{\citenamefont {Blum}(2012)}]{Blum_book}%
  \BibitemOpen
  \bibfield  {author} {\bibinfo {author} {\bibfnamefont {K.}~\bibnamefont
  {Blum}},\ }\href@noop {} {\emph {\bibinfo {title} {Density Matrix Theory and
  Applications}}},\ \bibinfo {edition} {3rd}\ ed.,\ \bibinfo {series} {Springer
  Series on Atomic, Optical, and Plasma Physics}, Vol.~\bibinfo {volume} {64}\
  (\bibinfo  {publisher} {Springer},\ \bibinfo {address} {Berlin, Heidelberg},\
  \bibinfo {year} {2012})\BibitemShut {NoStop}%
\bibitem [{\citenamefont {Labeau}\ \emph {et~al.}(2003)\citenamefont {Labeau},
  \citenamefont {Tamarat},\ and\ \citenamefont {Lounis}}]{Labeau2003}%
  \BibitemOpen
  \bibfield  {author} {\bibinfo {author} {\bibfnamefont {O.}~\bibnamefont
  {Labeau}}, \bibinfo {author} {\bibfnamefont {P.}~\bibnamefont {Tamarat}},\
  and\ \bibinfo {author} {\bibfnamefont {B.}~\bibnamefont {Lounis}},\
  }\bibfield  {title} {\bibinfo {title} {{Temperature dependence of the
  luminescence lifetime of single CdSe/ZnS quantum dots}},\ }\href
  {https://doi.org/10.1103/PhysRevLett.90.257404} {\bibfield  {journal}
  {\bibinfo  {journal} {Phys. Rev. Lett.}\ }\textbf {\bibinfo {volume} {90}},\
  \bibinfo {pages} {257404} (\bibinfo {year} {2003})}\BibitemShut {NoStop}%
\bibitem [{\citenamefont {Semina}\ \emph {et~al.}(2021)\citenamefont {Semina},
  \citenamefont {Golovatenko},\ and\ \citenamefont {Rodina}}]{Semina2021}%
  \BibitemOpen
  \bibfield  {author} {\bibinfo {author} {\bibfnamefont {M.~A.}\ \bibnamefont
  {Semina}}, \bibinfo {author} {\bibfnamefont {A.~A.}\ \bibnamefont
  {Golovatenko}},\ and\ \bibinfo {author} {\bibfnamefont {A.~V.}\ \bibnamefont
  {Rodina}},\ }\bibfield  {title} {\bibinfo {title} {Influence of the
  spin-orbit split-off valence band on the hole $g$ factor in semiconductor
  nanocrystals},\ }\href {https://doi.org/10.1103/PhysRevB.104.205423}
  {\bibfield  {journal} {\bibinfo  {journal} {Phys. Rev. B}\ }\textbf {\bibinfo
  {volume} {104}},\ \bibinfo {pages} {205423} (\bibinfo {year}
  {2021})}\BibitemShut {NoStop}%
\bibitem [{\citenamefont {Shornikova}\ \emph
  {et~al.}(2018{\natexlab{b}})\citenamefont {Shornikova}, \citenamefont
  {Biadala}, \citenamefont {Yakovlev}, \citenamefont {Feng}, \citenamefont
  {Sapega}, \citenamefont {Flipo}, \citenamefont {Golovatenko}, \citenamefont
  {Semina}, \citenamefont {Rodina}, \citenamefont {Mitioglu}, \citenamefont
  {Ballottin}, \citenamefont {Christianen}, \citenamefont {Kusrayev},
  \citenamefont {Nasilowski}, \citenamefont {Dubertret},\ and\ \citenamefont
  {Bayer}}]{Shornikova2018nl}%
  \BibitemOpen
  \bibfield  {author} {\bibinfo {author} {\bibfnamefont {E.~V.}\ \bibnamefont
  {Shornikova}}, \bibinfo {author} {\bibfnamefont {L.}~\bibnamefont {Biadala}},
  \bibinfo {author} {\bibfnamefont {D.~R.}\ \bibnamefont {Yakovlev}}, \bibinfo
  {author} {\bibfnamefont {D.}~\bibnamefont {Feng}}, \bibinfo {author}
  {\bibfnamefont {V.~F.}\ \bibnamefont {Sapega}}, \bibinfo {author}
  {\bibfnamefont {N.}~\bibnamefont {Flipo}}, \bibinfo {author} {\bibfnamefont
  {A.~A.}\ \bibnamefont {Golovatenko}}, \bibinfo {author} {\bibfnamefont
  {M.~A.}\ \bibnamefont {Semina}}, \bibinfo {author} {\bibfnamefont {A.~V.}\
  \bibnamefont {Rodina}}, \bibinfo {author} {\bibfnamefont {A.~A.}\
  \bibnamefont {Mitioglu}}, \bibinfo {author} {\bibfnamefont {M.~V.}\
  \bibnamefont {Ballottin}}, \bibinfo {author} {\bibfnamefont {P.~C.~M.}\
  \bibnamefont {Christianen}}, \bibinfo {author} {\bibfnamefont {Y.~G.}\
  \bibnamefont {Kusrayev}}, \bibinfo {author} {\bibfnamefont {M.}~\bibnamefont
  {Nasilowski}}, \bibinfo {author} {\bibfnamefont {B.}~\bibnamefont
  {Dubertret}},\ and\ \bibinfo {author} {\bibfnamefont {M.}~\bibnamefont
  {Bayer}},\ }\bibfield  {title} {\bibinfo {title} {{{Electron}} and {{Hole}}
  {\emph{g}}-{{Factors}} and {{Spin Dynamics}} of {{Negatively Charged
  Excitons}} in {{CdSe}}/{{CdS Colloidal Nanoplatelets}} with {{Thick
  Shells}}},\ }\href {https://doi.org/10.1021/acs.nanolett.7b04203} {\bibfield
  {journal} {\bibinfo  {journal} {Nano Lett.}\ }\textbf {\bibinfo {volume}
  {18}},\ \bibinfo {pages} {373} (\bibinfo {year}
  {2018}{\natexlab{b}})}\BibitemShut {NoStop}%
\bibitem [{\citenamefont {Ivchenko}\ \emph {et~al.}(1978)\citenamefont
  {Ivchenko}, \citenamefont {Lang},\ and\ \citenamefont
  {Pavlov}}]{Ivchenko1977}%
  \BibitemOpen
  \bibfield  {author} {\bibinfo {author} {\bibfnamefont {E.~L.}\ \bibnamefont
  {Ivchenko}}, \bibinfo {author} {\bibfnamefont {I.~G.}\ \bibnamefont {Lang}},\
  and\ \bibinfo {author} {\bibfnamefont {S.~T.}\ \bibnamefont {Pavlov}},\
  }\bibfield  {title} {\bibinfo {title} {Resonant secondary radiation in polar
  semiconductors},\ }\href
  {https://doi.org/https://doi.org/10.1002/pssb.2220850107} {\bibfield
  {journal} {\bibinfo  {journal} {Phys. Status Solidi B}\ }\textbf {\bibinfo
  {volume} {85}},\ \bibinfo {pages} {81} (\bibinfo {year} {1978})}\BibitemShut
  {NoStop}%
\bibitem [{\citenamefont {Long}(2002)}]{long2002raman}%
  \BibitemOpen
  \bibfield  {author} {\bibinfo {author} {\bibfnamefont {D.~A.}\ \bibnamefont
  {Long}},\ }\href@noop {} {\emph {\bibinfo {title} {The Raman Effect: A
  Unified Treatment of the Theory of Raman Scattering by Molecules and
  Crystals}}}\ (\bibinfo  {publisher} {John Wiley \& Sons},\ \bibinfo {address}
  {Chichester},\ \bibinfo {year} {2002})\BibitemShut {NoStop}%
\bibitem [{\citenamefont {Born}\ and\ \citenamefont
  {Wolf}(1999)}]{Born1999principles}%
  \BibitemOpen
  \bibfield  {author} {\bibinfo {author} {\bibfnamefont {M.}~\bibnamefont
  {Born}}\ and\ \bibinfo {author} {\bibfnamefont {E.}~\bibnamefont {Wolf}},\
  }\href@noop {} {\emph {\bibinfo {title} {Principles of Optics:
  Electromagnetic Theory of Propagation, Interference and Diffraction of
  Light}}},\ \bibinfo {edition} {7th}\ ed.\ (\bibinfo  {publisher} {Cambridge
  University Press},\ \bibinfo {address} {Cambridge},\ \bibinfo {year}
  {1999})\BibitemShut {NoStop}%
\bibitem [{\citenamefont {Ivchenko}(2005)}]{Ivchenko2005}%
  \BibitemOpen
  \bibfield  {author} {\bibinfo {author} {\bibfnamefont {E.~L.}\ \bibnamefont
  {Ivchenko}},\ }\href@noop {} {\emph {\bibinfo {title} {Optical Spectroscopy
  of Semiconductor Nanostructures}}}\ (\bibinfo  {publisher} {Alpha Science},\
  \bibinfo {address} {Harrow, UK},\ \bibinfo {year} {2005})\BibitemShut
  {NoStop}%
\bibitem [{\citenamefont {Anselm}(1981)}]{Anselm}%
  \BibitemOpen
  \bibfield  {author} {\bibinfo {author} {\bibfnamefont {A.}~\bibnamefont
  {Anselm}},\ }\href@noop {} {\emph {\bibinfo {title} {Introduction to
  Semiconductor Theory}}}\ (\bibinfo  {publisher} {Mir Publishers},\ \bibinfo
  {address} {Moscow},\ \bibinfo {year} {1981})\BibitemShut {NoStop}%
\bibitem [{\citenamefont {Rodina}\ \emph
  {et~al.}(2018{\natexlab{a}})\citenamefont {Rodina}, \citenamefont
  {Golovatenko}, \citenamefont {Shornikova},\ and\ \citenamefont
  {Yakovlev}}]{Rodina2018ftt}%
  \BibitemOpen
  \bibfield  {author} {\bibinfo {author} {\bibfnamefont {A.~V.}\ \bibnamefont
  {Rodina}}, \bibinfo {author} {\bibfnamefont {A.~A.}\ \bibnamefont
  {Golovatenko}}, \bibinfo {author} {\bibfnamefont {E.~V.}\ \bibnamefont
  {Shornikova}},\ and\ \bibinfo {author} {\bibfnamefont {D.~R.}\ \bibnamefont
  {Yakovlev}},\ }\bibfield  {title} {\bibinfo {title} {Spin physics of excitons
  in colloidal nanocrystals},\ }\href
  {https://doi.org/10.1134/S106378341808019X} {\bibfield  {journal} {\bibinfo
  {journal} {Phys. Solid State}\ }\textbf {\bibinfo {volume} {60}},\ \bibinfo
  {pages} {1537} (\bibinfo {year} {2018}{\natexlab{a}})}\BibitemShut {NoStop}%
\bibitem [{\citenamefont {Rodina}\ and\ \citenamefont
  {Efros}(2015)}]{Rodina2015}%
  \BibitemOpen
  \bibfield  {author} {\bibinfo {author} {\bibfnamefont {A.}~\bibnamefont
  {Rodina}}\ and\ \bibinfo {author} {\bibfnamefont {Al.~L.}\ \bibnamefont
  {Efros}},\ }\bibfield  {title} {\bibinfo {title} {{Magnetic properties of
  nonmagnetic nanostructures: dangling bond magnetic polaron in CdSe
  nanocrystals}},\ }\href {https://doi.org/10.1021/acs.nanolett.5b01566}
  {\bibfield  {journal} {\bibinfo  {journal} {Nano Lett.}\ }\textbf {\bibinfo
  {volume} {15}},\ \bibinfo {pages} {4214} (\bibinfo {year}
  {2015})}\BibitemShut {NoStop}%
\bibitem [{\citenamefont {Biadala}\ \emph {et~al.}(2017)\citenamefont
  {Biadala}, \citenamefont {Shornikova}, \citenamefont {Rodina}, \citenamefont
  {Yakovlev}, \citenamefont {Siebers}, \citenamefont {Aubert}, \citenamefont
  {Nasilowski}, \citenamefont {Hens}, \citenamefont {Dubertret}, \citenamefont
  {Efros},\ and\ \citenamefont {Bayer}}]{Biadala2017}%
  \BibitemOpen
  \bibfield  {author} {\bibinfo {author} {\bibfnamefont {L.}~\bibnamefont
  {Biadala}}, \bibinfo {author} {\bibfnamefont {E.~V.}\ \bibnamefont
  {Shornikova}}, \bibinfo {author} {\bibfnamefont {A.~V.}\ \bibnamefont
  {Rodina}}, \bibinfo {author} {\bibfnamefont {D.~R.}\ \bibnamefont
  {Yakovlev}}, \bibinfo {author} {\bibfnamefont {B.}~\bibnamefont {Siebers}},
  \bibinfo {author} {\bibfnamefont {T.}~\bibnamefont {Aubert}}, \bibinfo
  {author} {\bibfnamefont {M.}~\bibnamefont {Nasilowski}}, \bibinfo {author}
  {\bibfnamefont {Z.}~\bibnamefont {Hens}}, \bibinfo {author} {\bibfnamefont
  {B.}~\bibnamefont {Dubertret}}, \bibinfo {author} {\bibfnamefont {Al.~L.}\
  \bibnamefont {Efros}},\ and\ \bibinfo {author} {\bibfnamefont
  {M.}~\bibnamefont {Bayer}},\ }\bibfield  {title} {\bibinfo {title}
  {{{Magnetic}} polaron on dangling-bond spins in {{CdSe}} colloidal
  nanocrystals},\ }\href {https://doi.org/10.1038/nnano.2017.22} {\bibfield
  {journal} {\bibinfo  {journal} {Nat. Nanotechnol.}\ }\textbf {\bibinfo
  {volume} {12}},\ \bibinfo {pages} {569} (\bibinfo {year} {2017})}\BibitemShut
  {NoStop}%
\bibitem [{\citenamefont {Rodina}\ \emph
  {et~al.}(2018{\natexlab{b}})\citenamefont {Rodina}, \citenamefont
  {Golovatenko}, \citenamefont {Shornikova}, \citenamefont {Yakovlev},\ and\
  \citenamefont {Efros}}]{Rodina2018jem}%
  \BibitemOpen
  \bibfield  {author} {\bibinfo {author} {\bibfnamefont {A.~V.}\ \bibnamefont
  {Rodina}}, \bibinfo {author} {\bibfnamefont {A.~A.}\ \bibnamefont
  {Golovatenko}}, \bibinfo {author} {\bibfnamefont {E.~V.}\ \bibnamefont
  {Shornikova}}, \bibinfo {author} {\bibfnamefont {D.~R.}\ \bibnamefont
  {Yakovlev}},\ and\ \bibinfo {author} {\bibfnamefont {{\relax Al}.~L.}\
  \bibnamefont {Efros}},\ }\bibfield  {title} {\bibinfo {title} {{{Effect}} of
  {{Dangling Bond Spins}} on the {{Dark Exciton Recombination}} and {{Spin
  Polarization}} in {{CdSe Colloidal Nanostructures}}},\ }\href
  {https://doi.org/10.1007/s11664-018-6269-7} {\bibfield  {journal} {\bibinfo
  {journal} {J. Electron. Mater.}\ }\textbf {\bibinfo {volume} {47}},\ \bibinfo
  {pages} {4338} (\bibinfo {year} {2018}{\natexlab{b}})}\BibitemShut {NoStop}%
\end{thebibliography}
\end{document}